\documentclass[iop]{emulateapj}
\usepackage{lscape}
\usepackage{amsmath}

\newcommand{\CV}{\ion{C}{5}}
\newcommand{\CVI}{\ion{C}{6}}
\newcommand{\HI}{\ion{H}{1}}

\newcommand{\NVI}{\ion{N}{6}}
\newcommand{\NVII}{\ion{N}{7}}

\newcommand{\OI}{\ion{O}{1}}

\newcommand{\OVII}{\ion{O}{7}}
\newcommand{\OVIII}{\ion{O}{8}}

\newcommand{\Kalpha}{K$\alpha$}
\newcommand{\Kbeta}{K$\beta$}

\newcommand{\Kdelta}{K$\delta$}

\newcommand{\Lyalpha}{Ly$\alpha$}
\newcommand{\Lybeta}{Ly$\beta$}
\newcommand{\Lygamma}{Ly$\gamma$}
\newcommand{\Lydelta}{Ly$\delta$}

\newcommand{\NH}{\ensuremath{N_{\mathrm{H}}}}
\newcommand{\NHI}{\ensuremath{N(\mbox{\HI})}}

\newcommand{\angstrom}{\ensuremath{\mbox{\AA}}}
\newcommand{\nm}{\ensuremath{\mbox{\nm}}}
\newcommand{\cm}{\ensuremath{\mbox{cm}}}

\newcommand{\km}{\ensuremath{\mbox{km}}}

\newcommand{\pc}{\ensuremath{\mbox{pc}}}

\newcommand{\s}{\ensuremath{\mbox{s}}}

\newcommand{\ev}{\ensuremath{\mbox{eV}}}
\newcommand{\kev}{\ensuremath{\mbox{keV}}}

\newcommand{\erg}{\ensuremath{\mbox{erg}}}

\newcommand{\sr}{\ensuremath{\mbox{sr}}}

\newcommand{\K}{\ensuremath{\mbox{K}}}

\newcommand{\ph}{\ensuremath{\mbox{photons}}}
\newcommand{\counts}{\ensuremath{\mbox{counts}}}

\newcommand{\cmsq}{\ensuremath{\cm^2}}

\newcommand{\parcminsq}{\ensuremath{\mbox{arcmin}^{-2}}}
\newcommand{\pdegsq}{\ensuremath{\mbox{deg}^{-2}}}

\newcommand{\pcmsq}{\ensuremath{\cm^{-2}}}

\newcommand{\ps}{\ensuremath{\s^{-1}}}
\newcommand{\psr}{\ensuremath{\sr^{-1}}}

\newcommand{\emismeas}{\ensuremath{\cm^{-6}}\ \pc}

\newcommand{\flux}{\erg\ \pcmsq\ \ps}
\newcommand{\lineunit}{\ph\ \pcmsq\ \ps\ \psr}

\newcommand{\LU}{\ensuremath{\mbox{L.U.}}}
\newcommand{\kmps}{\km\ \ps}

\newcommand{\rassrate}{\counts\ \ps\ \parcminsq}

\newcommand{\chandra}{\textit{Chandra}}

\newcommand{\iras}{\textit{IRAS}}

\newcommand{\rosat}{\textit{ROSAT}}

\newcommand{\suzaku}{\textit{Suzaku}}

\newcommand{\xmm}{\textit{XMM-Newton}}

\newcommand{\citepossessive}[1]{\citeauthor{#1}'s \citeyearpar{#1}}

\newcommand{\chisq}{\ensuremath{\chi^2}}
\newcommand{\rchisq}{\ensuremath{\chi^2_\nu}}

\newcommand{\raymondsmith}{\citeauthor{raymond77} (\citeyear{raymond77} and updates)}
\providecommand{\eqref}[1]{equation~(\ref{#1})}

\newcommand{\esas}{\textit{XMM}-ESAS}

\newcommand{\Tfg}{\ensuremath{T_\mathrm{fg}}}
\newcommand{\EMfg}{\ensuremath{\mathcal{E}_\mathrm{fg}}}
\newcommand{\Sfg}{\ensuremath{S_\mathrm{fg}^{0.4-1.0}}}
\newcommand{\Th}{\ensuremath{T_\mathrm{h}}}
\newcommand{\EMh}{\ensuremath{\mathcal{E}_\mathrm{h}}}
\newcommand{\Sh}{\ensuremath{S_\mathrm{h}^{0.5-2.0}}}
\newcommand{\Son}{\ensuremath{S_\mathrm{on}}}
\newcommand{\Soff}{\ensuremath{S_\mathrm{off}}}
\newcommand{\SFG}{\ensuremath{S_\mathrm{fg}}}
\newcommand{\SH}{\ensuremath{S_\mathrm{h}}}
\newcommand{\SEG}{\ensuremath{S_\mathrm{eg}}}
\newcommand{\NHon}{\ensuremath{N_\mathrm{H,on}}}
\newcommand{\NHoff}{\ensuremath{N_\mathrm{H,off}}}
\newcommand{\Cplus}[1]{\ensuremath{\mathrm{C}^{#1+}}}
\newcommand{\Oplus}[1]{\ensuremath{\mathrm{O}^{#1+}}}
\newcommand{\pvalue}{\ensuremath{p~\mathrm{value}}}

\def\ApjSectionpenalty{10000}

\shorttitle{\textit{XMM-NEWTON} AND \textit{SUZAKU} SHADOWING OBSERVATIONS}
\shortauthors{HENLEY AND SHELTON}

\begin{document}

\title{\textit{XMM-Newton} and \textit{Suzaku} X-ray Shadowing Measurements of the \\
  Solar Wind Charge Exchange, Local Bubble, and Galactic Halo Emission}
\author{David B. Henley and Robin L. Shelton}
\affil{Department of Physics and Astronomy, University of Georgia, Athens, GA 30602; dbh@physast.uga.edu, rls@physast.uga.edu}

\begin{abstract}
We present results from a sample of \xmm\ and \suzaku\ observations of interstellar clouds that cast
shadows in the soft X-ray background (SXRB)---the first uniform analysis of such a sample
from these missions.  By fitting to the on- and off-shadow spectra, we separated
the foreground and Galactic halo components of the SXRB.
We tested different foreground models---two solar wind charge exchange (SWCX)
models and a Local Bubble (LB) model. We also examined different abundance tables.
We found that \citet{anders89} abundances, commonly used in previous SXRB studies, may result in
overestimated foreground brightnesses and halo temperatures. We also found that assuming
a single solar wind ionization temperature for a SWCX model can lead to unreliable results.
We compared our measurements of the foreground emission with predictions of the SWCX emission from a
smooth solar wind, finding only partial agreement. Using available observation-specific SWCX
predictions and various plausible assumptions, we placed an upper limit on the LB's
\OVII\ intensity of $\sim$0.8~\lineunit\ (90\% confidence).
Comparing the halo results obtained with SWCX and LB foreground models implies that, if the
foreground is dominated by SWCX and is brighter than $\sim$$1.5\times10^{-12}~\flux\ \pdegsq$
(0.4--1.0~\kev), then using an LB foreground model may bias the halo temperature upward and the
0.5--2.0~\kev\ surface brightness downward by $\sim$$(\mbox{0.2--0.3})\times10^6~\K$ and
$\sim$$(\mbox{1--2})\times10^{-12}~\flux\ \pdegsq$, respectively.
Similarly, comparing results from different observatories implies that there may be
uncertainties in the halo temperature and surface brightness of up to $\sim$$0.2\times10^6~\K$ and
$\sim$25\%, respectively, in addition to the statistical uncertainties.
These uncertainties or biases may limit the ability of X-ray measurements to discriminate between
Galactic halo models.
\end{abstract}


\section{INTRODUCTION}
\label{sec:Introduction}

Several emission components contribute to the diffuse soft X-ray background (SXRB) emission that is
observed in all directions on the sky \citep{mccammon90}. The closest, and most recently discovered,
source is solar wind charge exchange (SWCX) emission, which arises from charge exchange (CX) reactions
between solar wind ions and neutral H and He in the heliosphere or the Earth's exosphere
\citep{cravens00,robertson03a,robertson03b,koutroumpa06}. Beyond that, there may be emission from
the Local Bubble (LB), a cavity in the local interstellar medium (ISM) thought to be filled with
$\sim$$1 \times 10^6~\K$ gas (\citealt{sanders77,snowden90}; although see also
\citealt{welsh09}). Further out still, beyond the main Galactic disk, there is emission from
$\sim$$\mbox{(1--3)} \times 10^6~\K$ gas in the Galactic halo (\citealt{burrows91,wang95,kuntz00};
\citealt{yoshino09}, hereafter Yosh09; \citealt{henley13}, hereafter HS13)---this gas is also
observed via X-ray absorption lines
\citep[e.g.,][]{nicastro02,rasmussen03,yao07a,gupta12,miller13}. The most distant contributor to the
SXRB is the extragalactic background of unresolved active galactic nuclei \citep[AGN;
  e.g.,][]{brandt05}.

In order to test models for the local (SWCX and/or LB) and halo contributions to the SXRB, it is
necessary to separate their emission.\footnote{Dealing with the extragalactic background component
  of the SXRB is relatively simple, as its spectrum is well characterized as a power law
  \citep[e.g.,][]{chen97}, albeit with evidence of steepening below
  $\sim$1~\kev\ \citep{roberts01}.} To this end, one can use observations of interstellar clouds
that cast shadows in the SXRB by partially absorbing the distant emission
\citep{burrows91,snowden91}. By comparing the X-ray emission observed toward and to the side of such
a shadow, one can determine the contributions to the emission from the foreground and from the hot
halo.

Such shadowing observations were first carried out with the low-spectral-resolution proportional
counters on board \rosat\ \citep{burrows91,snowden91,snowden93,wang95,kuntz97,snowden00}. More
recently, the CCD spectrometers on board \xmm\ and \suzaku\ have been used for such observations
\citep{smith07a,galeazzi07,henley07,henley08a,gupta09b,lei09,henley15a}. These instruments have
higher spectral resolution, allowing some line emission features (e.g., \OVII\ and \OVIII) to be
resolved in the spectra. However, the CCD data are only usable above $\sim$0.3 or $\sim$0.4~\kev,
whereas \rosat\ could observe in the so-called 1/4~\kev\ band ($\sim$0.1--0.284~\kev). This means
fewer shadowing targets are available to \xmm\ and \suzaku---as the photoelectric absorption cross
section decreases with increasing photon energy, fewer clouds are sufficiently optically thick in
the \xmm/\suzaku\ band that the contrast between the on- and off-cloud spectra allows one to
separate the foreground and halo emission.

The shadowing clouds that have been observed with \xmm\ and/or \suzaku\ have typically yielded halo
temperatures and emission measures of $\sim$$2 \times 10^6$~K and a few times $10^{-3}~\emismeas$,
respectively \citep{smith07a,galeazzi07,gupta09b,lei09,henley15a}. However, these observations were
not analyzed in a homogeneous fashion. In addition, the sensitivity of the halo results to the
assumed foreground model was not tested. Furthermore, some of these studies used older abundance
tables \citep[e.g.,][]{anders89}, in which the carbon, nitrogen, and oxygen abundances are now known
to be too high for the sun and the local interstellar medium \citep{asplund09,wilms00}. Using these
older abundances may bias the results (see Section~\ref{subsec:ResultsDifferentAbundances}).

Here, we analyze as large a set as possible of \xmm\ and \suzaku\ shadowing observations in a
uniform fashion. We test different foreground models in order to examine the sensitivity of the halo
results to the choice of foreground model. Recent studies have argued that a combination of SWCX and
LB emission is needed to explain the foreground emission in the 1/4~\kev\ \citep{smith14,galeazzi14}
and \xmm/\suzaku\ \citep{koutroumpa11} bands. However, the relative contributions of the LB and of
SWCX to the foreground in an arbitrary \xmm\ or \suzaku\ observation are not known.  Therefore, we
follow \citet{henley15a}, and consider limits in which either LB or SWCX emission dominate the
foreground. (Note that in most previous shadowing studies, an LB-like spectral model was used for
the foreground.) We also examine the sensitivity of our results to the choice of abundance table
used in the spectral analysis.

\begin{deluxetable*}{lllrcrrccc}
\tablecaption{Observation Details\label{tab:Observations}}
\tablehead{
                 &                       & \colhead{On or} &                   & \colhead{Start} &                 &                 & \colhead{Nominal}  & \colhead{Usable}   & \\
\colhead{Shadow} & \colhead{Observatory} & \colhead{off}   & \colhead{ObsID}   & \colhead{date}  & \colhead{$l$}   & \colhead{$b$}   & \colhead{exposure} & \colhead{exposure} & \colhead{\NH} \\
                 &                       &                 &                   &                 & \colhead{(deg)} & \colhead{(deg)} & \colhead{(ks)}     & \colhead{(ks)}     & \colhead{($10^{20}~\pcmsq$)} \\
\colhead{(1)}    & \colhead{(2)}         & \colhead{(3)}   & \colhead{(4)}     & \colhead{(5)}   & \colhead{(6)}   & \colhead{(7)}   & \colhead{(8)}      & \colhead{(9)}      & \colhead{(10)}
}
\startdata
G048+37  & \xmm    &  On & 0670900501 & 2012-02-01 &  47.74 & $+37.28$ &  33.0 & 19.9,21.6,12.8 &  9.3 \\
         &         & Off & 0670900201 & 2012-01-30 &  48.97 & $+40.38$ &  33.0 & 19.1,19.4,10.8 &  2.2 \\
MBM 12   & \suzaku &  On &  500015010 & 2006-02-03 & 159.21 & $-34.47$ & 102.9 &           68.6 & 32.2 \\
         &         & Off &  501104010 & 2006-02-06 & 157.36 & $-36.82$ &  75.3 &           51.1 &  3.8 \\
MBM 16   & \suzaku &  On &  508078010 & 2013-08-07 & 170.58 & $-37.28$ &  82.3 &           49.5 & 24.1 \\
         &         & Off &  508073010 & 2013-08-09 & 165.84 & $-38.39$ &  83.1 &           50.6 &  6.3 \\
MBM 20   & \xmm    &  On & 0203900201 & 2004-08-23 & 211.37 & $-36.57$ & 101.1 & 54.1,59.7,29.3 & 17.6 \\
         &         & Off & 0203900101 & 2004-08-09 & 213.40 & $-39.11$ & 109.4 & 72.8,74.3,62.8 &  2.3 \\
MBM 20   & \suzaku &  On &  502075010 & 2008-02-11 & 211.41 & $-36.56$ & 107.1 &           71.8 & 19.1 \\
         &         & Off &  502076010 & 2007-07-30 & 213.42 & $-39.10$ & 103.8 &           83.6 &  2.2 \\
G236+38  & \suzaku &  On &  506055010 & 2011-06-01 & 235.95 & $+38.21$ &  69.8 &           41.8 & 12.2 \\
         &         & Off &  506056010 & 2011-06-07 & 237.09 & $+41.11$ &  70.8 &           41.6 &  1.4 \\
Filament & \xmm    &  On & 0084960201 & 2002-05-03 & 278.67 & $-45.32$ &  10.0 &  11.8,11.8,7.9 &  7.6 \\
         &         & Off & 0084960101 & 2002-05-03 & 278.73 & $-47.09$ &  25.1 &    4.5,4.1,0.9 &  2.0 \\
Filament & \suzaku &  On &  501002010 & 2006-03-03 & 278.64 & $-45.30$ & 101.5 &           72.4 &  8.3 \\
         &         & Off &  501001010 & 2006-03-01 & 278.69 & $-47.07$ &  80.1 &           60.9 &  2.0
\enddata
\tablecomments{The shadows are tabulated in order of increasing Galactic longitude.
  Columns 6 and 7 contain the observation pointing direction in Galactic coordinates.
  The nominal exposure (column~8) is the exposure time taken from the header of the unfiltered events list.
  The usable exposure (column~9) is the exposure time taken from the header of the spectrum file, extracted
  after the data have been filtered (Sections~\ref{subsec:XMMReduction} and \ref{subsec:SuzakuReduction}).
  For the \xmm\ observations, the listed values are the usable MOS1, MOS2, and pn exposures, respectively.
  The absorbing column density, \NH\ (column~10), was calculated from the average 100-micron intensity within
  the field of view (see Section~\ref{sec:SpectralModel}).}
\end{deluxetable*}

The remainder of the paper is arranged as follows. In Section~\ref{sec:Reduction} we describe our
observation selection and data reduction, and we describe our spectral model in
Section~\ref{sec:SpectralModel}.
We present our results in Section~\ref{sec:Results}---in
particular, we examine the results obtained with different foreground models
(Section~\ref{subsec:ResultsDifferentForegrounds}) and with different abundance tables
(Section~\ref{subsec:ResultsDifferentAbundances}). We also compare our results with those from
previously published shadowing analyses (Section~\ref{subsec:CompareWithPrevious}).
We discuss our results in Section~\ref{sec:Discussion}. We first discuss which of the foreground
models we examined is our preferred model, and why (Section~\ref{subsec:ForegroundChoice}).  We then
compare our foreground measurements with heliospheric SWCX model predictions
(Section~\ref{subsec:SWCX}). We also use our foreground measurements to place limits on the emission
from the LB (Section~\ref{subsec:LB}).
Finally, we discuss our halo measurements in Section~\ref{subsec:HaloDiscussion}, including
discussing possible uncertainties or biases that may arise from uncertainties in the foreground
emission and from other aspects of the spectral modeling
(Section~\ref{subsubsec:HaloUncertaintyBias}), and discussing the use of such halo measurements in
testing Galactic halo models (Section~\ref{subsubsec:TestingHaloModels}). We conclude with a summary
in Section~\ref{sec:Summary}.

\section{OBSERVATION SELECTION AND DATA REDUCTION}
\label{sec:Reduction}

Our sample includes, but is not limited to, all previously published \xmm\ and \suzaku\ shadowing
observations that consist of separate on- and off-shadow pointings. The targets of these
observations are MBM~12 (observed with \suzaku; \citealt{smith07a}), MBM~20 (observed with \xmm\ and
\suzaku; \citealt{galeazzi07,gupta09b}), and an unnamed dusty filament in the southern Galactic
hemisphere, hereafter dubbed ``filament'' (observed with \xmm\ and \suzaku;
\citealt{henley07,henley08a,lei09}).  Although MBM~12 has also been observed with
\xmm\ \citep{koutroumpa11}, we did not include these data in our sample because there is not a
corresponding blank-sky off-cloud pointing.  Furthermore, we have previously found that the
on-MBM~12 observation analyzed by \citet{koutroumpa11} was too badly contaminated by soft protons to
be included in our \xmm\ survey of the SXRB \citep{henley12b}.

To the above set of observations, we added a pair of unpublished \suzaku\ observations of MBM~16
(preliminary results from these observations have been presented by \citealt{ursino14}).  Our sample
also includes previously unpublished observations of two clouds in the northern Galactic hemisphere,
which we refer to as G048+37 (observed with \xmm) and G236+38 (observed with \suzaku). These clouds
were identified as viable shadowing targets from the DIRBE-corrected \iras\ maps of the diffuse
100-\micron\ intensity, $I_{100}$ \citep{schlegel98}. In particular, we sought pairs of directions
within $\approx$3\degr\ of each other with absorbing column densities, \NH, of $\ga$$10 \times
10^{20}$ and $\la$$2 \times 10^{20}~\pcmsq$ (from the \citealt{snowden00}
$I_{100}$-to-\NH\ conversion relation). With such a column density contrast we expected to be able
to adequately separate the foreground and halo emission with reasonable-length pointings.  These
observations are the first published CCD-resolution shadowing observations from the northern
Galactic hemisphere.

Another potential target in the northern hemisphere is MBM~36 ($(l,b)=(4\fdg2,+35\fdg8)$), for which
a pair of on- and off-cloud observations is available in the \suzaku\ archive (ObsIDs 508079020 and
508074010, respectively). However, this cloud lies in the direction of Scorpius-Centaurus (Sco-Cen)
superbubble \citep{egger95}, the X-ray emission from which would complicate the analysis of the SXRB
spectra, making it difficult to isolate the foreground and halo components. We therefore excluded
MBM~36 from our sample.

\begin{figure}
\centering
\plotone{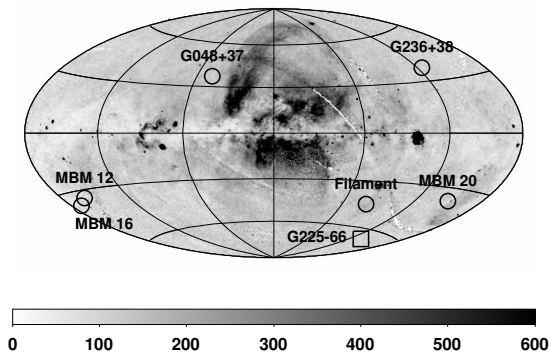}
\caption{\rosat\ All-Sky Survey map of the 3/4~\kev\ SXRB (units: $10^{-6}~\rassrate$;
  \citealt{snowden97}), centered on the Galactic Center, showing the locations of the shadows
  analyzed in this paper (circles), and of shadow G225$-$66 (square; \citealt{henley15a}).
  \label{fig:ShadowMap}}
\end{figure}

\begin{figure*}
\centering
\plottwo{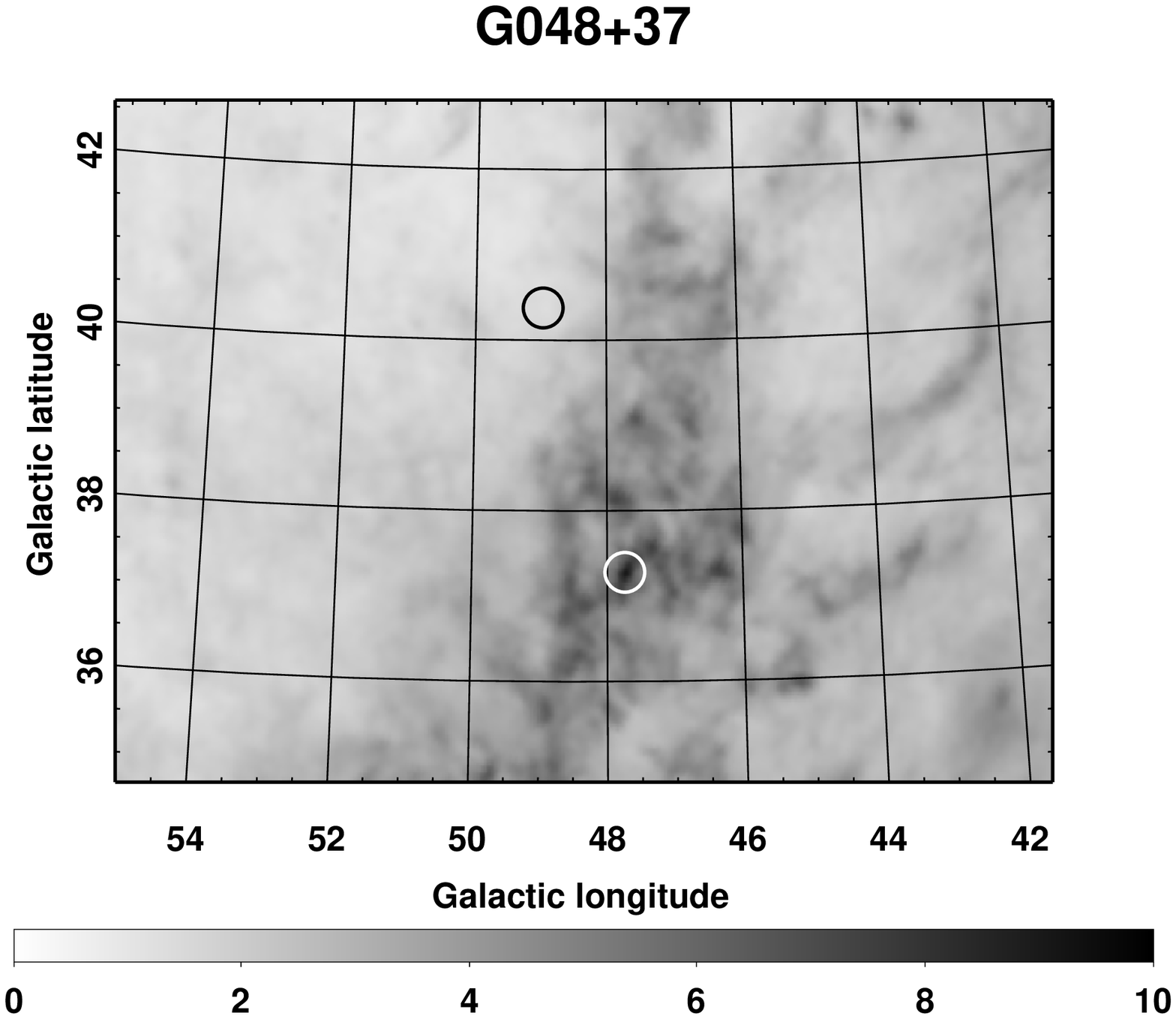}{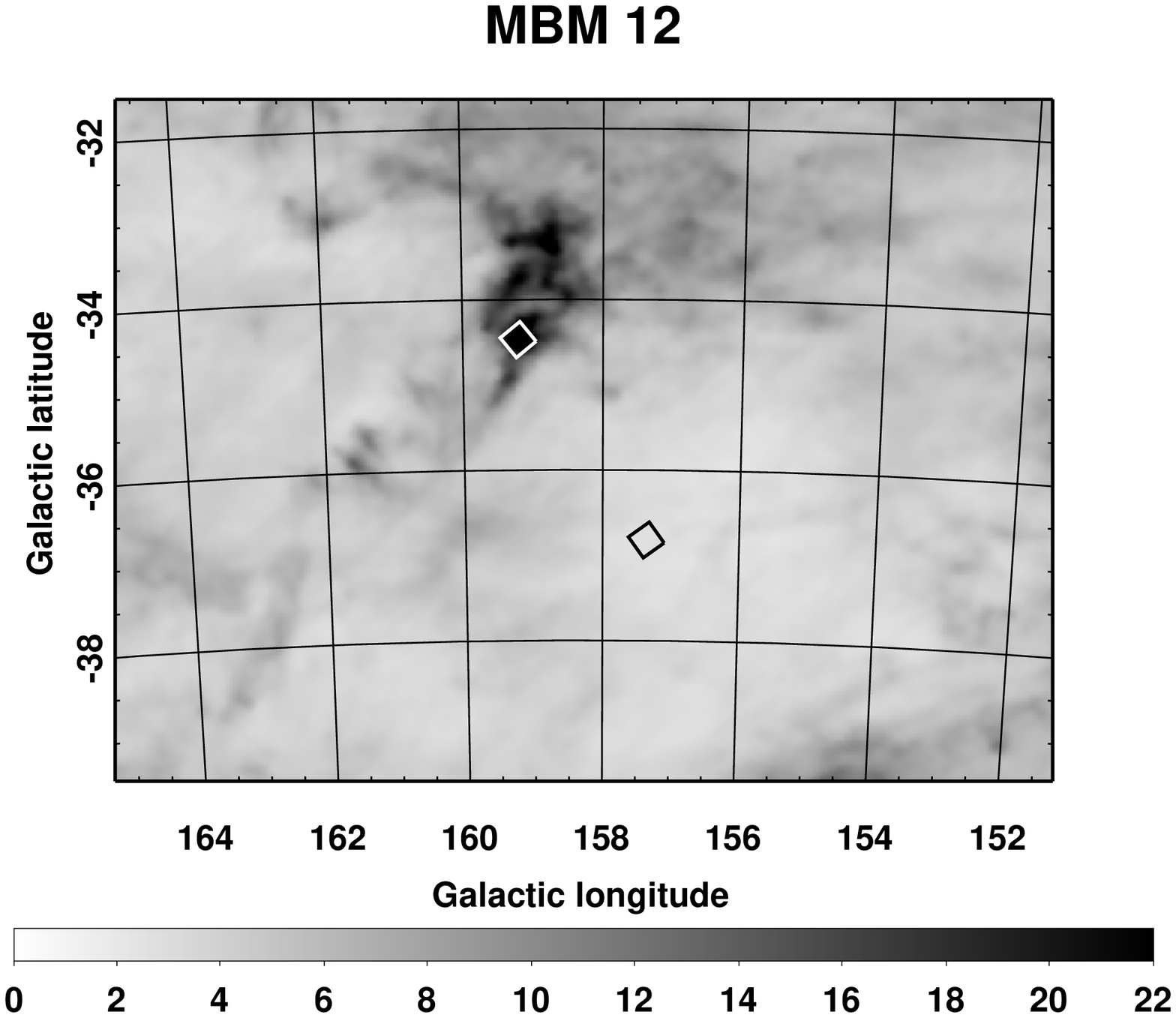} \\
\plottwo{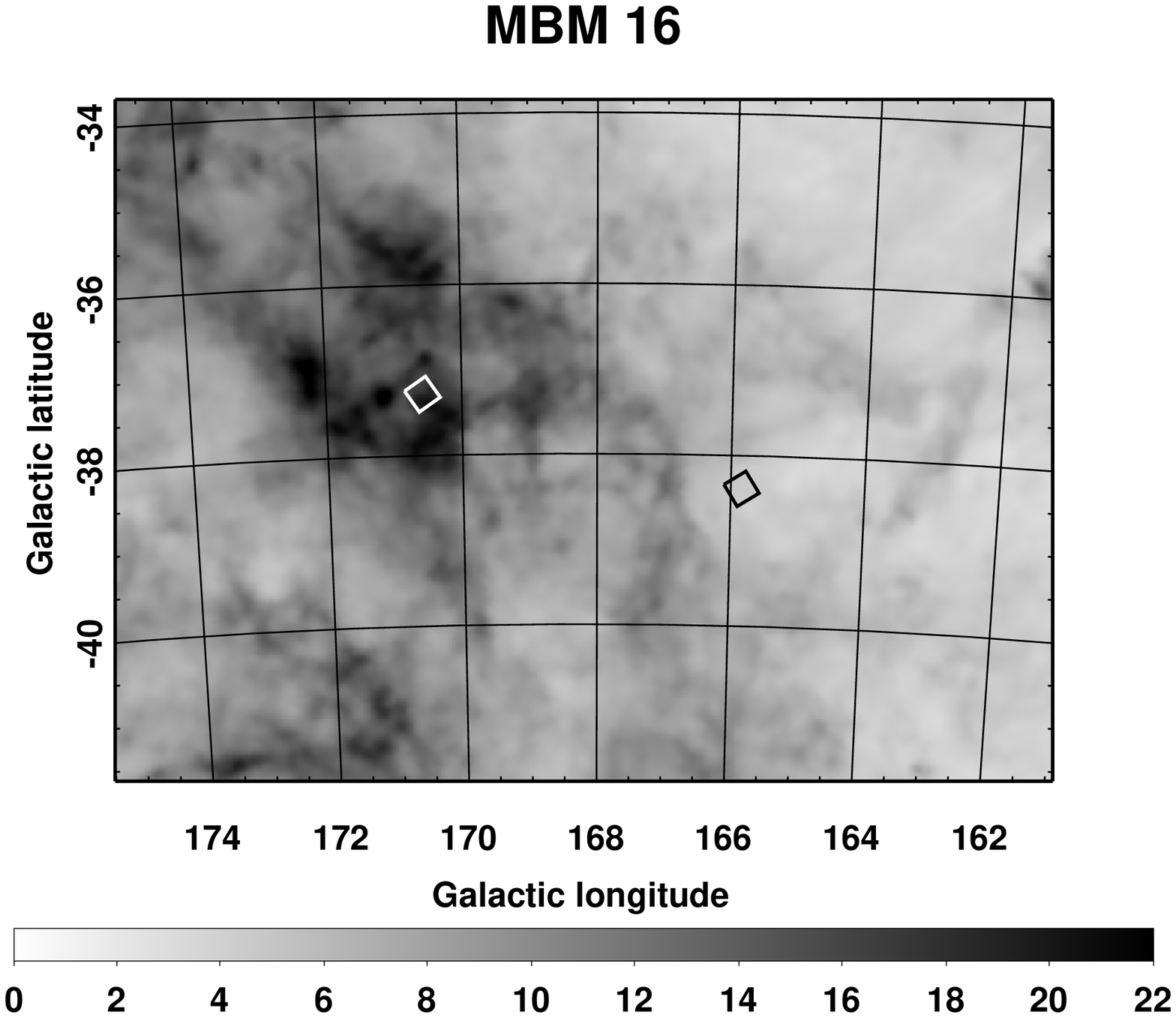}{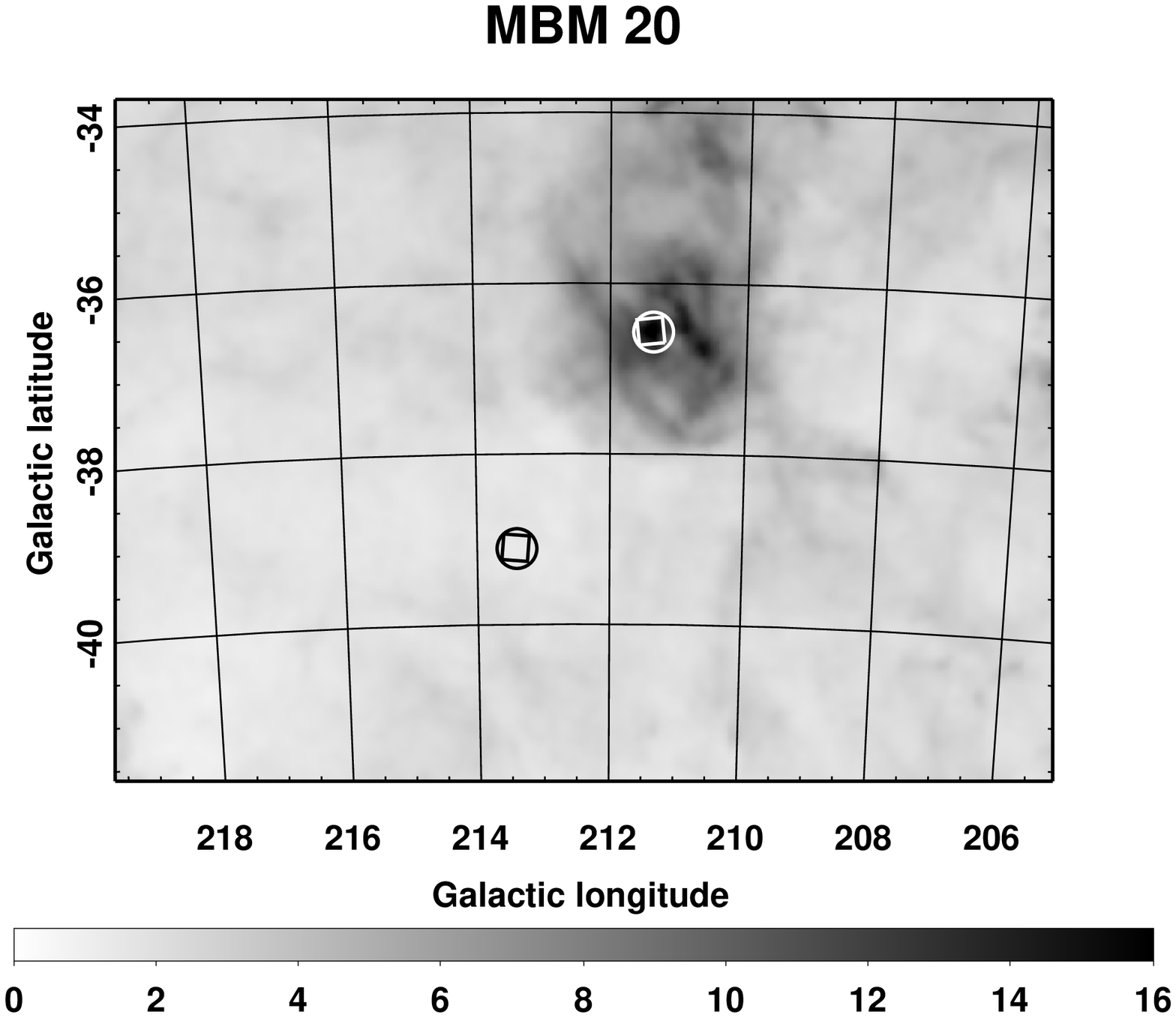} \\
\plottwo{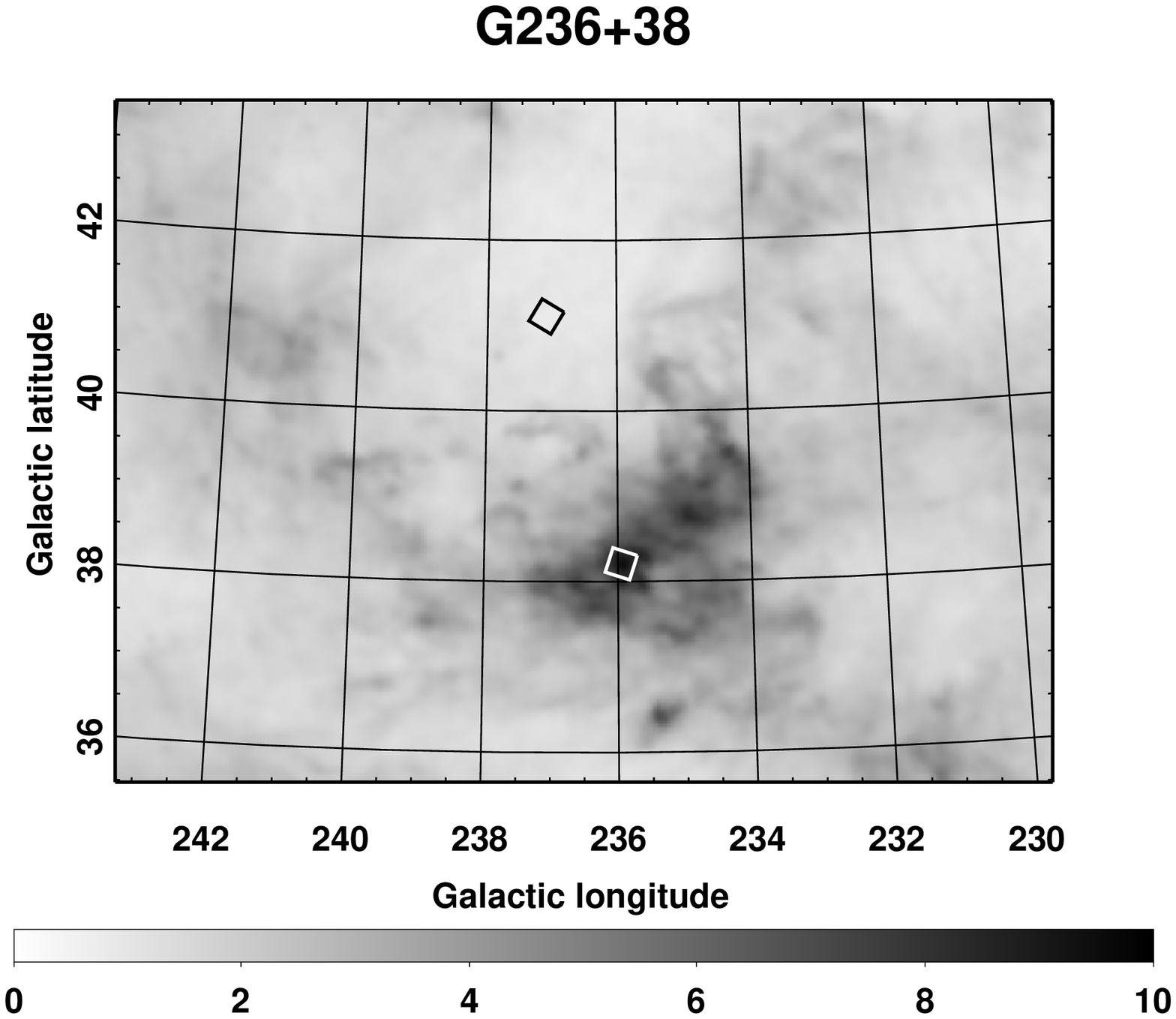}{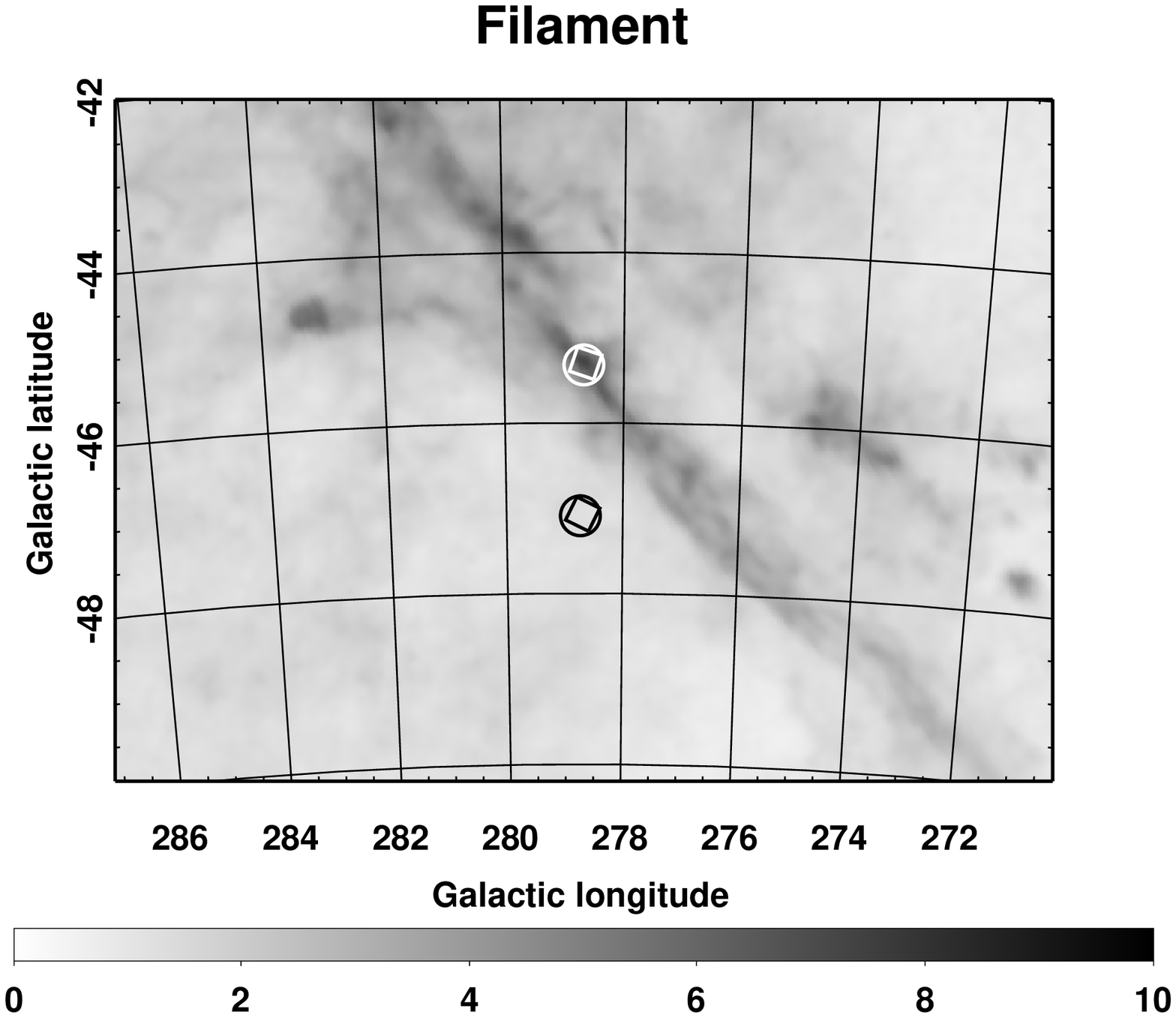}
\caption{DIRBE-corrected \iras\ 100-micron maps of the shadowing clouds studied here
  \citep{schlegel98}. The grayscales are in MJy \psr. The circles ($\mathrm{radius} = 14\arcmin$)
  and squares ($18\arcmin \times 18\arcmin$) indicate the approximate \xmm\ and \suzaku\ fields of
  view, respectively.
  \label{fig:ShadowImages}}
\end{figure*}

The details of our \xmm\ and \suzaku\ shadowing observations are shown in
Table~\ref{tab:Observations}. Figure~\ref{fig:ShadowMap} shows the locations of these shadows on the
sky. This figure also shows the location of shadow G225.60$-$66.40 (G225$-$66 hereafter), analyzed
by \citet{henley15a}.  Unlike the shadows analyzed here, the on- and off-shadow spectra for
G225$-$66 were extracted from a single \xmm\ field. However, the subsequent analysis of the spectra
was essentially the same as that employed here (Section~\ref{sec:SpectralModel}). We will include
the results from G225$-$66 in our discussion (Section~\ref{sec:Discussion}). 100-micron maps of the
shadowing clouds are shown in Figure~\ref{fig:ShadowImages}. The following two subsections describe
the data reduction for the \xmm\ and \suzaku\ observations, respectively.

\subsection{\xmm\ Data Reduction}
\label{subsec:XMMReduction}

The \xmm\ data reduction and spectral extraction, which we carried out with the \xmm\ Extended
Source Analysis Software\footnote{http://heasarc.gsfc.nasa.gov/docs/xmm/xmmhp\_xmmesas.html} (\esas;
\citealt{snowden13}) as distributed with version 13.5.0 of the Science Analysis
System\footnote{http://xmm.esac.esa.int/sas/} (SAS), was generally similar to that described in
\citet{henley12b}. However, note that here we used data from the pn camera \citep{struder01}, in
addition to data from the MOS cameras \citep{turner01}.  For each observation, we first used the
standard SAS \texttt{emchain} and \texttt{epchain} scripts to produce calibrated events lists from
each camera, and then used the \esas\ \texttt{mos-filter} and \texttt{pn-filter} scripts to remove
from the data periods of elevated count rate, due to soft-proton flaring. The amounts of good time
remaining after this filtering are shown in column~9 of Table~\ref{tab:Observations}. Note that the
resulting pn exposures are shorter than the corresponding MOS exposures. As discussed in
\citet{henley14b}, this is likely due to the greater sensitivity of the pn detector, meaning that
relatively smaller departures from the mean count rate are flagged as soft proton flares.

For most observations we used data from the Second \xmm\ Serendipitous Source Catalog
(2XMM\footnote{Specifically, we used data from the 2XMMi DR3 data release;
  http://xmmssc-www.star.le.ac.uk/Catalogue/2XMMi-DR3/ \label{fn:2XMM}}; \citealt{watson09}) to
identify point sources with 0.5--2.0~keV fluxes exceeding $1 \times 10^{-14}~\flux$. This is a lower
source removal threshold than that used in \citet{henley12b}, but it is the threshold used in our
\xmm\ survey of the halo emission (HS13). However, the observations of G048+37 are not included in
the 2XMM catalog, and so for these observations we ran the source detection ourselves, using the SAS
\texttt{edetect\_chain} script. In all cases, the detected sources were excised from the data using
circles of radius 50\arcsec. A visual inspection indicated that there were no bright or extended
sources in the fields that would have needed larger source removal regions.

We identified and excluded MOS CCDs that exhibited the anomalous state identified by
\citet{kuntz08a}.  Also, we found that, for the on-MBM~20 observation (0203900201), the exposure
time for the first (upper left, in detector coordinates) quadrant of the pn detector is
$\sim$40\%\ of those for the other three quadrants. It is unclear why this is the case. We decided
to exclude this quadrant of the pn detector from the subsequent processing and analysis of this
observation.

Having excluded point sources and anomalous CCDs, we used the \esas\ \texttt{mos-spectra} and
\texttt{pn-spectra} scripts to extract SXRB spectra from the full remaining fields of view of the
MOS and pn cameras, respectively, and used the \texttt{mos\_back} and \texttt{pn\_back} programs to
calculate corresponding quiescent particle background (QPB) spectra. The QPB spectra were calculated
from a database of filter-wheel-closed data, scaled to our observations using data from the
unexposed regions of the cameras \citep{kuntz08a}.  The spectral extraction scripts also calculated
the response files---the redistribution matrix file (RMF) and ancillary response file (ARF)---needed
for each spectrum, using the SAS \texttt{rmfgen} and \texttt{arfgen} programs, respectively. Before
carrying out our spectral analysis, we grouped each SXRB spectrum such that there were at least 50
counts per bin, and then subtracted the corresponding QPB spectrum.

\subsection{\suzaku\ Data Reduction}
\label{subsec:SuzakuReduction}

For the \suzaku\ observations, we used only the data from the back-illuminated (BI) XIS1 camera,
which is more sensitive to soft X-rays than the front-illuminated (FI) XIS cameras \citep{koyama07}.
We processed the \suzaku\ data using HEASoft\footnote{http://heasarc.gsfc.nasa.gov/lheasoft/}
version 6.15.1. For each observation, we first ran the \texttt{aepipeline} script to produce a
calibrated and screened XIS1 events list. We then combined the data taken in the $3 \times 3$ and $5
\times 5$ modes, and applied further screening to the data, in addition to the standard screening
described in the \suzaku\ ABC
Guide.\footnote{http://heasarc.gsfc.nasa.gov/docs/suzaku/analysis/abc/abc.html} In particular, in
order to reduce the non-X-ray background (NXB), we kept only times when the geomagnetic cut-off
rigidity (COR) exceeded 8~GV, and in order to reduce contamination from solar X-rays scattered off
the Earth's atmosphere, we kept only times when the angle between \suzaku's sight line and the limb
of the Earth, ELV, exceeded 10\degr\ (cf.\ the standard screening criterion is $\mathrm{ELV} >
5\degr$). For most observations, we applied the standard screening for the angle between \suzaku's
sight line and the limb of the sunlit Earth, DYE\_ELV, and kept only times when $\mathrm{DYE\_ELV} >
20\degr$. However, \citet{sekiya14} showed that, with this DYE\_ELV threshold, \OI\ emission at
$E=0.525~\kev$ from the fluorescence of atmospheric oxygen can significantly contaminate the
astrophysical \OVII\ \Kalpha\ emission at $E\approx0.57~\kev$ in \suzaku\ spectra taken after
2011. This contamination can be removed by increasing the DYE\_ELV threshold from 20\degr\ to
60\degr. As our G236+38 observations were taken in mid-2011, near the time of the increased
atmospheric \OI\ contamination reported by \citet{sekiya14}, we decided to err on the side of
caution and applied this stricter DYE\_ELV screening criterion to the G236+38 observations.  We also
applied this stricter screening to the MBM~16 observations, which were taken in 2013.  For both sets
of observations, this stricter screening reduced the usable exposure time by about one quarter.  The
usable exposures remaining after the screening of the \suzaku\ data are shown in column~9 of
Table~\ref{tab:Observations}.

We have previously found that automated source detection software does not work well on
\suzaku\ images, presumably because of \suzaku's broad point-spread function \citep{henley08a}.  We
therefore excluded sources from the \suzaku\ data by hand. For the MBM~12 observations, we used the
same source exclusion regions described in \citet{smith07a}. For the filament and MBM~20
observations, we used a similar procedure to that described in \citet{henley08a}, though here we
used more stringent source exclusion thresholds. We used data from the 2XMM source catalog
\citep{watson09}\footnote{See footnote~\ref{fn:2XMM}.} to identify sources in the \suzaku\ fields
with 0.5--2.0~\kev\ fluxes exceeding $2 \times 10^{-14}~\flux$. Sources with fluxes of
$\mbox{(2--10)} \times 10^{-14}$ and $>$$1 \times 10^{-13}~\flux$ were excised from the data using
circles of radius 1\farcm5 and 2\farcm5, respectively. The MBM~16 and G236+38 observations are not
covered by the 2XMM catalog. We identified by eye one bright source each in the on- and off-MBM~16
observations, at $(\alpha,\delta) = (03^\mathrm{h}19^\mathrm{m}09^\mathrm{s},
+11\degr43\arcmin15\arcsec)$ and $(03^\mathrm{h}04^\mathrm{m}12^\mathrm{s},
+13\degr07\arcmin42\arcsec)$, respectively. We excluded these sources with 2\arcmin\ circles.  We
identified by eye two bright sources in the on-G236+38 observation, at $(\alpha,\delta) =
(09^\mathrm{h}46^\mathrm{m}02^\mathrm{s}, +00\degr35\arcmin25\arcsec)$ and
$(09^\mathrm{h}46^\mathrm{m}34^\mathrm{s}, +00\degr28\arcmin07\arcsec)$. We excluded these sources
with 4\arcmin\ circles (for both MBM~16 and G236+38, the source exclusion radii were chosen by
eye). No bright point sources were apparent in the off-G236+38 observation.

Although we identified the sources to be excluded from our \suzaku\ observations using sky (J2000)
coordinates, we carried out the subsequent spectral extraction, NXB calculation, and ARF calculation
(see below) using XIS detector coordinates. Therefore, we converted the sources' coordinates to
detector coordinates using \texttt{aecoordcalc}. The conversion from sky to detector coordinates
varies as a function of time, depending on the spacecraft's attitude. We carried out the conversion
using the mid-time of each observation. For a few sources, the resulting source exclusion region in
detector coordinates was not perfectly aligned with the image of the source. In such cases, we
adjusted the location of the exclusion region by hand---these adjustments were typically
$\la$20\arcsec.

We extracted SXRB spectra from the full field of view of the XIS1 camera, minus the excluded sources
described above. We did not exclude the regions of the detector illuminated by the Fe-55 calibration
sources, as the manganese K lines produced by these sources lie above the energy band we
are interested in.\footnote{Manganese also produces L-shell lines at
    $\sim$640~\ev\ \citep{kortright09}, within our energy band of interest.  However, while the
    \suzaku\ NXB spectra clearly exhibit the manganese K lines, the L lines are not apparent,
    implying that contamination from these lines is not significant.}  As with the \xmm\ spectra,
we grouped each \suzaku\ SXRB spectrum such that were at least 50 counts per bin.

From each SXRB spectrum we subtracted a corresponding NXB spectrum, calculated using
\texttt{xisnxbgen} \citep{tawa08}.  The NXB spectra were constructed from night-Earth observations
with the same COR distributions as the SXRB observations. We also calculated the RMF and ARF needed
for each spectrum, using \texttt{xisrmfgen} and \texttt{xissimarfgen} \citep{ishisaki07},
respectively. The latter tool takes into account the contamination on the XIS1 detector's optical
blocking filter, which reduces the detector's low-energy sensitivity \citep{koyama07}. For the ARF
calculation, we assumed a uniform source of radius 20\arcmin.

\section{SPECTRAL MODEL DESCRIPTION}
\label{sec:SpectralModel}

In order to separate the foreground and halo emission, we used XSPEC version 12.8.1l
\citep{arnaud96} to fit an SXRB spectral model to the on- and off-shadow spectra from each pair of
observations in Table~\ref{tab:Observations}. Our spectral model is essentially the same as that
used by \citet{henley15a}, and consists of components representing (1) the foreground emission, (2)
the Galactic halo emission, (3) the extragalactic background emission, and (4) (for \xmm\ spectra
only) components of the particle background not removed by the filtering and QPB subtraction
described in Section~\ref{subsec:XMMReduction}. In most regards, the on- and off-shadow models for
each pair of observations were the same, apart from the absorbing column used to attenuate the halo
and extragalactic components---this difference between the on- and off-shadow models is the key to
separating the foreground and halo emission.

We briefly describe the model components below, referring the reader to \citet{henley15a} for more
details. We also describe a modification to our basic spectral model that was necessary for
the on-MBM~12 \suzaku\ spectrum.

(1) We examined three different foreground models, representing limits in which SWCX emission or LB
emission dominate the foreground. For our reference spectral model, we used the C14-SWCX model from
\citet{henley15a}, which is based on CX line ratio data from \citet{cumbee14}.  This model consisted
of \Kalpha--$\delta$ lines from \CV\footnote{\citet{henley15a} did not include these lines in their
  version of the model---they cut off their spectra at 0.4~\kev, whereas these lines are at
  0.299--0.379~\kev.} and \Kalpha--$\epsilon$ lines from \CVI, \OVII, and \OVIII. For each ion, the
lines' intensities were tied to those of a reference line using the \citet{cumbee14} line intensity
ratios. The overall normalization for each ion's line emission was a free parameter. Note that,
because of the relatively poor spectral resolution of the \xmm\ and \suzaku\ CCD spectrometers at
low energies, we did not include nitrogen lines in this foreground model (the \NVI\ and
\NVII\ \Kalpha\ lines lie between those of \CVI\ and \OVII).

We also investigated a second SWCX foreground model, based on the AtomDB Charge Exchange code
\citep[ACX;][]{smith14}, which uses analytical expressions to calculate the distributions of
principal quantum number, $n$, and orbital angular momentum, $l$, for the electron that transfers
from the donor atom to the receiving ion (the user has a number of such expressions to choose
from). Following \citet{henley15a}, we refer to this model as the ACX-SWCX foreground model. The
relative strengths of lines from different ions are controlled via an ionization temperature
parameter, assuming collisional ionization equilibrium (CIE). This temperature was a free parameter
in the fitting. We used the same settings for the ACX model component as \citet{henley15a}, i.e.,
$\mathtt{swcx}=1$ (suitable for studying SWCX emission) and $\mathtt{model}=8$ (meaning that
captured electrons are distributed between the two nearest $n$ shells to the most-probable shell,
and that the Separable distribution is used for the $l$ levels; \citealt{smith14,smith14a}). These
are also the settings favored by \citet{smith14} for modeling SWCX emission. However, it should be
noted that the expressions used by the ACX model cannot be expected to accurately model the $n$ and
$l$ distributions for all relevant CX reactions. We will discuss the reliability of the ACX model in
Section~\ref{subsec:ForegroundChoice}.

The third and final foreground model assumed that the emission was from hot plasma in the LB, rather
than from SWCX. While such a model may not be realistic, as SWCX emission likely dominates the
foreground emission in the \xmm/\suzaku\ band \citep{koutroumpa07,koutroumpa09a,koutroumpa11}, it
does allow our results to be compared with previous shadowing studies that have used such a model
\citep{galeazzi07,lei09,gupta09b}. For this LB model, we modeled the foreground emission with a
single-temperature ($1T$) APEC thermal plasma model, assuming CIE \citep{smith01a,foster12}. The
temperature and normalization of this model were free parameters.

Note that our spectral modeling assumes that the foreground emission is identical in the on- and
off-shadow spectra. For most of our shadowing clouds, the on- and off-shadow pointings took place
within a few days of each other (Table~\ref{tab:Observations}). Assuming that the foreground
emission is approximately constant over such a time scale is not unreasonable (although it should be
noted that variations in the SWCX emission have been observed over shorter times;
\citealt{snowden04,fujimoto07,carter08,carter10,ezoe10,wargelin14}).  For the \xmm\ and
\suzaku\ observations of MBM~20, the two pointings were two weeks and $\sim$six months apart,
respectively (Table~\ref{tab:Observations}). Over such periods of time, the solar wind conditions
and/or the viewing geometry through the heliosphere could change significantly, resulting in large
changes in the SWCX emission, which would adversely affect the spectral analysis.  We will discuss
this in Section~\ref{subsubsec:HaloUncertaintyBias}.

(2) We modeled the halo emission using a $1T$ APEC model, assuming CIE. In general, the temperature
and normalization of this component were free parameters. However, \citet{ursino14} reported the
non-detection of halo emission in the direction of MBM~16, with an upper limit on the emission
measure of $\sim$$10^{-3}~\emismeas$. In our survey of the halo emission with \xmm\ (HS13), we did
not detect halo emission on $\sim$1/5 of our sight lines; for these sight lines, the lower and upper
quartiles of the emission measure upper limits were $0.8 \times 10^{-3}$ and $1.4 \times
10^{-3}~\emismeas$, respectively. Here, we found that, if the halo temperature was a free parameter
in the MBM~16 fits, we typically could not constrain this parameter. Therefore, for MBM~16 only, we
fixed the halo temperature at $2.1 \times 10^6~\K$ (a typical value; HS13).

The halo component was subject to absorption, using the XSPEC \texttt{phabs} model
\citep{balucinska92,yan98}. For each observation, we calculated the absorbing column density, \NH,
by first calculating the average value of $I_{100}$ in the field view \citep{schlegel98}, and then
converting this value to \NH\ using the conversion relations from \citet{snowden00}. The resulting
column densities are shown in column~10 of Table~\ref{tab:Observations}. The advantage of
the $I_{100}$ data over \NHI\ data \citep[e.g.,][]{kalberla05} is that the former have higher
spatial resolution, allowing us to account for variations in the amount of absorbing material over
the X-ray fields of view. Using the HEASARC
\NH\ tool\footnote{http://heasarc.gsfc.nasa.gov/cgi-bin/Tools/w3nh/w3nh.pl} results in on-cloud
column densities that are typically $\sim$2--4 times smaller than those tabulated in
Table~\ref{tab:Observations}, due to its averaging over an area larger than that covered by the
X-ray detectors. Note that the column densities for the filament observations are different from
those used in \citet{henley07}, \citet{henley08a}, and \citet{lei09}, as in those studies
we did not average $I_{100}$ over the fields of view.

(3) We modeled the extragalactic emission using a double broken power-law \citep{smith07a}. For the
\xmm\ spectra, the presence of soft proton contamination (see (4), below) meant that we could not
independently constrain the normalization of this model, and so it was necessary to fix this
normalization. We rescaled \citepossessive{smith07a} nominal extragalactic model so that its
0.5--2.0~\kev\ surface brightness matched that expected from sources below the source removal flux
threshold (HS13). For the \suzaku\ spectra, which do not suffer from soft proton contamination, we
allowed the overall normalization of this model to be a free parameter. For a given shadow, the
normalizations for the on- and off-shadow directions were independent.  The extragalactic model was
attenuated by the same absorbing columns as the halo component (see above).

(4) For the \xmm\ spectra only, we added Gaussians at $\sim$1.49 (pn and MOS spectra) and
$\sim$1.75~keV (MOS spectra only) representing the Al and Si instrumental lines, which were not
removed by the QPB subtraction \citep{kuntz08a}. As the filtering described in
Section~\ref{subsec:XMMReduction} does not remove all the soft proton contamination from the
\xmm\ spectra, we also added a power-law that was not folded through the instrumental response to
model any residual contamination \citep{kuntz08a,snowden13}. The parameters of these additional
model components were independent for each individual exposure that was analyzed.

The on-MBM~12 \suzaku\ field includes the intermediate polar XY~Ari, which we excised from the data
using a circle of radius 2\arcmin\ (following \citealt{smith07a}). However, \citet{smith07a} found
that this source is so bright that X-rays from the wings of the point spread function significantly
contaminate the SXRB spectrum above $\sim$1~\kev. Like \citeauthor{smith07a}, we modeled this
contamination by adding an absorbed bremmstrahlung component to the on-MBM~12 spectral model.  This
component typically had a best-fit temperature of $kT = 200~\kev$ (this is the default upper limit
of the temperature in the XSPEC \texttt{bremss} model), and a hydrogen column density of
$\sim$$(\mbox{5--7}) \times 10^{22}~\pcmsq$ (similar to the values found by \citealt{smith07a}). As
the XY~Ari contamination component contributes very little emission below 1~\kev, the details of
this model should not significantly affect our measurements of the foreground and halo SXRB
emission.

Many previous \xmm\ or \suzaku\ studies of the SXRB emission have assumed abundances
\citep[e.g.,][hereafter AG89]{anders89} in which the oxygen abundance relative to hydrogen is
$\sim$$8 \times 10^{-4}$ (\citealt{smith07a,galeazzi07,gupta09b}; Yosh09;
\citealt{henley10b,henley15a}; HS13). This choice was often in part to allow easier comparisons with
preceding studies. However, recent measurements of the solar oxygen abundance yield a value about
0.2~dex lower: $4.90 \times 10^{-4}$ \citep[hereafter Aspl09]{asplund09}. Since oxygen emission
dominates the halo emission in the \xmm/\suzaku\ band, the assumed oxygen abundance will affect the
halo emission measures inferred from the spectral analysis. Here, we decided to use the Aspl09 solar
abundances as our reference abundances.  However, to allow our results to be compared with earlier
studies, we also carried out fits assuming AG89 solar abundances. Note that, because of its higher
carbon, nitrogen, and oxygen abundances, using the AG89 abundance table not only results in more
soft X-ray emission per unit emission measure from the hot gas, but also in greater photoelectric
absorption cross-sections in the cold gas.  In addition, to further test the sensitivity of our
results to the assumed abundances, we carried out fits using the \citet[hereafter Wilm00]{wilms00}
interstellar abundances. (Note that the Wilm00 oxygen abundance is the same as the Aspl09
value. However, the carbon and nitrogen abundances, which are also relevant to emission in the
\xmm/\suzaku\ band, are 0.05~dex lower and 0.05~dex higher in Wilm00 than in Aspl09, respectively.)

\newcounter{ResultsTable}
\setcounter{ResultsTable}{\value{table}}
\addtocounter{table}{1}
{
  \tabletypesize{\scriptsize}
  \begin{deluxetable*}{lrr@{,}lrr@{,}lcrr@{,}lrr@{,}lcrr@{,}lrr@{,}l}
\tablewidth{0pt}
\tabletypesize{\scriptsize}
\tablecaption{Foreground and Halo Surface Brightnesses\label{tab:SB}}
\tablehead{
                                  & \multicolumn{6}{c}{C14-SWCX model}                 && \multicolumn{6}{c}{ACX-SWCX model}                 && \multicolumn{6}{c}{LB model} \\
\cline{2-7} \cline{9-14} \cline{16-21}
\colhead{Shadow\tablenotemark{a}} & \multicolumn{3}{c}{\Sfg} & \multicolumn{3}{c}{\Sh} && \multicolumn{3}{c}{\Sfg} & \multicolumn{3}{c}{\Sh} && \multicolumn{3}{c}{\Sfg} & \multicolumn{3}{c}{\Sh} \\
\colhead{(1)}                     & \multicolumn{3}{c}{(2)}  & \multicolumn{3}{c}{(3)} && \multicolumn{3}{c}{(4)}  & \multicolumn{3}{c}{(5)} && \multicolumn{3}{c}{(6)}  & \multicolumn{3}{c}{(7)}
}
\startdata
\cutinhead{\citet{asplund09} abundances}
G048+37 (X)  & 1.25 & (1.14 & 1.65) & 6.28 & (5.83 & 6.84) & & 2.18 & (1.89 & 2.82) & 5.03 & (4.64 & 5.58) & & 2.08 & (1.75 & 2.54) & 5.50 & (4.90 & 5.78) \\
MBM 12 (S)   & 1.09 & (0.90 & 1.25) & 1.04 & (0.85 & 1.65) & & 1.26 & (1.07 & 1.43) & 1.26 & (0.96 & 1.83) & & 1.31 & (1.13 & 1.48) & 0.94 & (0.53 & 1.45) \\
MBM 16 (S)   & 3.11 & (2.60 & 3.51) & 0.04 & (0.00 & 0.93) & & 3.29 & (2.89 & 3.58) & 0.00 & (0.00 & 0.62) & & 2.66 & (2.27 & 3.05) & 1.15 & (0.48 & 1.81) \\
MBM 20 (X)   & 0.10 & (0.09 & 0.13) & 2.98 & (2.89 & 3.02) & & 0.21 & (0.18 & 0.25) & 3.02 & (2.94 & 3.19) & & 0.16 & (0.14 & 0.20) & 2.88 & (2.82 & 2.94) \\
MBM 20 (S)   & 0.52 & (0.11 & 0.78) & 2.19 & (1.69 & 3.49) & & 0.34 & (0.19 & 0.49) & 2.51 & (2.05 & 3.00) & & 0.64 & (0.45 & 0.83) & 2.25 & (1.90 & 2.59) \\
G236+38 (S)  & 2.21 & (1.05 & 3.38) & 4.63 & (2.84 & 6.54) & & 1.44 & (1.02 & 1.86) & 5.70 & (4.84 & 6.56) & & 3.79 & (2.24 & 4.91) & 3.43 & (1.96 & 6.61) \\
Filament (X) & 1.79 & (1.30 & 2.14) & 7.37 & (6.57 & 7.66) & & 1.72 & (1.19 & 2.45) & 7.60 & (6.78 & 8.64) & & 3.78 & (2.90 & 4.31) & 5.21 & (4.80 & 6.93) \\
Filament (S) & 0.13 & (0.04 & 0.48) & 5.80 & (5.15 & 6.03) & & 0.22 & (0.09 & 0.35) & 5.76 & (5.48 & 6.04) & & 0.39 & (0.19 & 0.93) & 5.56 & (4.72 & 5.99) \\
\cutinhead{\citet{anders89} abundances}
G048+37 (X)  & 2.28 & (2.05 & 2.59) & 5.78 & (5.29 & 6.34) & & 3.48 & (3.10 & 3.76) & 4.17 & (3.88 & 4.68) & & 3.30 & (2.97 & 3.68) & 4.80 & (4.42 & 5.63) \\
MBM 12 (S)   & 1.17 & (1.02 & 1.32) & 1.26 & (0.78 & 1.86) & & 1.47 & (1.30 & 1.64) & 1.27 & (0.94 & 2.11) & & 1.40 & (1.24 & 1.55) & 1.17 & (0.65 & 1.74) \\
MBM 16 (S)   & 3.28 & (2.84 & 3.62) & 0.25 & (0.00 & 1.16) & & 3.53 & (3.19 & 3.82) & 0.00 & (0.00 & 0.60) & & 2.74 & (2.38 & 3.10) & 1.75 & (1.03 & 2.52) \\
MBM 20 (X)   & 0.49 & (0.45 & 0.53) & 2.98 & (2.94 & 3.09) & & 0.28 & (0.24 & 0.33) & 3.24 & (3.13 & 3.28) & & 0.62 & (0.58 & 0.67) & 2.85 & (2.79 & 2.93) \\
MBM 20 (S)   & 0.86 & (0.60 & 1.10) & 2.22 & (1.94 & 2.49) & & 0.49 & (0.34 & 0.65) & 2.88 & (2.58 & 3.18) & & 0.92 & (0.72 & 1.28) & 2.43 & (2.02 & 2.85) \\
G236+38 (S)  & 3.41 & (2.43 & 4.48) & 3.96 & (2.31 & 6.53) & & 1.83 & (1.40 & 2.27) & 6.12 & (5.11 & 7.13) & & 4.67 & (3.58 & 5.63) & 3.13 & (2.38 & 5.14) \\
Filament (X) & 1.70 & (1.35 & 1.90) & 8.76 & (8.23 & 9.51) & & 3.16 & (2.48 & 3.70) & 6.77 & (6.03 & 7.70) & & 3.82 & (3.48 & 4.17) & 6.18 & (5.08 & 6.98) \\
Filament (S) & 0.58 & (0.18 & 0.99) & 6.03 & (5.73 & 6.26) & & 0.38 & (0.25 & 0.51) & 6.41 & (6.09 & 6.72) & & 1.16 & (0.64 & 1.67) & 5.46 & (4.52 & 6.46) \\
\cutinhead{\citet{wilms00} abundances}
G048+37 (X)  & 1.48 & (1.36 & 1.73) & 6.14 & (5.47 & 6.50) & & 2.48 & (2.17 & 2.89) & 4.75 & (4.36 & 5.20) & & 2.39 & (2.01 & 2.77) & 5.28 & (4.89 & 5.95) \\
MBM 12 (S)   & 1.08 & (0.92 & 1.23) & 1.10 & (0.73 & 1.66) & & 1.30 & (1.12 & 1.49) & 1.24 & (0.93 & 1.92) & & 1.34 & (1.16 & 1.51) & 0.97 & (0.57 & 1.50) \\
MBM 16 (S)   & 3.15 & (2.65 & 3.43) & 0.02 & (0.00 & 0.90) & & 3.33 & (2.94 & 3.61) & 0.00 & (0.00 & 0.60) & & 2.73 & (2.33 & 3.12) & 1.14 & (0.47 & 1.80) \\
MBM 20 (X)   & 0.11 & (0.10 & 0.14) & 3.04 & (2.95 & 3.11) & & 0.18 & (0.15 & 0.21) & 2.92 & (2.87 & 2.99) & & 0.26 & (0.23 & 0.30) & 2.94 & (2.84 & 2.98) \\
MBM 20 (S)   & 0.55 & (0.25 & 0.83) & 2.18 & (1.68 & 3.51) & & 0.37 & (0.22 & 0.53) & 2.52 & (2.06 & 3.01) & & 0.71 & (0.52 & 0.91) & 2.24 & (1.88 & 2.59) \\
G236+38 (S)  & 2.30 & (1.47 & 3.45) & 4.59 & (2.79 & 6.52) & & 1.54 & (1.11 & 1.97) & 5.70 & (4.82 & 6.58) & & 3.74 & (2.49 & 5.07) & 3.67 & (1.90 & 6.32) \\
Filament (X) & 1.33 & (1.07 & 1.67) & 8.06 & (7.21 & 8.66) & & 2.41 & (1.83 & 2.97) & 6.79 & (6.01 & 7.32) & & 3.76 & (3.39 & 4.39) & 5.53 & (4.58 & 5.98) \\
Filament (S) & 0.15 & (0.10 & 0.29) & 5.82 & (5.43 & 6.04) & & 0.27 & (0.14 & 0.40) & 5.76 & (5.47 & 6.04) & & 0.52 & (0.30 & 1.08) & 5.49 & (4.62 & 5.94)
\enddata
\tablecomments{Surface brightness units: $10^{-12}~\flux\ \pdegsq$. The values in parentheses are the 90\%\ confidence intervals.}
\tablenotetext{a}{The X or S in parentheses indicates whether the shadow was observed with \xmm\ or \suzaku, respectively.}
\end{deluxetable*}

}

For each pair of \xmm\ observations, we fitted the above-described model simultaneously to the
0.3--5.0~\kev\ MOS and 0.4--5.0~\kev\ pn spectra from the on and off-shadow directions (the pn data
are generally not usable below 0.4~\kev; \citealt{snowden13}). In general, for each pair of
\suzaku\ observations, we fitted the model simultaneously to the on- and off-shadow
0.3--5.0~\kev\ XIS1 spectra.  The exceptions are the G236+38 and MBM~16 observations.  In the
G236+38 XIS1 spectra, we found that the count rates increased unexpectedly below $\sim$0.35~\kev,
resulting in poor fits. This increase is unexpected because the XIS1 camera's sensitivity is small
below $\sim$0.35~\kev---because of the build up of contamination on the camera's optical blocking
filter, the XIS1 low-energy sensitivity is smaller for the G236+38 observations than for
observations taken earlier in the mission. It is possible that the NXB is inaccurately modeled in
the G236+38 spectra at low energies.  Therefore, for G236+38, we cut off the XIS1 spectra at
0.35~\kev, rather than 0.3~\kev. Similarly, we found that the MBM~16 XIS1 count rates increased
unexpectedly below $\sim$0.4~\kev, and so for these spectra we placed the low-energy cut-off at
0.4~\kev. Because of this higher cut-off, we did not include the \CV\ lines in the C14-SWCX model
when fitting to the MBM~16 spectra.

\section{SPECTRAL ANALYSIS RESULTS}
\label{sec:Results}

\begin{figure*}[t]
\centering
\includegraphics[width=0.33\linewidth]{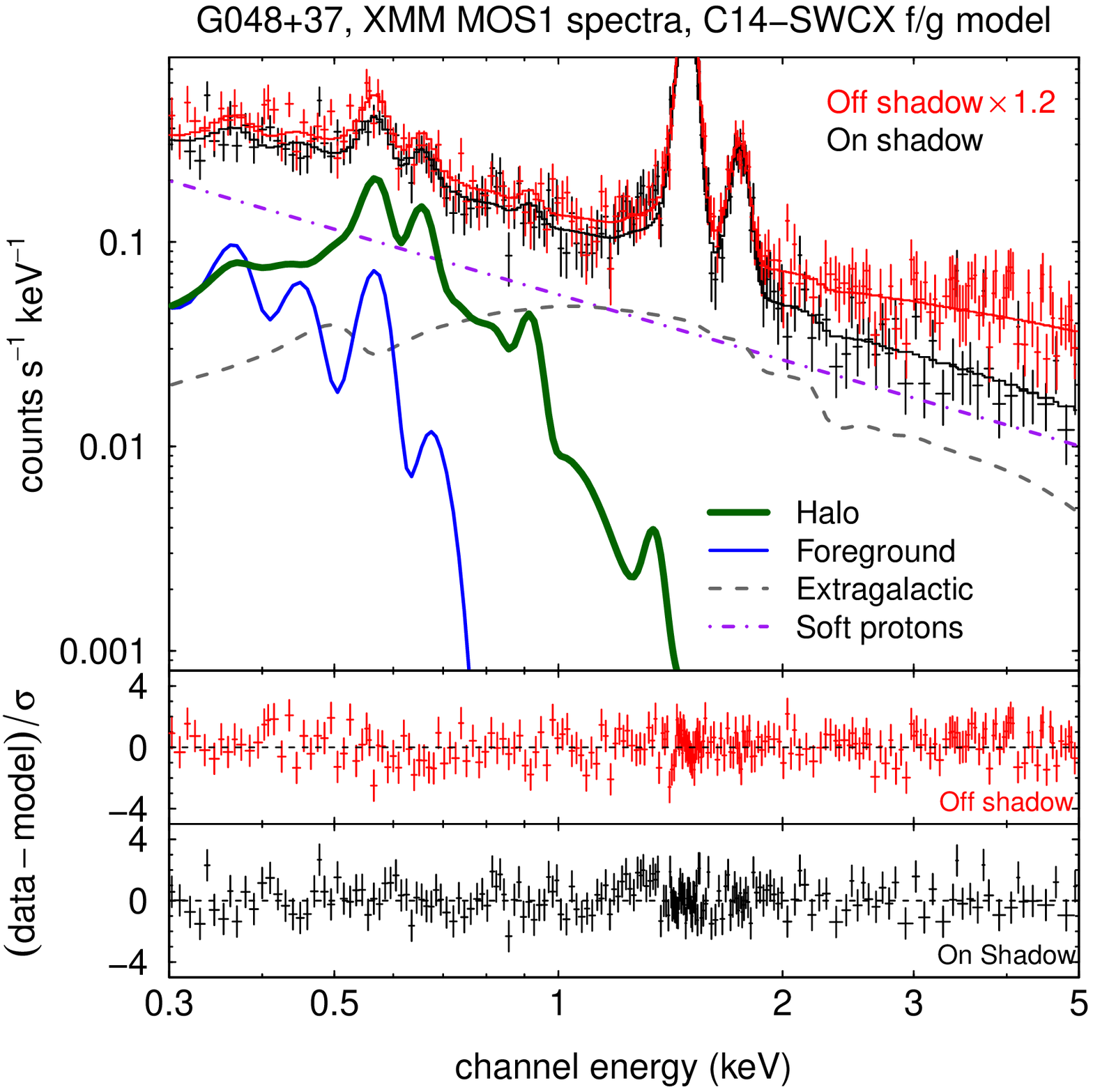}
\includegraphics[width=0.33\linewidth]{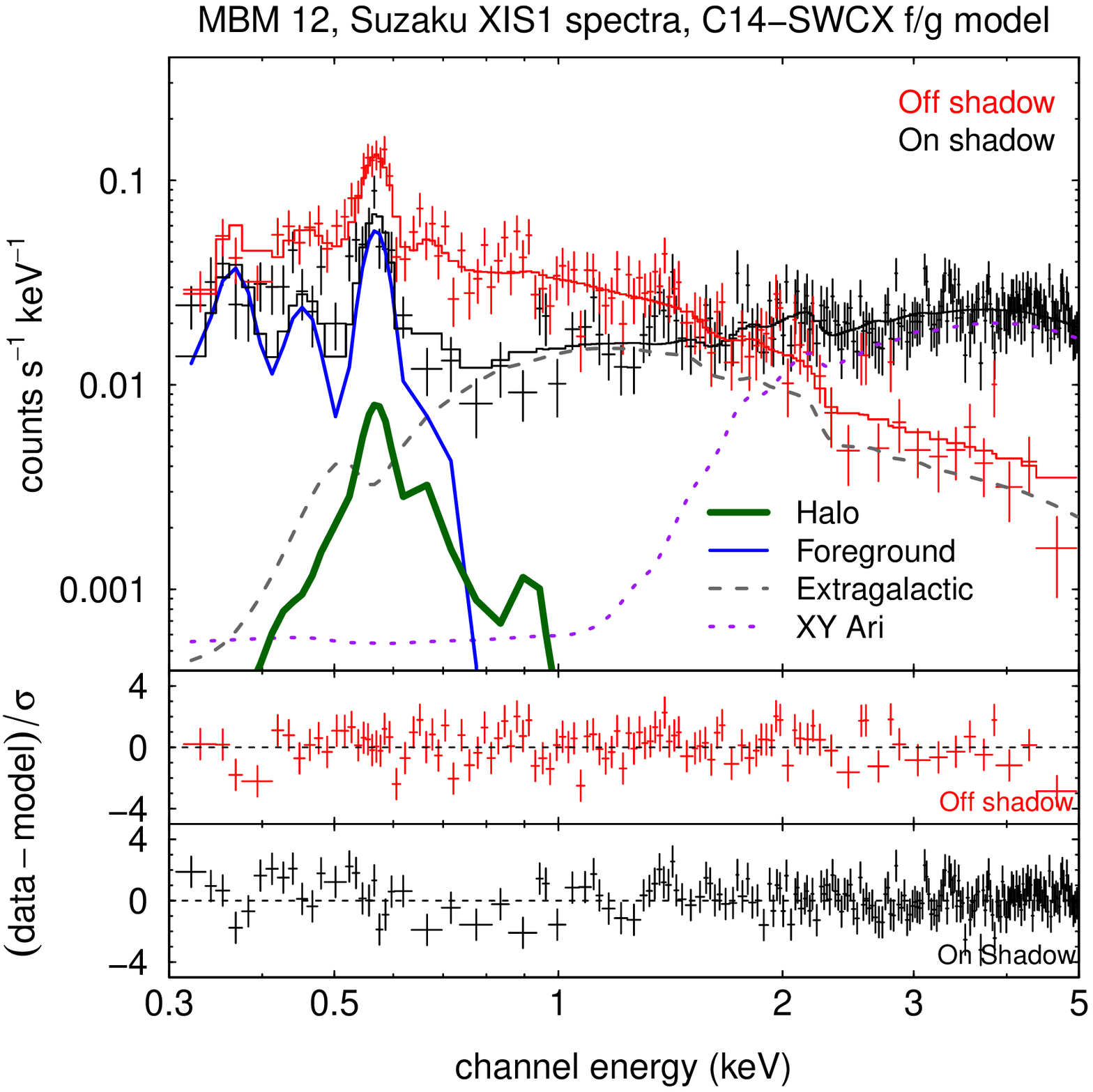}
\includegraphics[width=0.33\linewidth]{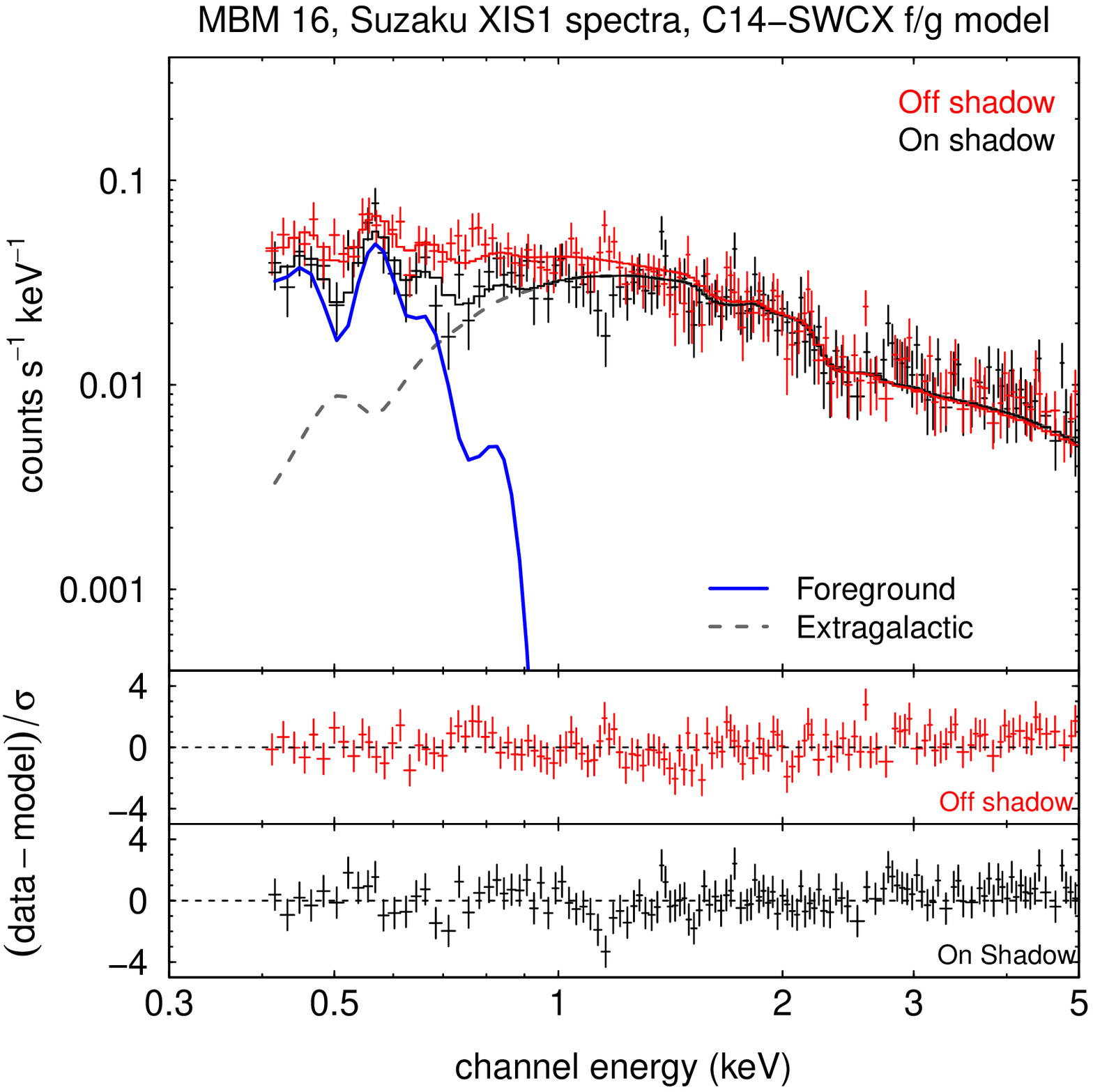} \\
\includegraphics[width=0.33\linewidth]{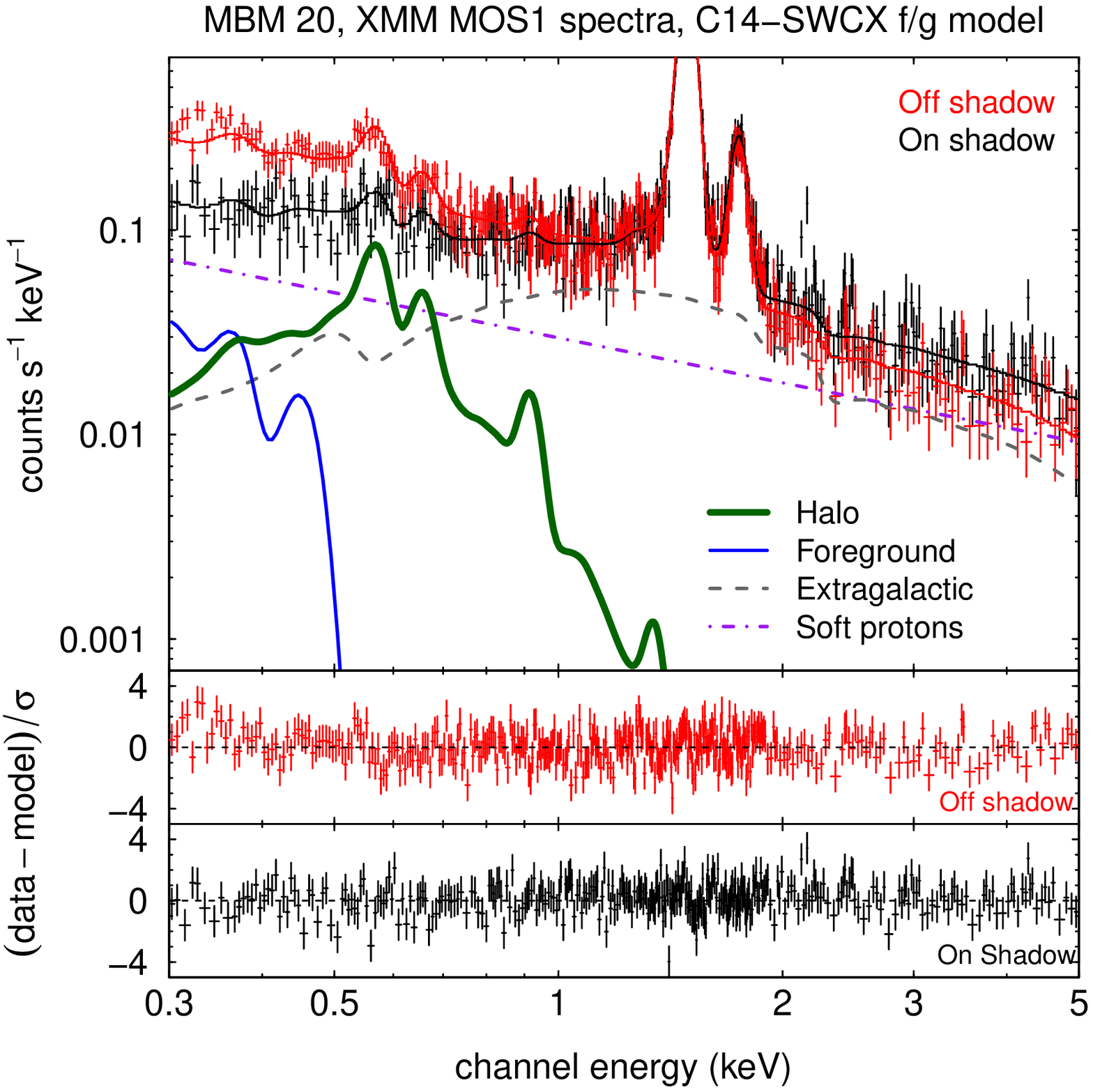}
\includegraphics[width=0.33\linewidth]{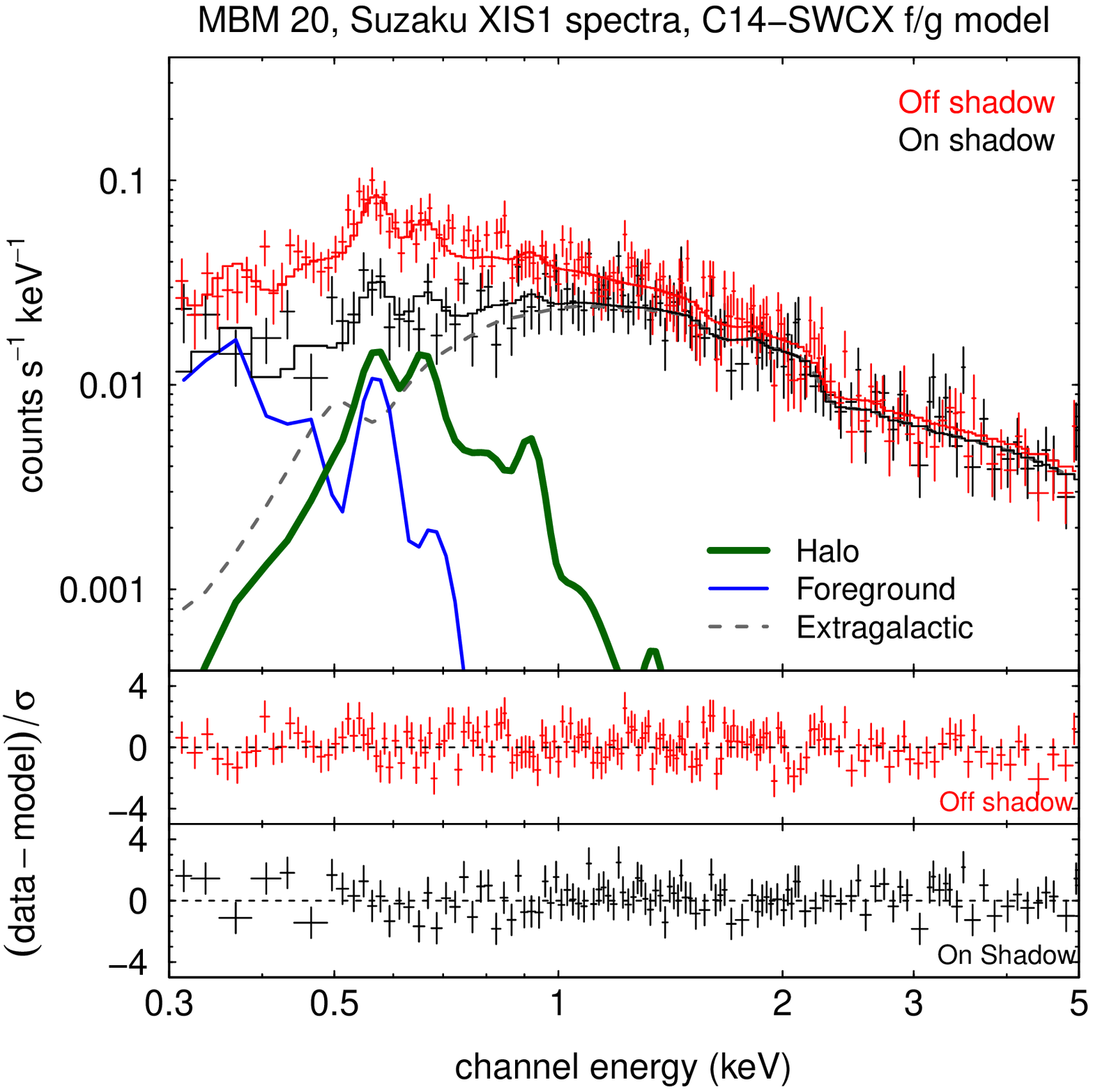}
\includegraphics[width=0.33\linewidth]{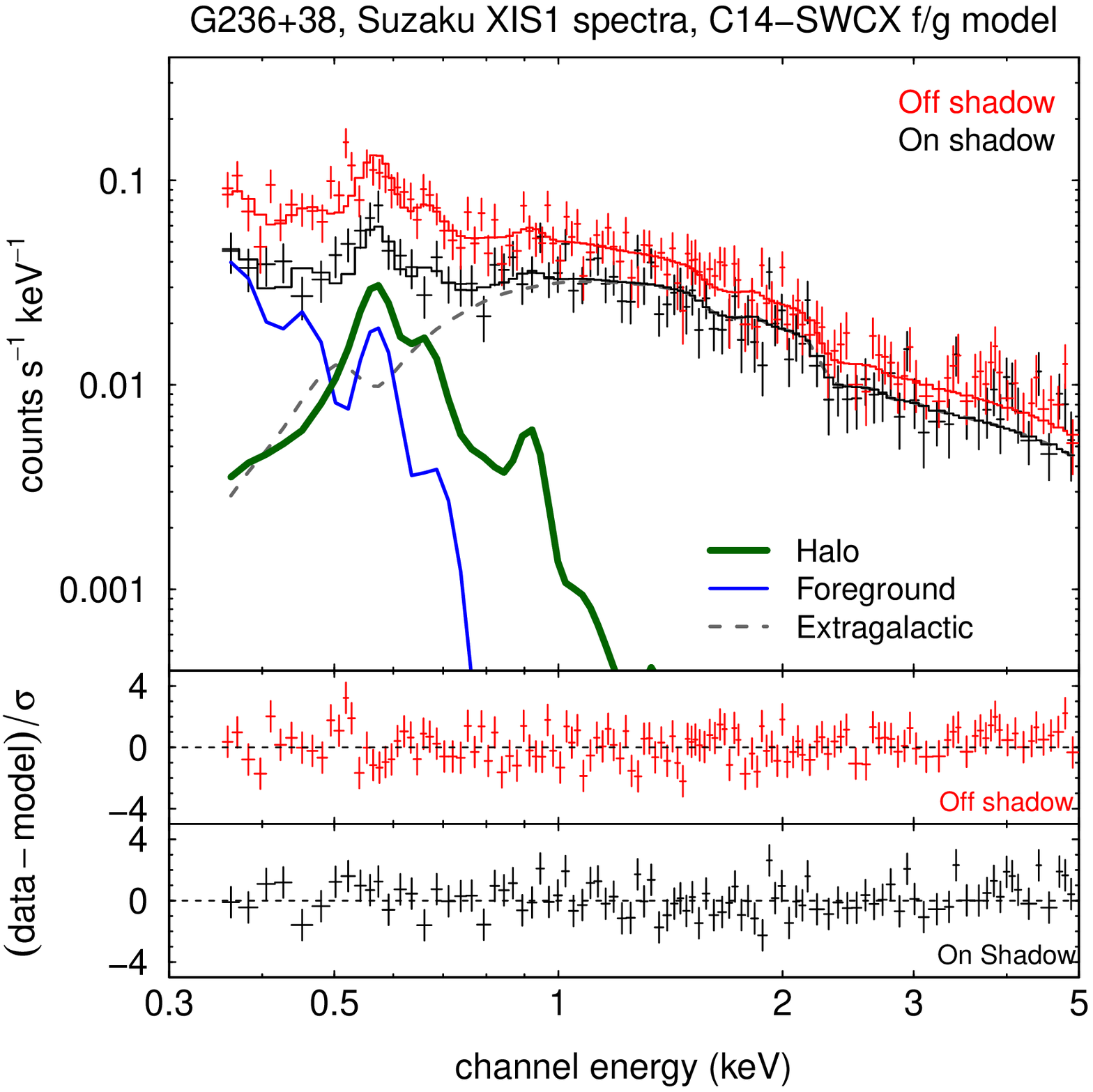} \\
\includegraphics[width=0.33\linewidth]{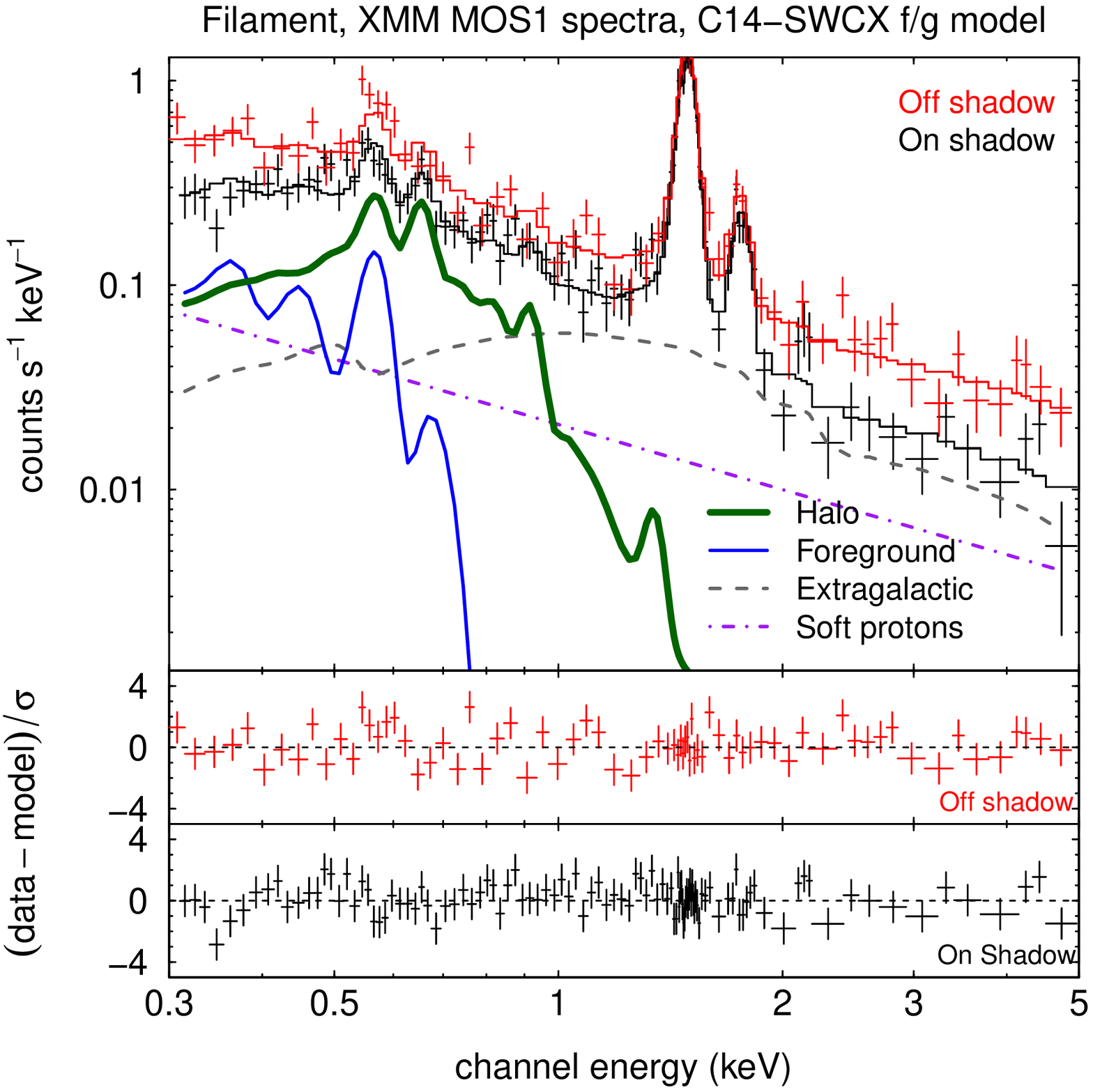}
\includegraphics[width=0.33\linewidth]{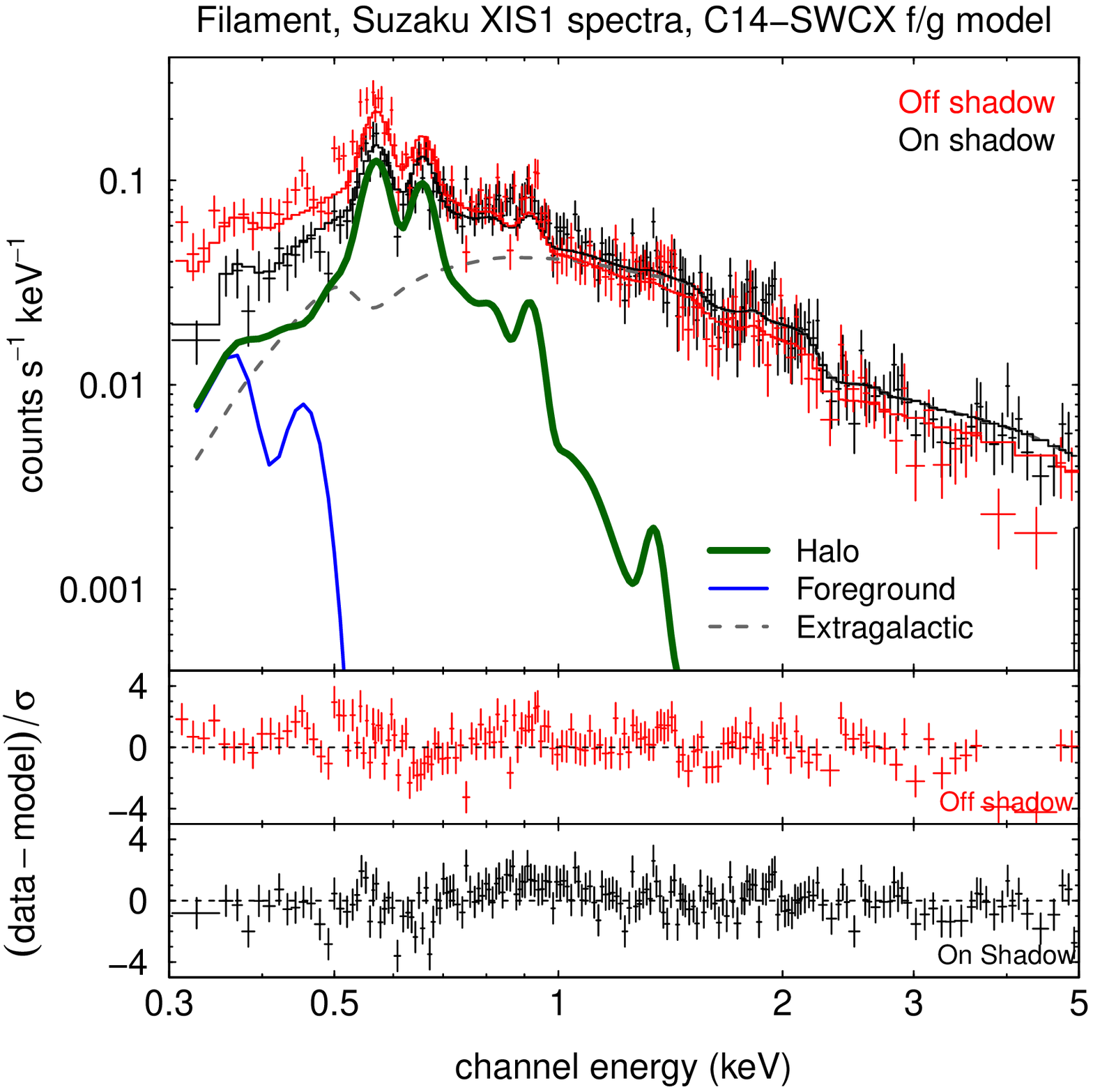}
\caption{Observed on- and off-shadow spectra for each of our shadows, with the best-fit models
  obtained using the C14-SWCX foreground model, assuming Aspl09 abundances.
  For the \xmm\ observations we show the MOS1 spectra, while for the \suzaku\ observations we show
  the XIS1 spectra. For plotting purposes only, the data have been regrouped such that each bin has
  a signal-to-noise ratio of at least 3. Note that the off-shadow G048+37 data points have been
  scaled up by 20\% because one of the CCDs was in an anomalous state \citep{kuntz08a} and its data
  could not be used, which also resulted in a relatively brighter soft proton component at low
  energies than in the on-shadow observation.
  For the on-shadow spectra, we also show the individual model components, as indicated in the
  key. Note that, for the \xmm\ spectra, we show the soft proton component, but not the components
  representing the instrumental Al and Si lines. For the on-MBM~12 XIS1 spectrum, we show the
  component representing the residual emission from XY~Ari (see Section~\ref{sec:SpectralModel}).
  The smaller panels beneath the main panel show the residuals.
  \label{fig:SpectraSWCX}}
\end{figure*}

\begin{figure*}[t]
\centering
\includegraphics[width=0.33\linewidth]{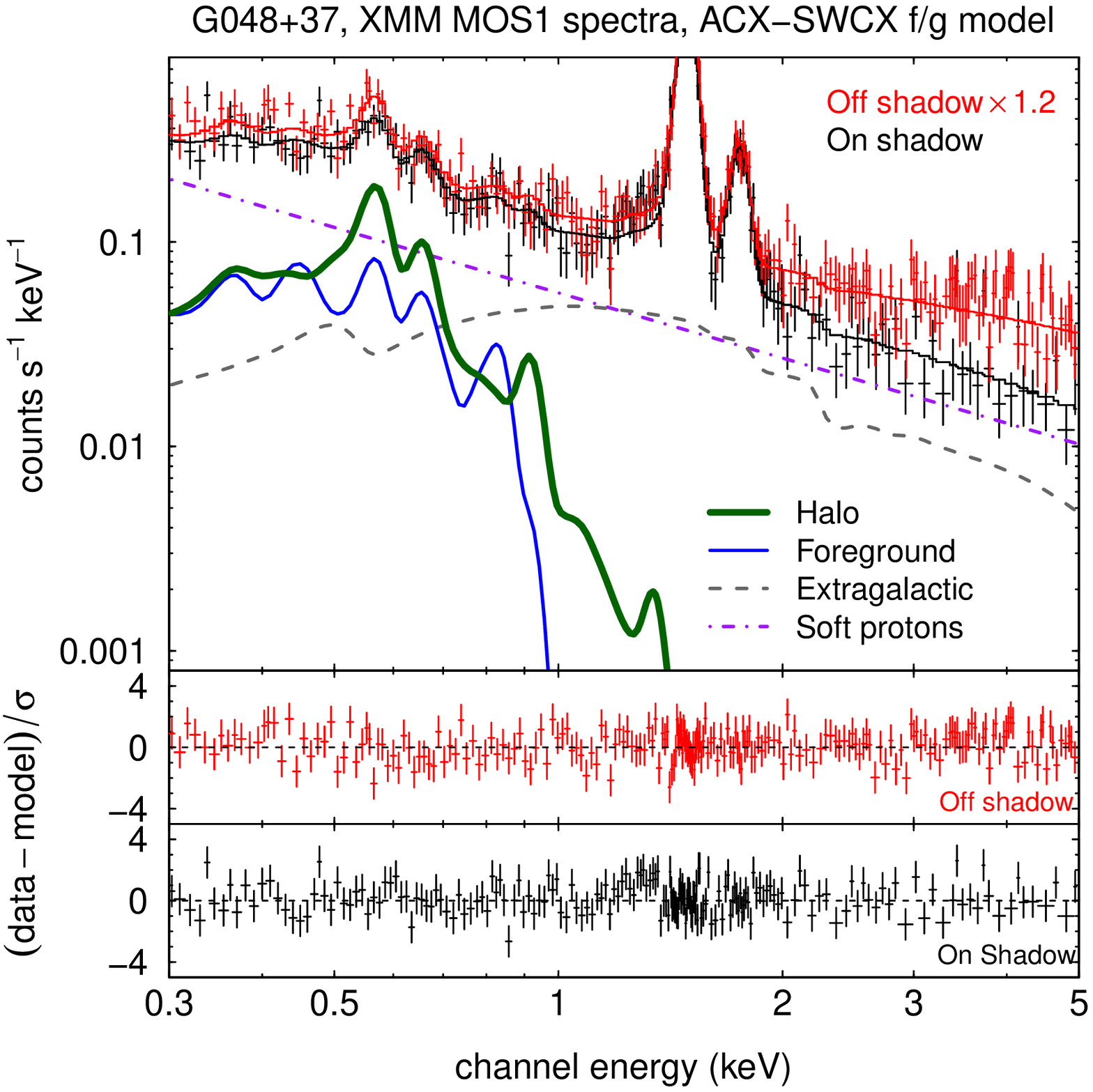}
\includegraphics[width=0.33\linewidth]{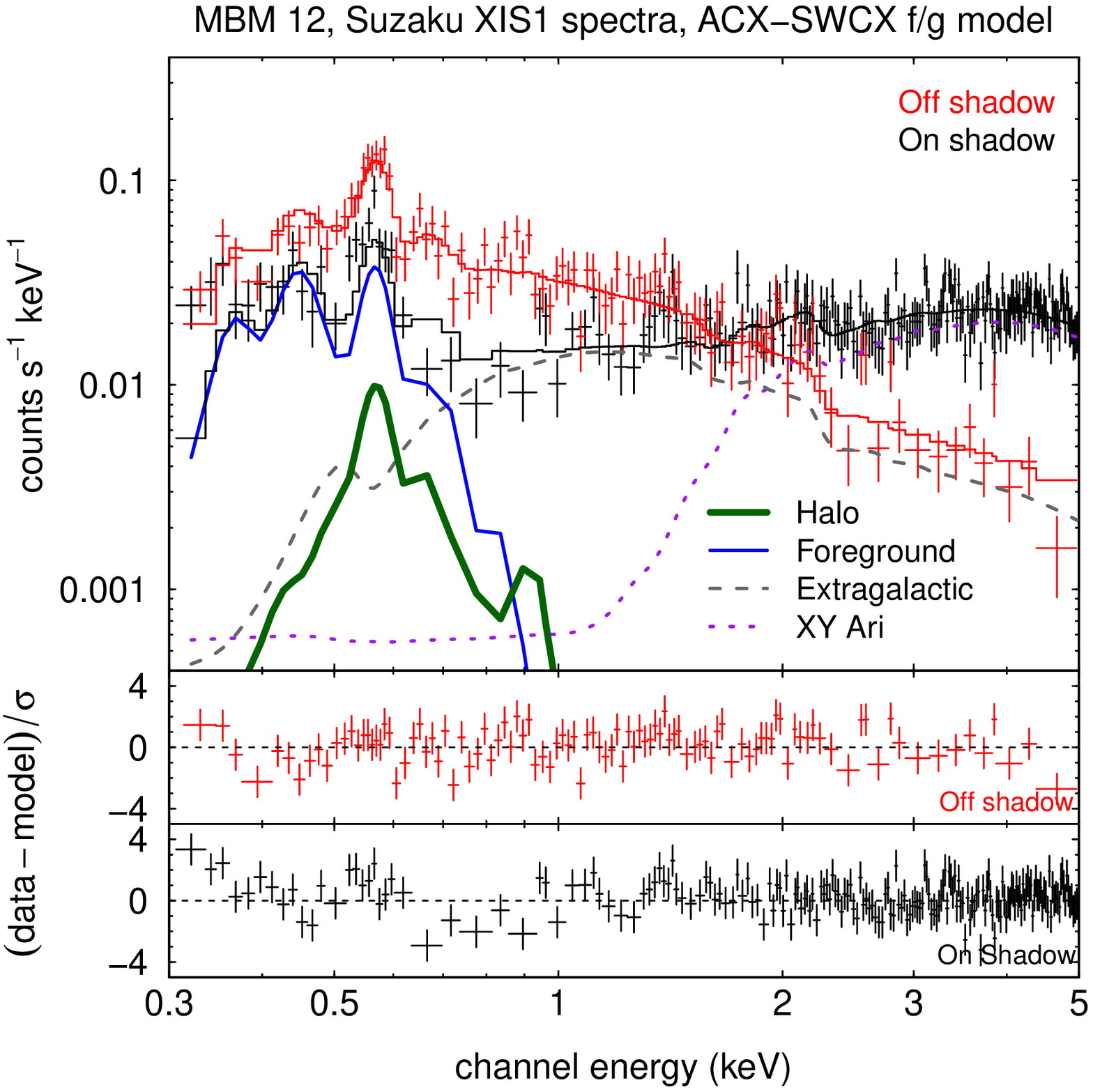}
\includegraphics[width=0.33\linewidth]{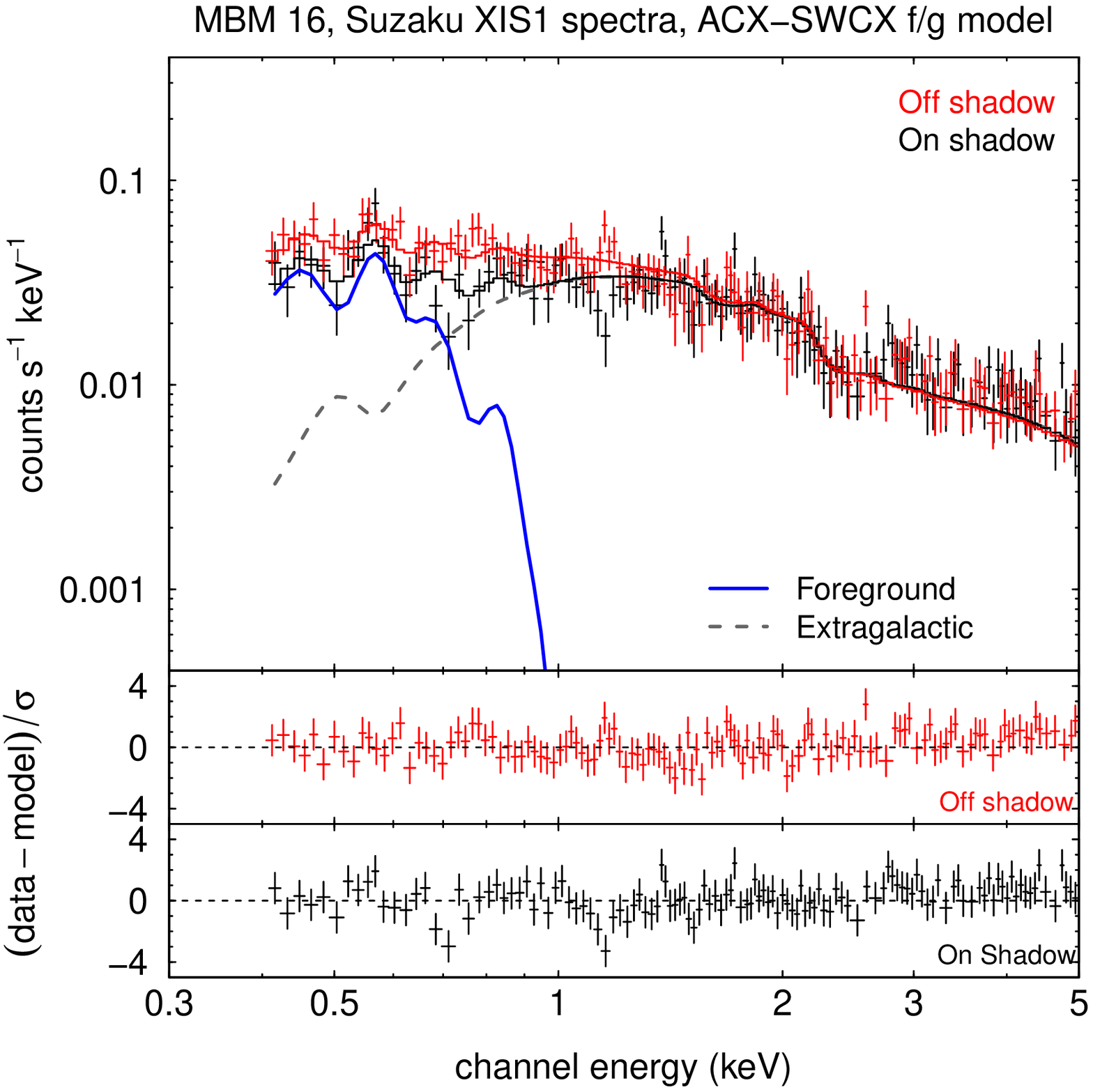} \\
\includegraphics[width=0.33\linewidth]{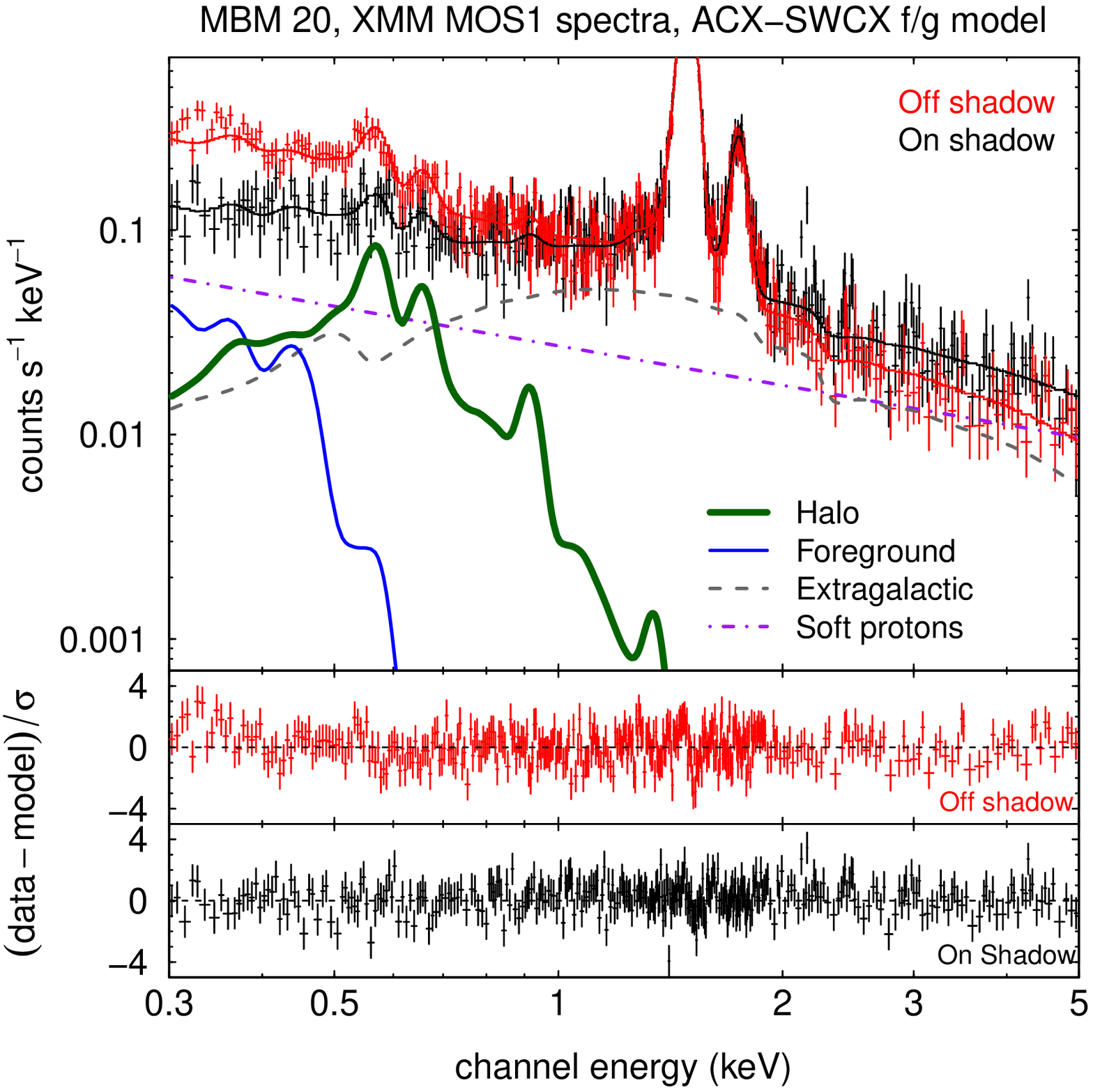}
\includegraphics[width=0.33\linewidth]{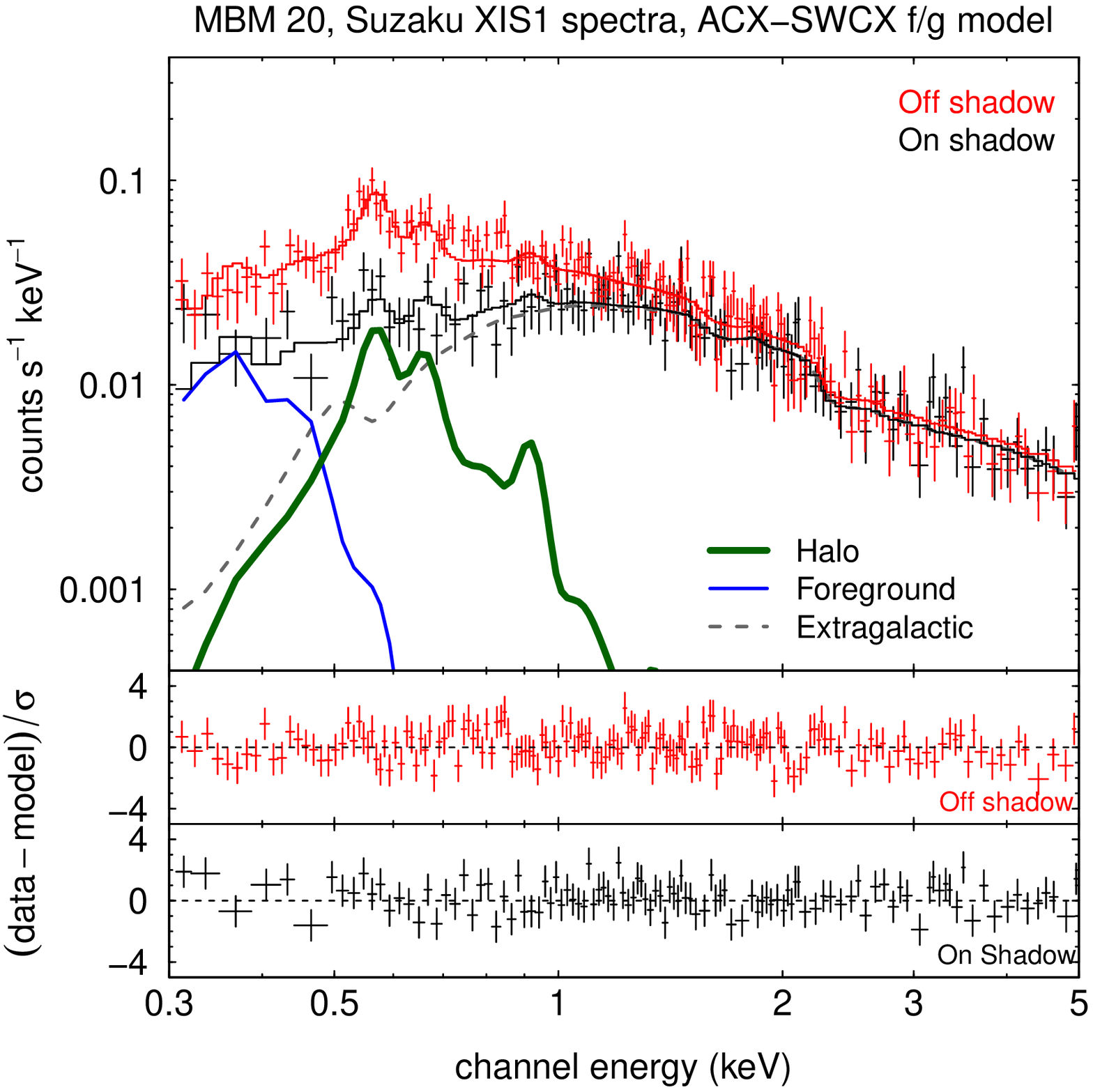}
\includegraphics[width=0.33\linewidth]{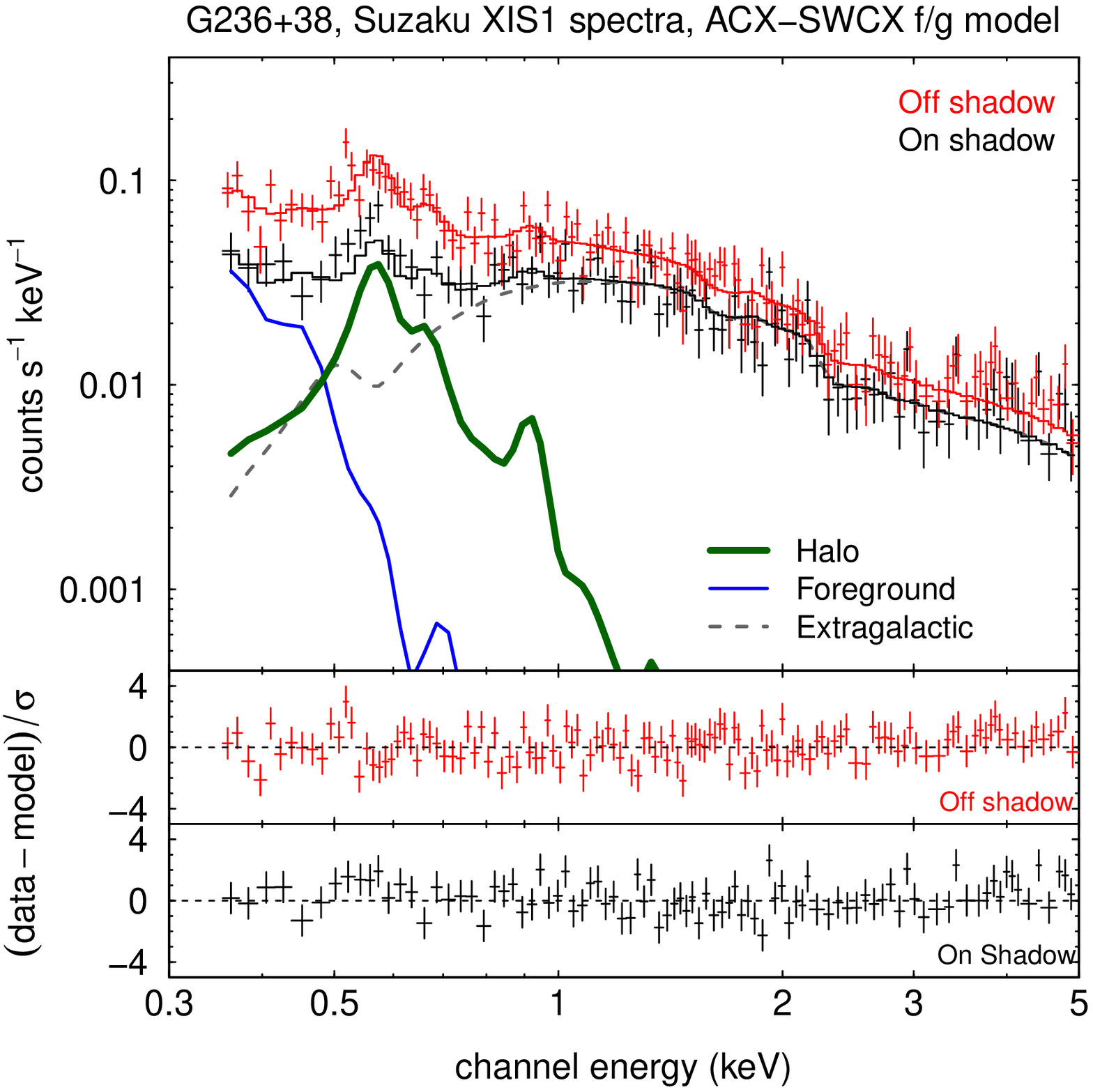} \\
\includegraphics[width=0.33\linewidth]{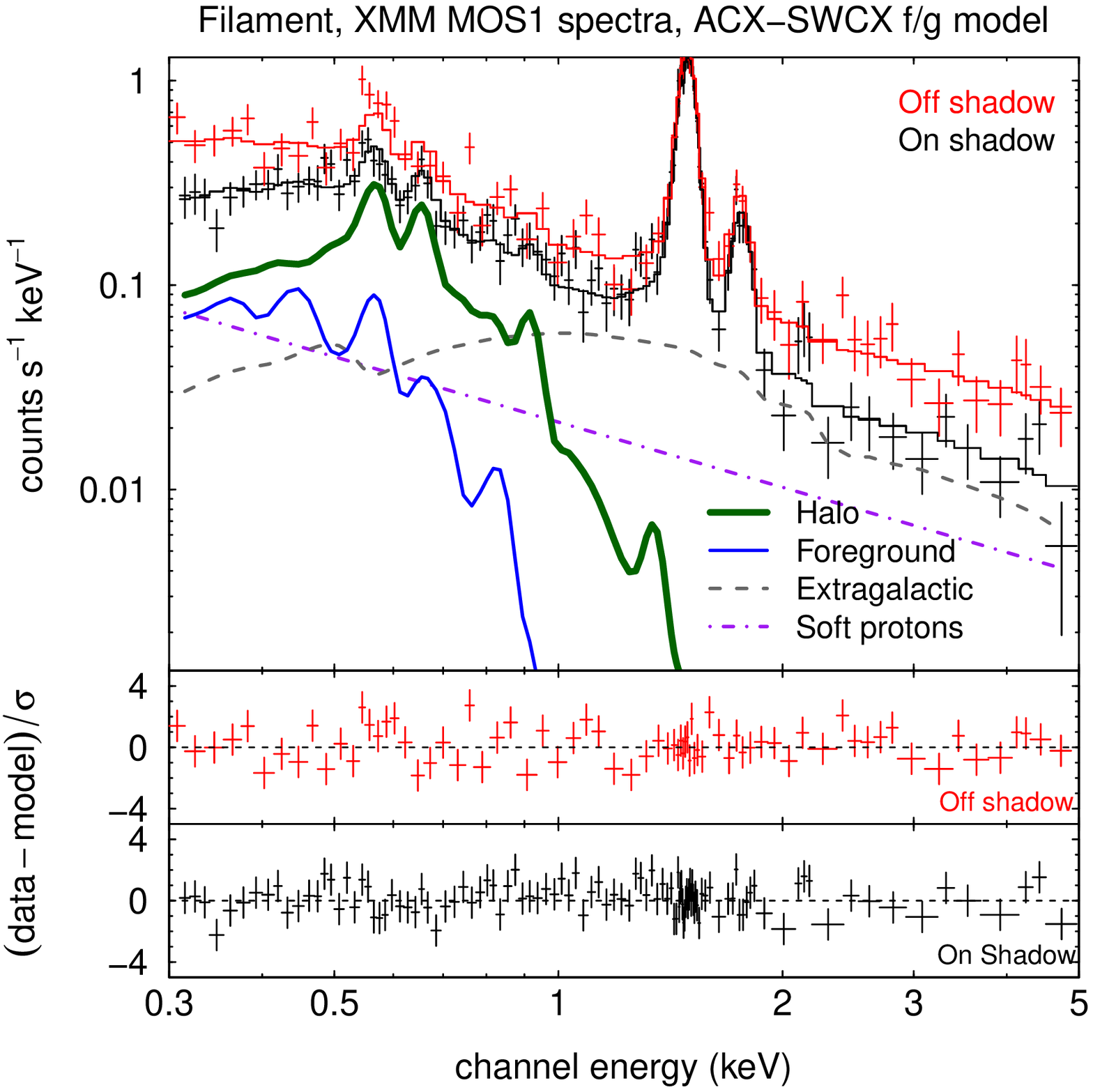}
\includegraphics[width=0.33\linewidth]{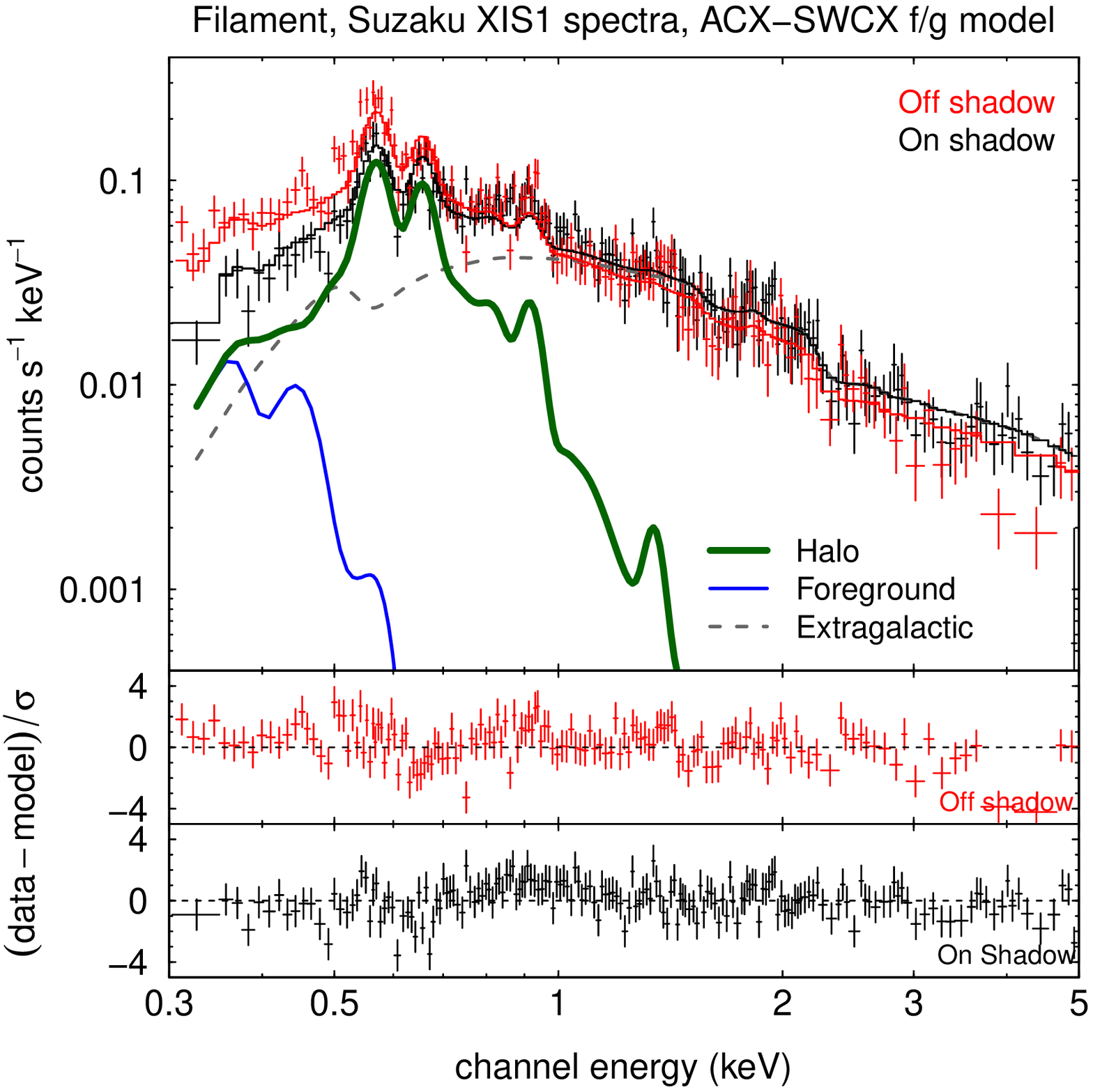}
\caption{Same as Figure~\ref{fig:SpectraSWCX}, but for the ACX-SWCX foreground model.
  \label{fig:SpectraACX}}
\end{figure*}

\begin{figure*}[t]
\centering
\includegraphics[width=0.33\linewidth]{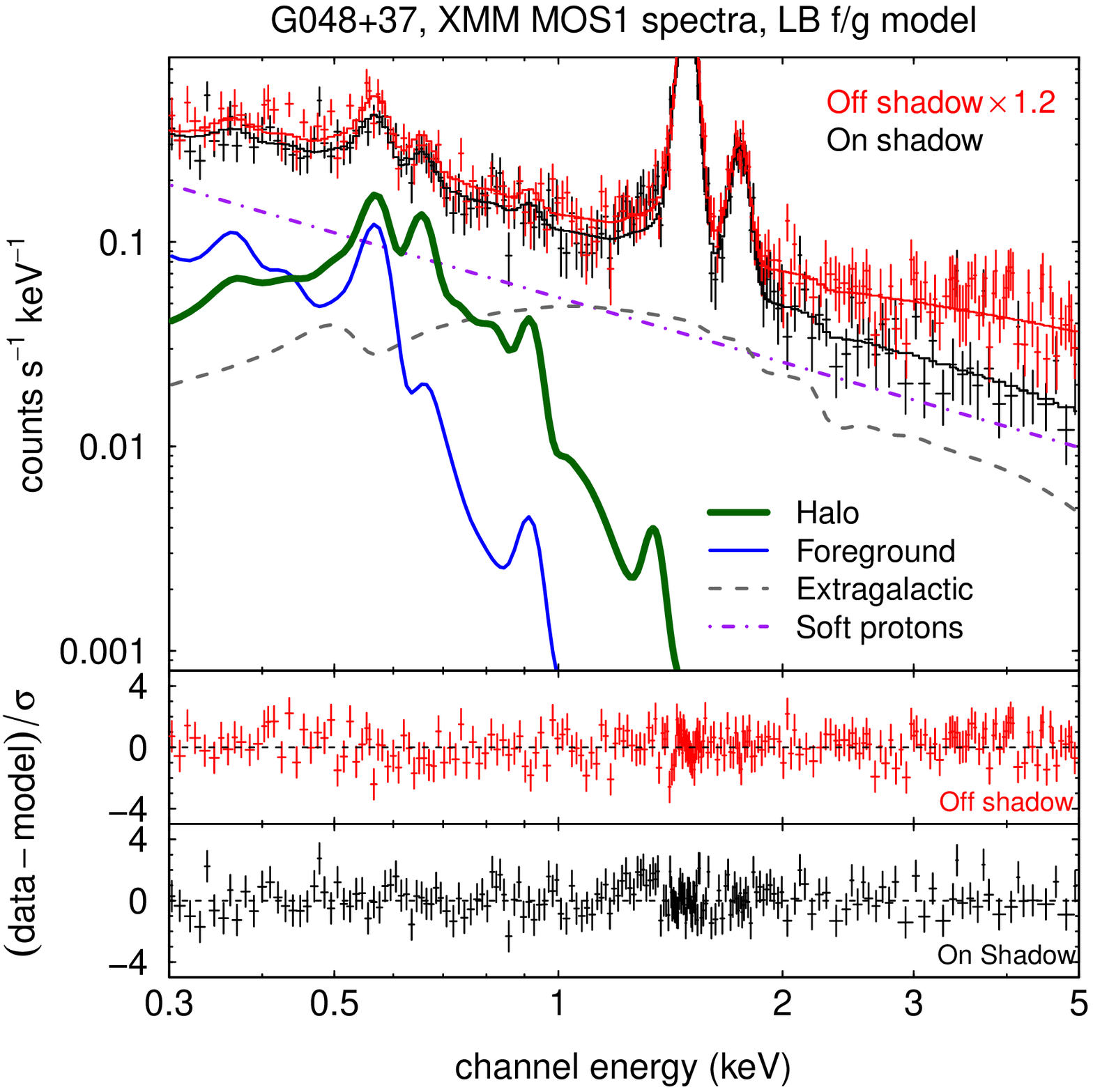}
\includegraphics[width=0.33\linewidth]{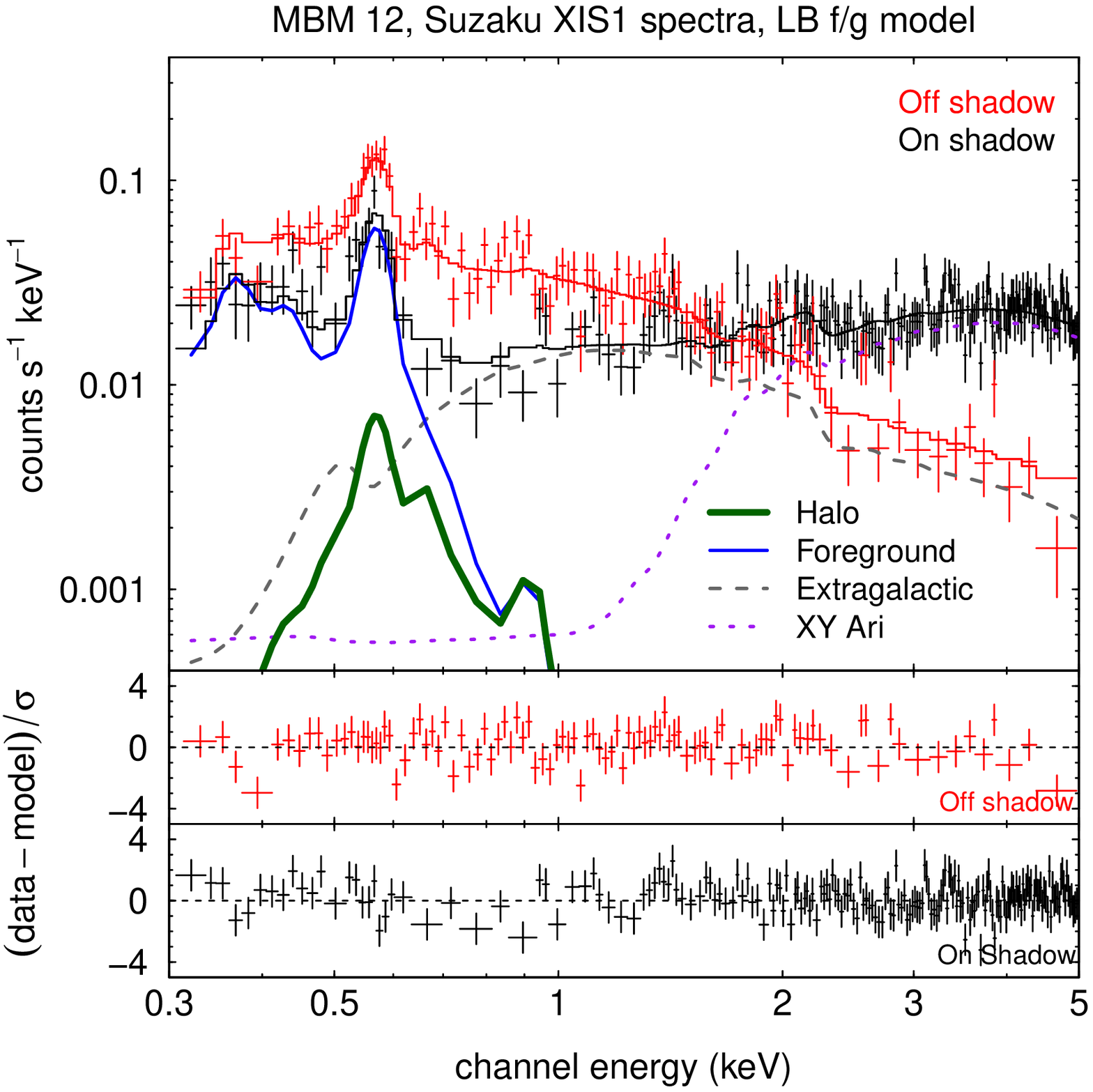}
\includegraphics[width=0.33\linewidth]{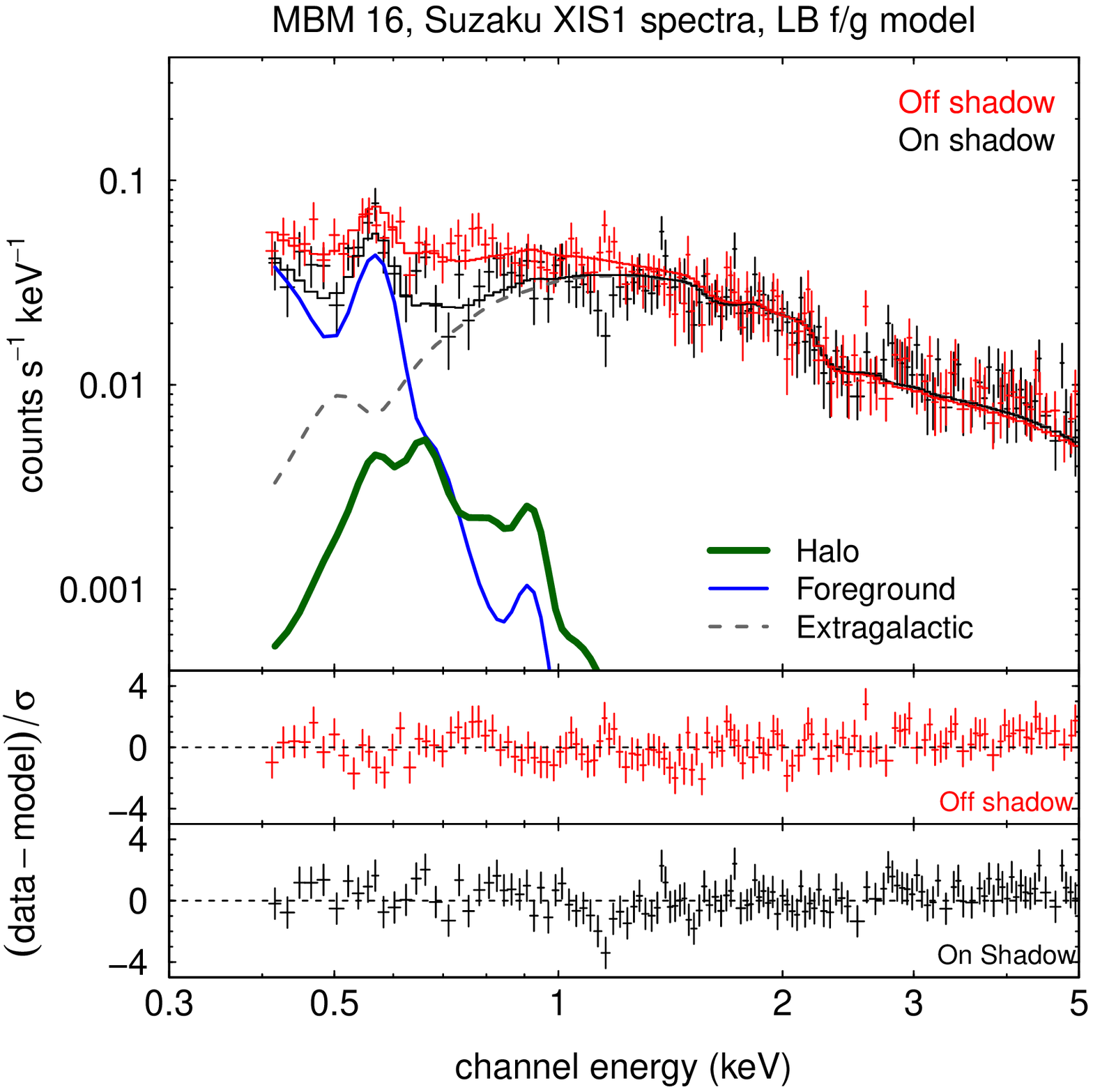} \\
\includegraphics[width=0.33\linewidth]{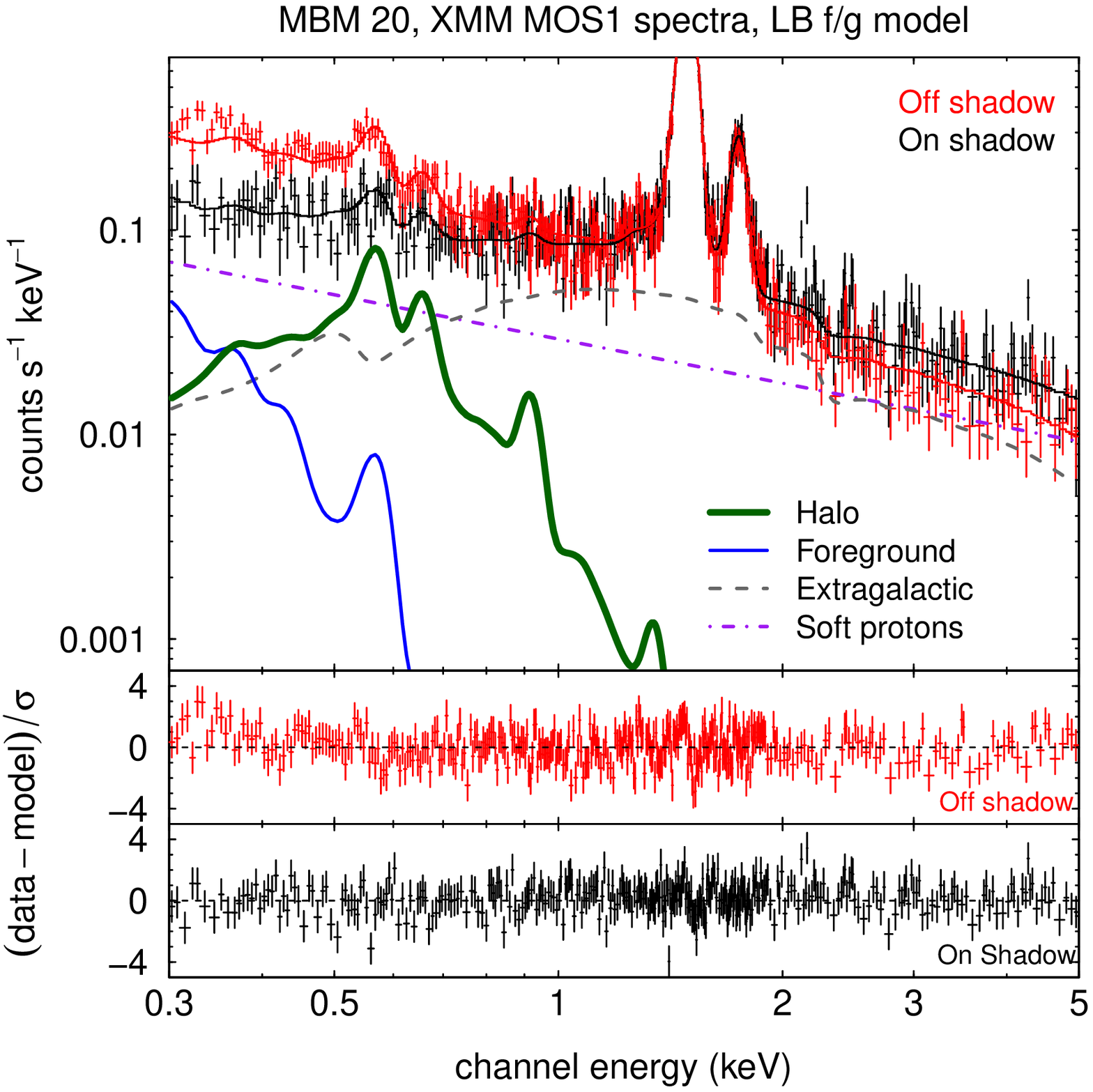}
\includegraphics[width=0.33\linewidth]{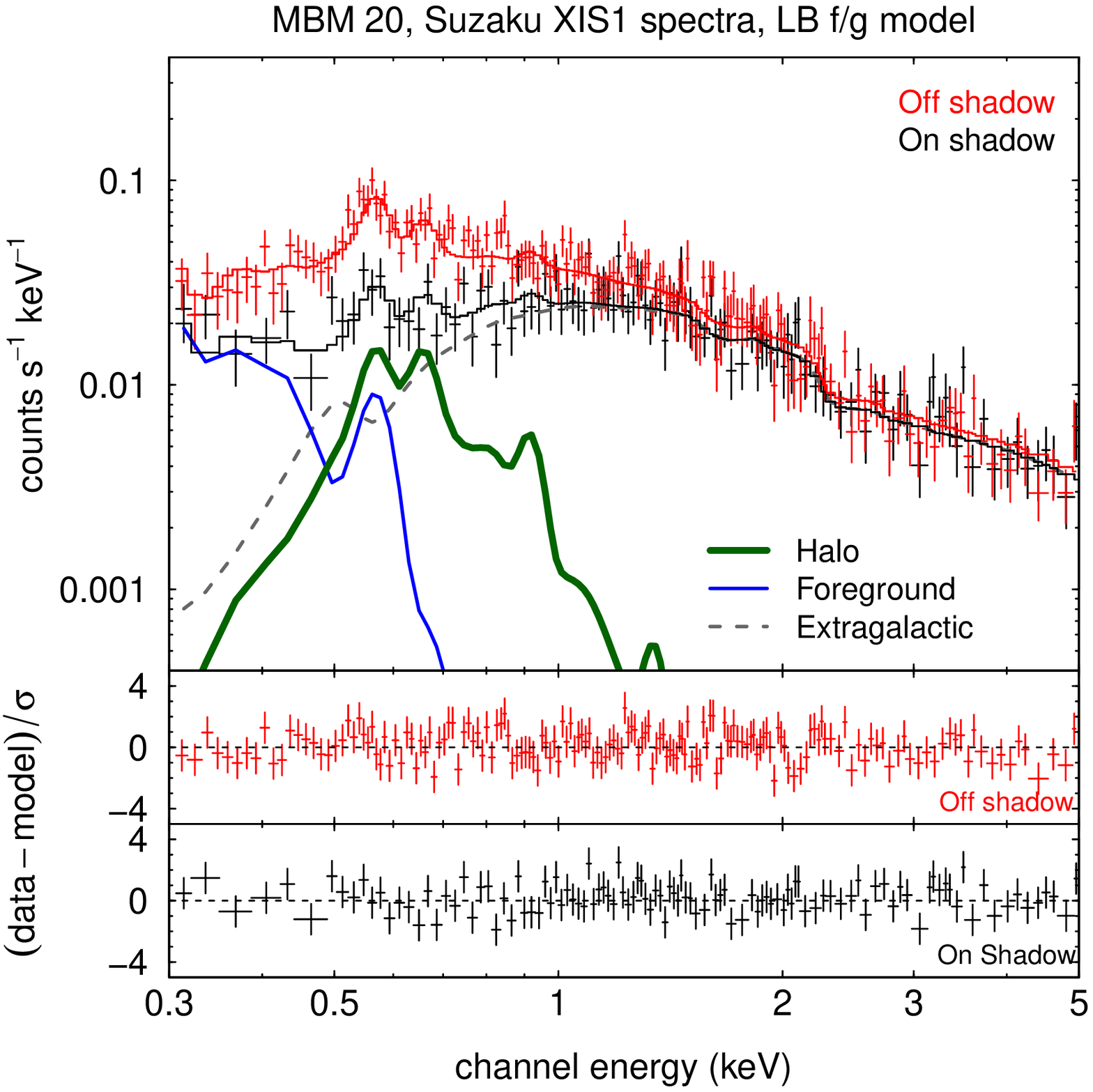}
\includegraphics[width=0.33\linewidth]{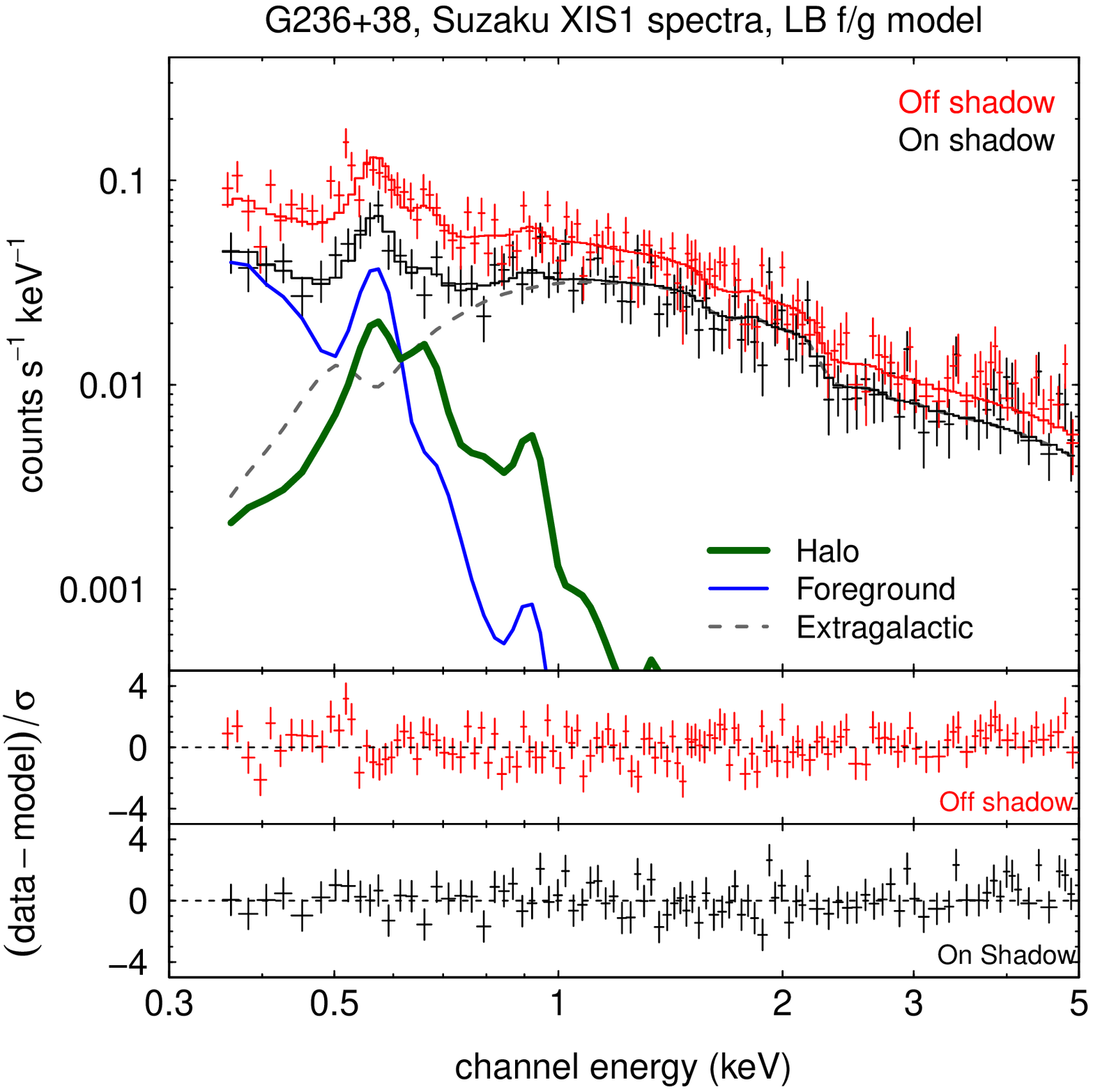} \\
\includegraphics[width=0.33\linewidth]{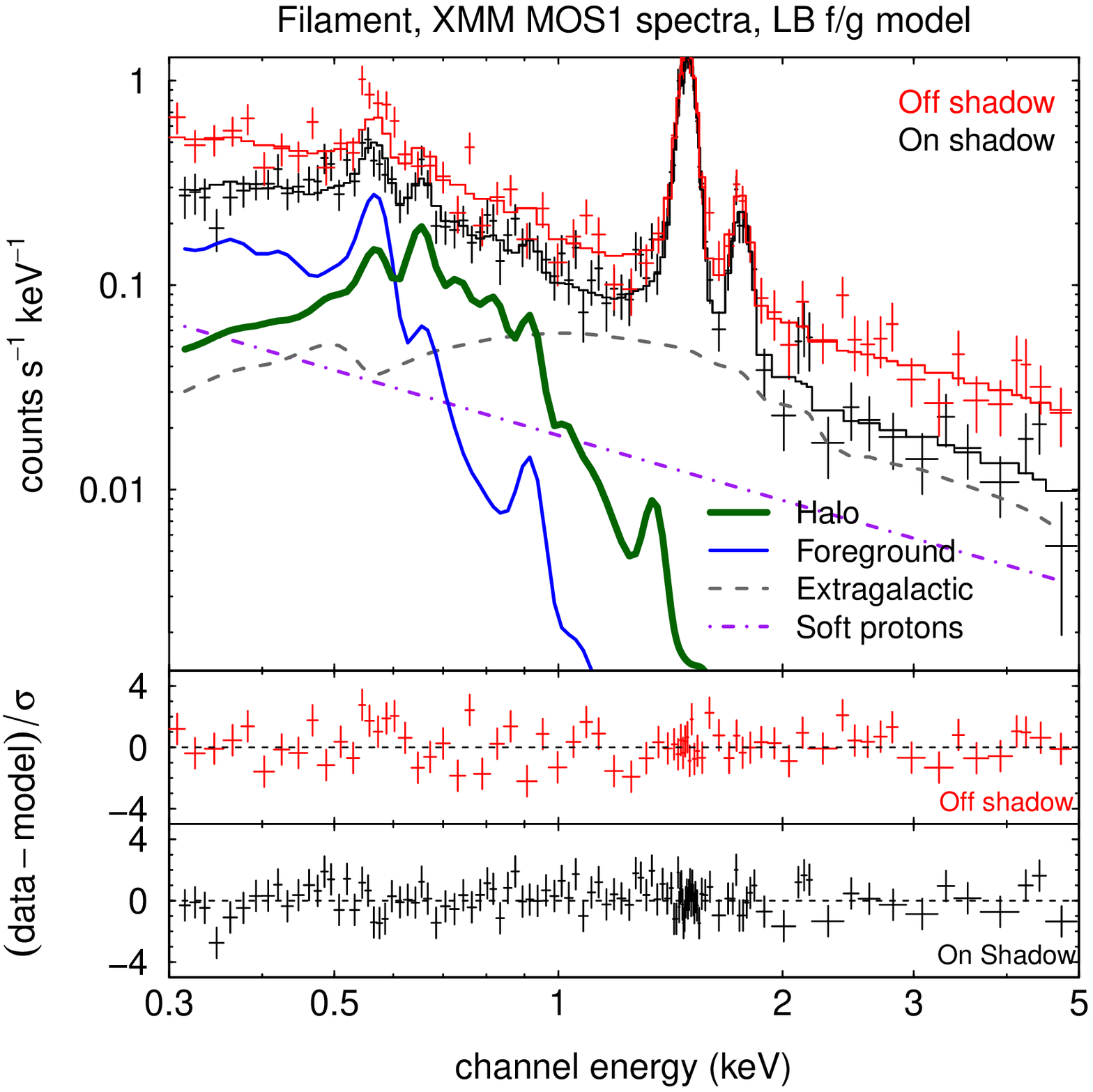}
\includegraphics[width=0.33\linewidth]{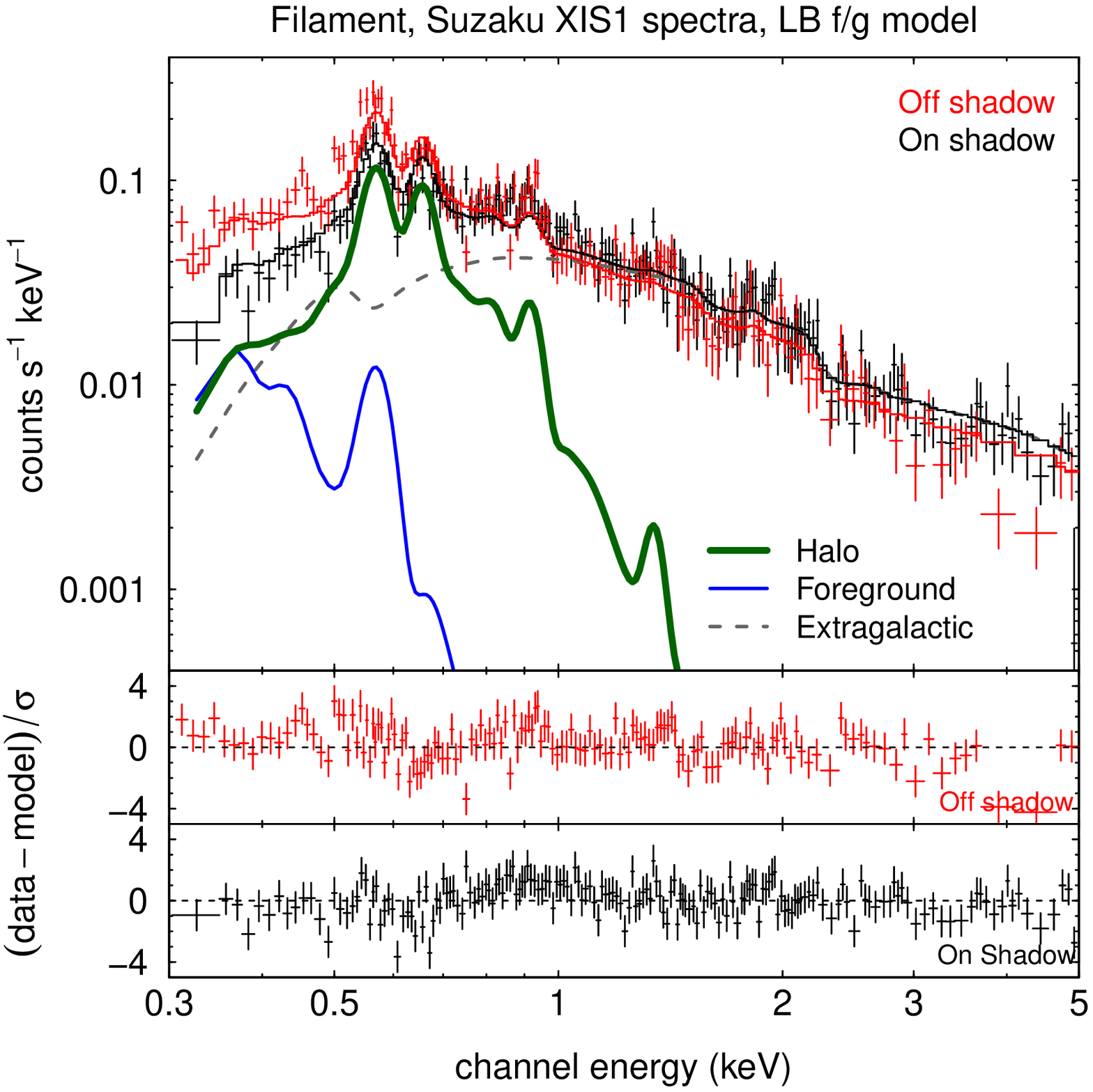}
\caption{Same as Figure~\ref{fig:SpectraSWCX}, but for the LB foreground model.
  \label{fig:SpectraLB}}
\end{figure*}

Our spectral fit results are presented in Table~\ref{tab:Results}. The headings within the body of
the table indicate the foreground model and abundances used for each set of results. Note in
particular that the results from our reference model are in Section~(a) of the table. For the
C14-SWCX model, the foreground \Kalpha\ intensities for \CV\ (if included), \CVI, \OVII, and
\OVIII\ are shown in Columns~4--7, respectively, in line units ($\LU = \lineunit$).  For the
ACX-SWCX and LB models, the temperature and normalization of the foreground component are shown in
Columns~2 and 3, respectively. In all cases, the halo temperature, \Th, and emission measure, \EMh,
are shown in Columns~8 and 9, respectively. Column~10 contains \chisq\ and the number of degrees of
freedom (dof) for each fit.  The intrinsic surface brightnesses of the foreground (0.4--1.0~\kev)
and halo (0.5--2.0~\kev) components of our models, \Sfg\ and \Sh, respectively, are shown in
Table~\ref{tab:SB}. Note that, because the foreground components are generally softer than the halo
components, the foreground surface brightnesses are for a lower energy band than the halo surface
brightnesses.

Figures~\ref{fig:SpectraSWCX}--\ref{fig:SpectraLB} show the observed on- and off-shadow spectra
along with the best fit spectral models obtained with the C14-SWCX, ACX-SWCX, and LB spectral
models, respectively (assuming Aspl09 abundances). In general the fits are good, typically with
$\rchisq \approx 1.0$--1.4, where \rchisq\ is the reduced \chisq.

\begin{figure}
\centering
\plotone{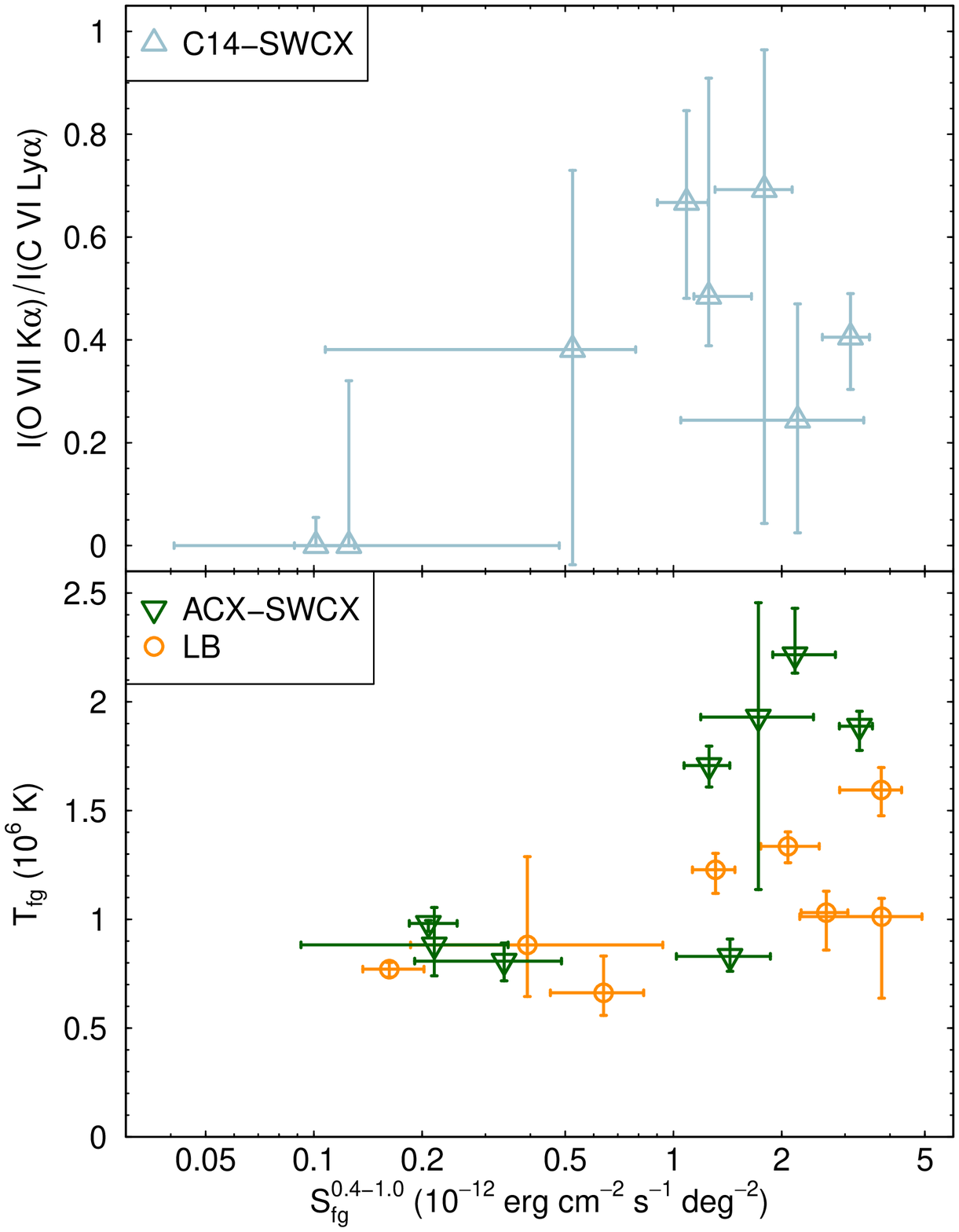}
\caption{Measures of the foreground hardness against the 0.4--1.0~\kev\ foreground surface
  brightness. The upper panel shows results for the C14-SWCX model, for which we use the
  \OVII/\CVI\ \Kalpha\ intensity ratio as our measure of the foreground hardness. The lower panel
  shows results for the ACX-SWCX (green triangles) and LB (orange circles) models, for which we use
  the temperature as our measure of the foreground hardness. All results were obtained using Aspl09
  abundances.
  \label{fig:ForegroundResults}}
\end{figure}
\begin{figure}
\centering
\plotone{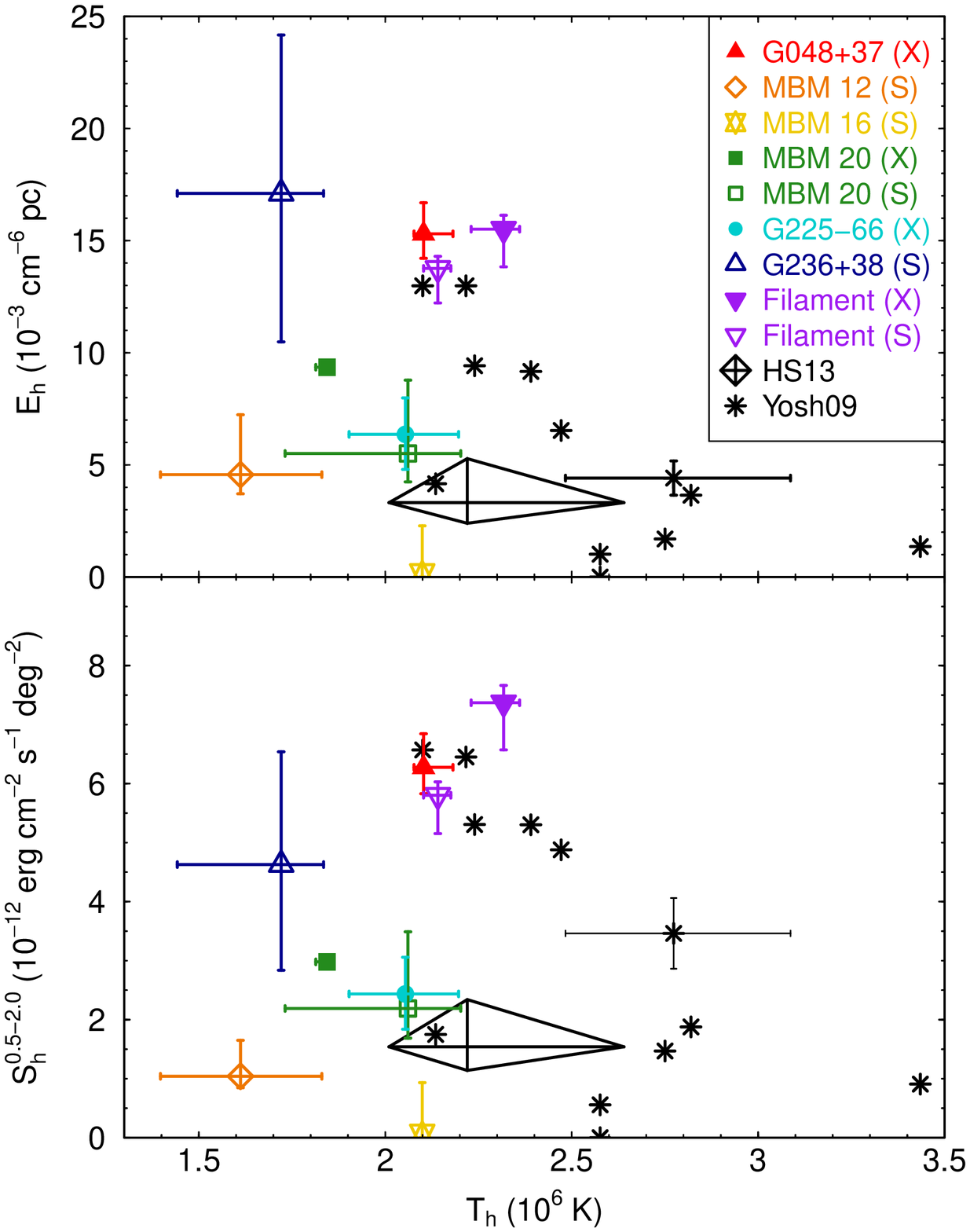}
\caption{Halo emission measure, \EMh, and intrinsic 0.5--2.0~\kev\ surface brightness, \Sh, against
  halo temperature, \Th, obtained with our reference model (C14-SWCX foreground model with Aspl09
  abundances). Each observation is plotted with its own symbol--see key for details (note that
  \xmm\ (X) and \suzaku\ (S) results are plotted with filled and open symbols, respectively, and
  that different observations of the same shadow are plotted in the same color).
  In addition, we show results for G225$-$66 obtained using the C14-SWCX model (\citealt{henley15a};
  note that we have rescaled their emission measure, as they assumed AG89 abundances; see
  footnote~\ref{fn:OxygenAbundanceG225}).
  The large black crossed diamonds indicate the median values and the lower and upper quartiles from
  HS13's \xmm\ survey of the halo emission, for sight lines on which such emission was detected
  (first row of their Table~2).
  The black stars indicate results from the Yosh09 \suzaku\ study of the halo (from their Table~6;
  we calculated the surface brightnesses using their best-fit temperature and emission measures,
  taking into account non-solar abundance ratios if necessary). To reduce clutter, we only show
  errors for a single, typical Yosh09 observation.
  As with G225$-$66, we have rescaled the HS13 and Yosh09 emission measures, because these results
  were also obtained using AG89 abundances.
  We will compare our results with those from these other studies in
  Section~\ref{subsubsec:HaloOtherStudies}.
  \label{fig:HaloResults}}
\end{figure}

The foreground results are plotted in Figure~\ref{fig:ForegroundResults}. For each foreground model,
we show a measure of the foreground hardness (the \OVII/\CVI\ ratio for C14-SWCX, the temperature
for ACX-SWCX and LB) against the 0.4--1.0~\kev\ surface brightness. The foreground surface
brightnesses span more than a order of magnitude, from $\sim$$10^{-13}$ to $\sim$$4 \times
10^{-12}~\flux\ \pdegsq$. There is a slight tendency for the brighter foreground models to be harder
(although this correlation is not statistically significant).  This is almost entirely due to the
\OVII\ \Kalpha\ emission---a harder foreground model means brighter \OVII\ emission, which means a
higher total surface brightness. The various foreground models do not always yield consistent
results---we will examine this further in Section~\ref{subsec:ResultsDifferentForegrounds}.

Figure~\ref{fig:HaloResults} shows the halo results obtained using our reference model. In addition,
we show results for G225$-$66 \citep{henley15a}.\footnote{\citet{henley15a} used AG89 abundances in
  their analysis of G225$-$66. As they discussed, because the halo emission in the
  \xmm/\suzaku\ band is dominated by oxygen \Kalpha\ emission, the best fit halo emission measure is
  approximately inversely proportional to the assumed oxygen abundance. We must allow for this if we
  wish to fairly compare the emission measures obtained with different abundance tables. Therefore,
  in Figure~\ref{fig:HaloResults} we have multiplied the emission measures obtained using AG89
  abundances by the ratio of the AG89 and Aspl09 oxygen abundances,
  $(8.51\times10^{-4})/(4.90\times10^{-4})=1.74$.\label{fn:OxygenAbundanceG225}} The measured
temperatures range from $1.6 \times 10^6$ to $2.3 \times 10^6~\K$, with detected emission measures
in the range $(\mbox{5--17}) \times 10^{-3}~\emismeas$ and intrinsic 0.5--2.0~\kev\ surface
brightness in the range $(\mbox{1--7}) \times 10^{-12}~\flux\ \pdegsq$.  With this model, we do not
detect halo emission in the direction of MBM~16, consistent with the result previously reported by
\citet{ursino14}.

Our sample includes two shadows---the filament and MBM~20---that have been observed more than once,
albeit with different observatories. For each shadow, while the foreground emission may be different
in the two observations,\footnote{For brevity, from here on we will use the word ``observation'' to
  refer jointly to the on- and off-shadow pointings.} due to changes in the SWCX emission, we should
obtain consistent halo measurements from the two observations. With our reference model, we obtain
consistent halo temperatures and surface brightnesses from the two MBM~20 observations (green
squares), in that the error bars overlap. The errors on the emission measures from these two
observations do not quite overlap.  For the two filament observations (purple triangles) we obtain
consistent halo emission measures, but not temperatures or surface brightnesses.

Our other spectral models yield broadly similar results for the halo---temperatures around $2 \times
10^6~\K$, and emission measures and surface brightnesses generally in the above-quoted ranges (after
allowing for different oxygen abundances if necessary; see
Section~\ref{subsec:ResultsDifferentAbundances}). However, there are differences in the details
among some of the models. In the following sections, we will examine the foreground and halo results
obtained using different foreground models (Section~\ref{subsec:ResultsDifferentForegrounds}) and
different abundances (Section~\ref{subsec:ResultsDifferentAbundances}). Finally in this section, we
will compare our results with those from previous studies of the same shadows, where available
(Section~\ref{subsec:CompareWithPrevious}).

\subsection{Results Obtained with Different Foreground Models}
\label{subsec:ResultsDifferentForegrounds}

\begin{figure*}
\centering
\plottwo{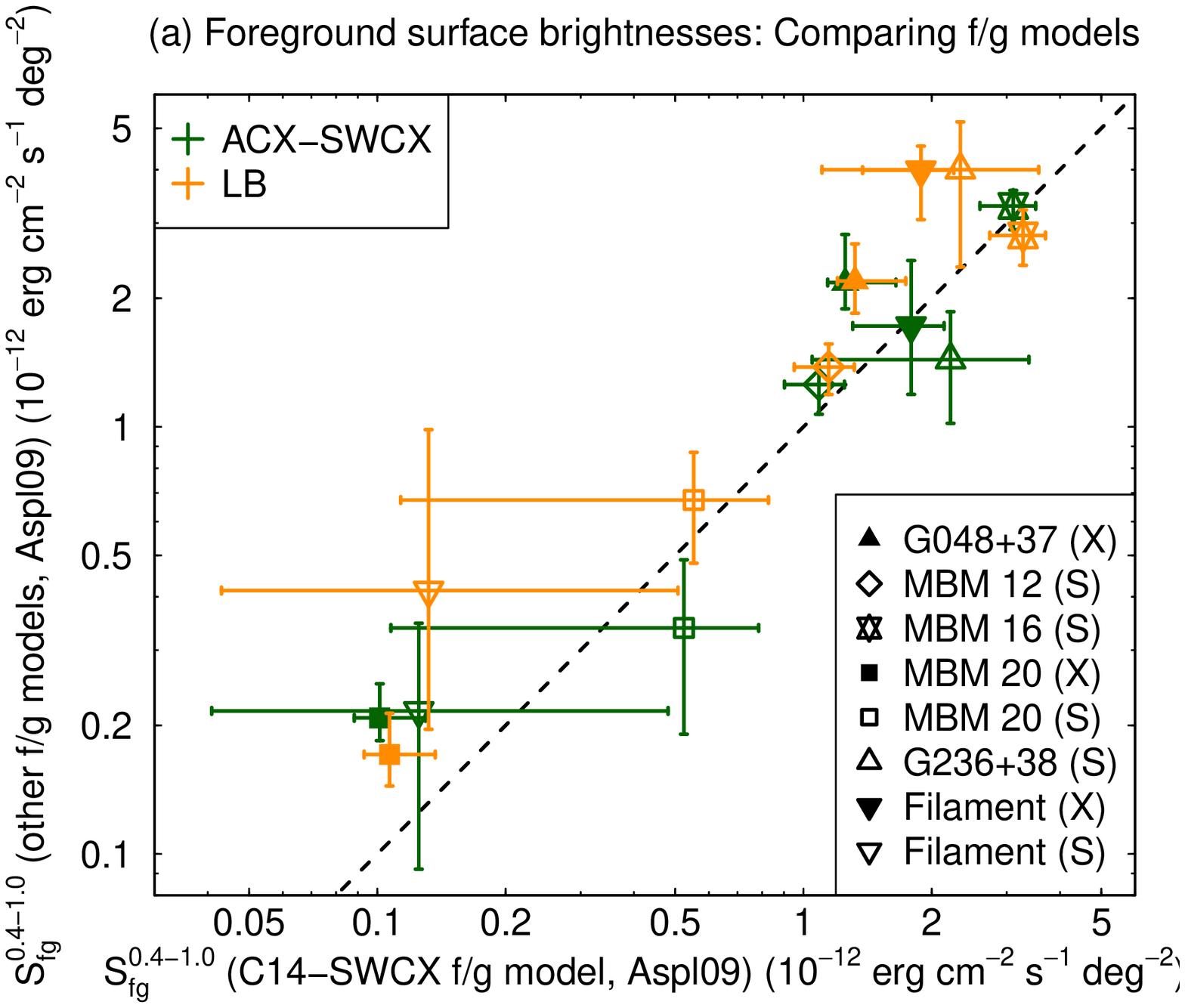}{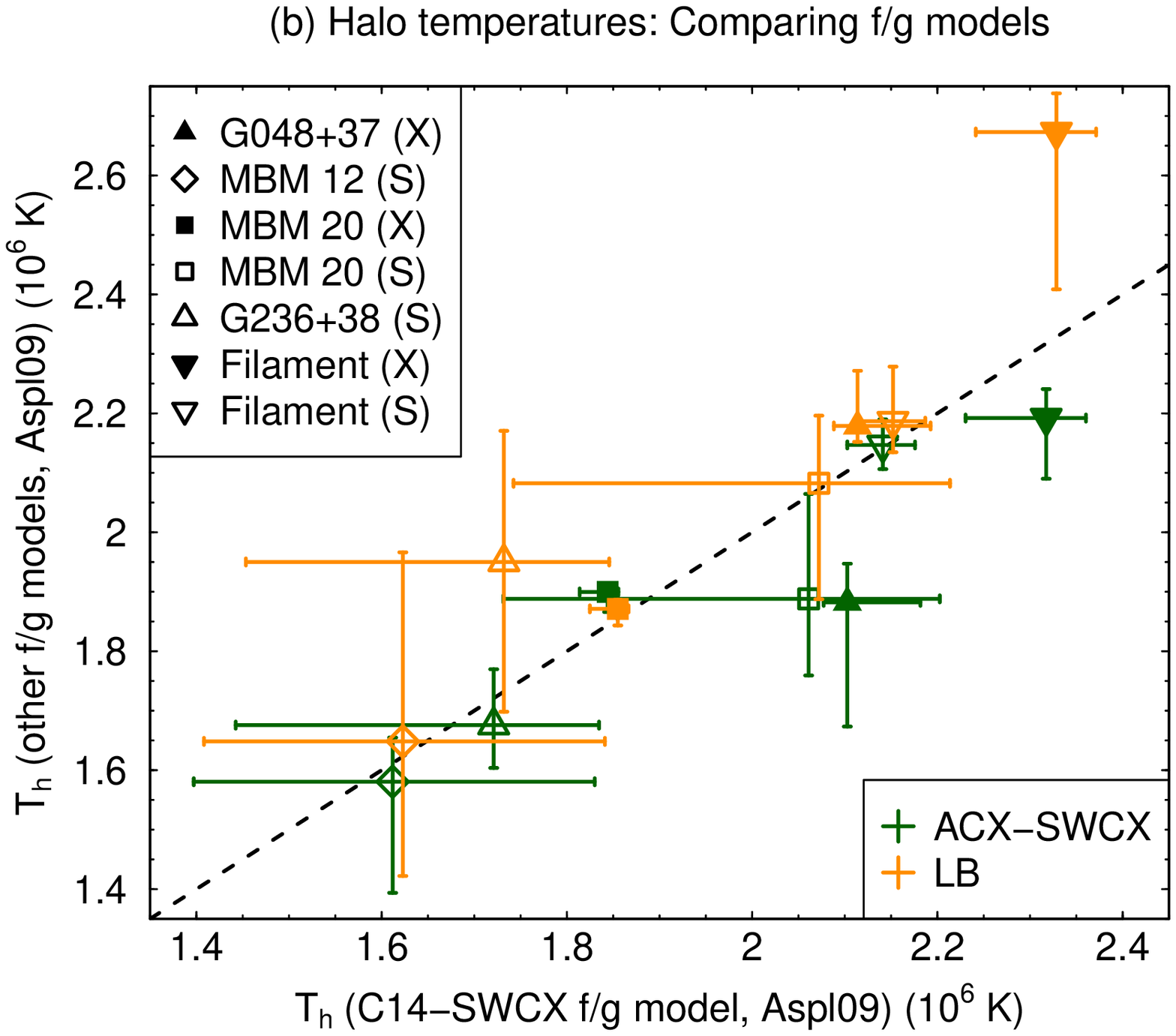} \\
\plottwo{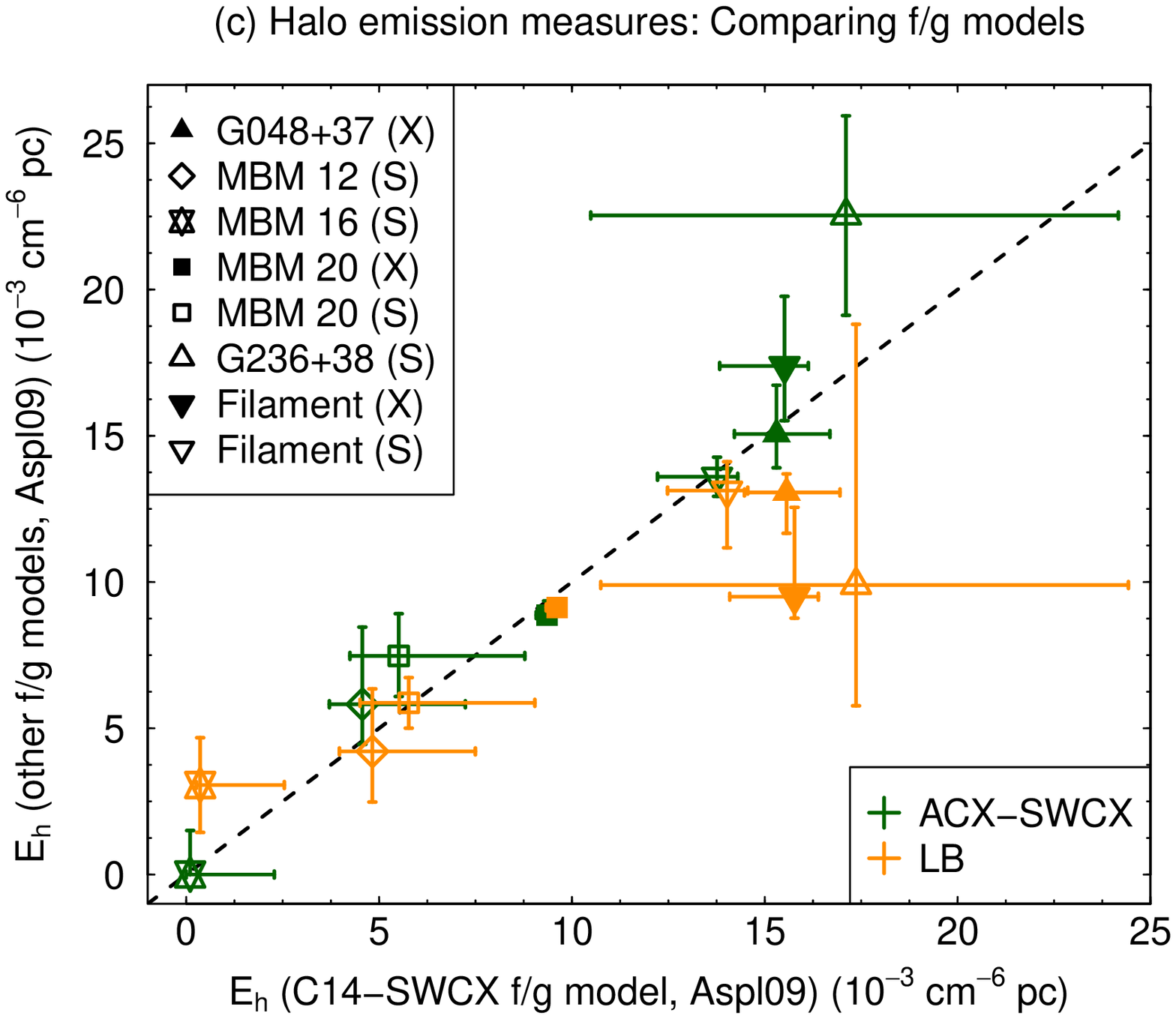}{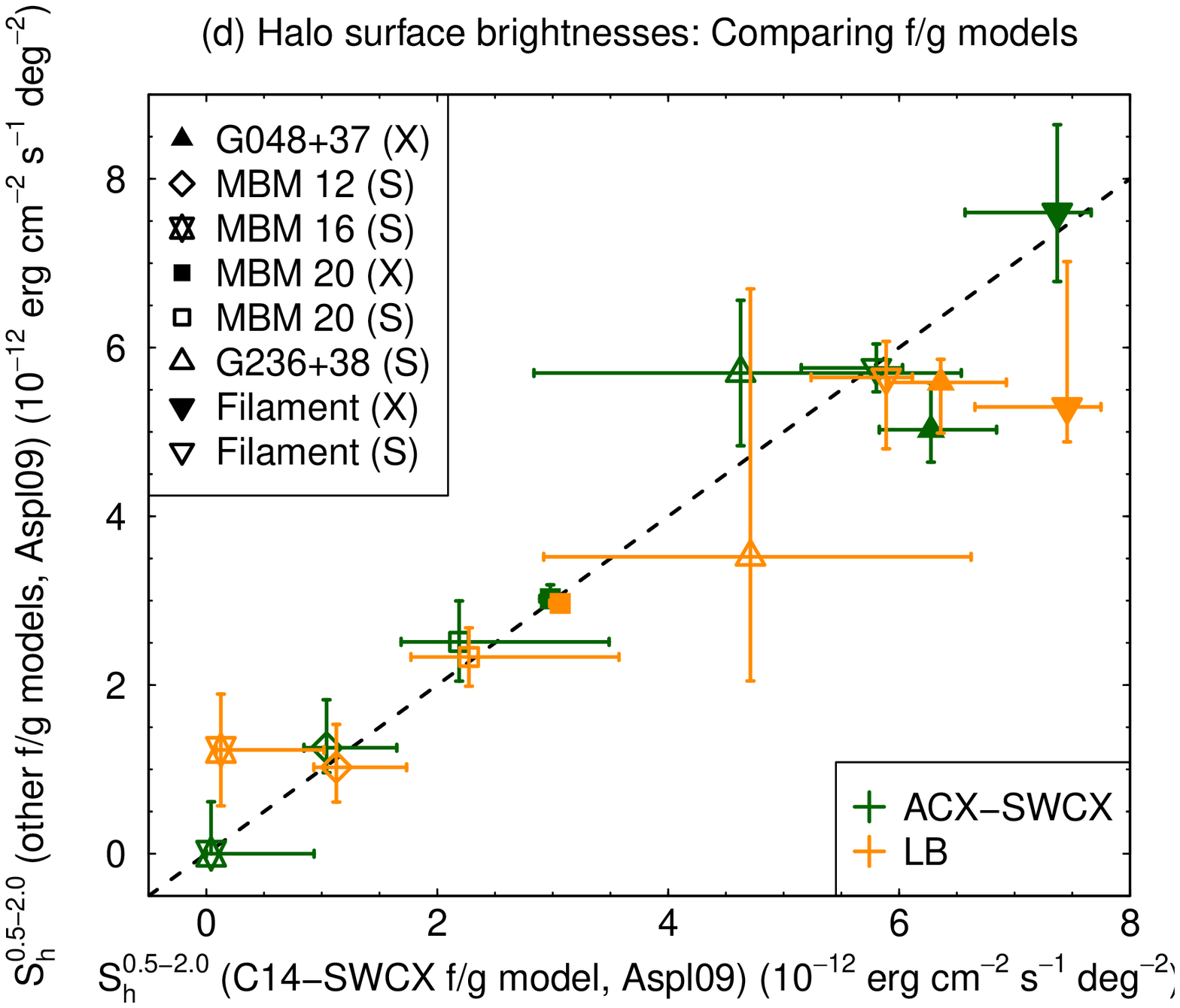}
\caption{Comparison of (a) the foreground 0.4--1.0~\kev\ surface brightnesses, (b) the halo
  temperatures, (c) the halo emission measures, and (d) the halo intrinsic 0.5--2.0~\kev\ surface
  brightnesses obtained using different foreground models.
  Each panel compares the values obtained using the ACX-SWCX and LB foreground models (green
  and orange symbols, respectively) with those obtained using our reference model (C14-SWCX
  foreground model).
  The various shadows' results are plotted with different symbols, using the same symbols
  (but not colors) used in Figure~\ref{fig:HaloResults}.
  All results were obtained using Aspl09 abundances.
  To avoid overlapping error bars, we have shifted the points comparing the LB results with the
  C14-SWCX results upward and to the right (by 1\%\ of the range of each horizontal axis).
  Note that panel~(a) has logarithmic axes, while the other panels have linear axes.
  Note also that the halo temperature plot (panel~(b)) does not include MBM~16, as the halo
  temperature was not a free parameter for this shadow (see Section~\ref{sec:SpectralModel}).
  In all panels, the dashed line indicates equality.
  \label{fig:CompareFGModels}}
\end{figure*}

Figure~\ref{fig:CompareFGModels} compares the foreground and halo results obtained using our
different foreground models. In general, the various foreground models yield consistent foreground
surface brightnesses (Figure~\ref{fig:CompareFGModels}(a)). However, in a couple of cases, the
ACX-SWCX model yields a lower best-fit foreground surface brightness than the C14-SWCX model
(specifically G236+38 and the \suzaku\ observation of MBM~20, although it should be noted that the
differences are not statistically significant). If we compare the relevant panels in
Figure~\ref{fig:SpectraSWCX} (C14-SWCX) and Figure~\ref{fig:SpectraACX} (ACX-SWCX), we see that
these lower ACX-SWCX surface brightnesses are due to the absence of foreground
\OVII\ \Kalpha\ emission at $\sim$0.57~\kev\ in the ACX-SWCX models.  The foreground
\OVII\ \Kalpha\ emission is also fainter with the ACX-SWCX model than with the C14-SWCX model for
MBM~12 and the \xmm\ observation of the filament. In contrast, for G048+37, the ACX-SWCX model
and C14-SWCX models attribute similar amounts of \OVII\ emission to the foreground, but the
foreground \OVIII\ emission is brighter in the ACX-SWCX model.

If using the ACX-SWCX model results in less \OVII\ emission being attributed to the foreground, then
naturally more \OVII\ emission is attributed to the halo. The amount of \OVIII\ emission attributed
to the halo, meanwhile, is typically unaffected, and so the halo \OVII/\OVIII\ ratio tends to
increase. In the case of G048+37, attributing more \OVIII\ to the foreground results in less being
attributed to the halo, which also increases the halo \OVII/\OVIII\ ratio. Thus, in some cases, we
find that the ACX-SWCX foreground model yields lower halo temperatures than the C14-SWCX model
(Figure~\ref{fig:CompareFGModels}(b)). Correspondingly, there is a slight tendency for the ACX-SWCX
model to yield higher halo emission measures (Figure~\ref{fig:CompareFGModels}(c)). This is because
the \OVIII\ emissivity is an increasing function of temperature around $T=2\times10^6~\K$. If the
halo \OVIII\ emission is unaffected by a change in the foreground model, a decrease in the best-fit
halo temperature must be compensated by an increase in the halo emission measure.

In general the ACX-SWCX foreground model yields similar halo surface brightnesses to the C14-SWCX
model (Figure~\ref{fig:CompareFGModels}(d)), because any changes in the foreground surface
brightness are typically small compared with the halo surface brightness. However, for G048+37
(G236+38) the ACX-SWCX foreground surface brightness is $\sim$$1\times10^{-12}~\flux\ \pdegsq$
higher (lower) than that obtained with the C14-SWCX model, which is compensated for by a
corresponding decrease (increase) in the halo surface brightness.

For seven out of eight sight lines, using a LB foreground model resulted in a larger foreground
surface brightness than did using a C14-SWCX foreground model. For six of these seven sight lines
the LB model also yielded a lower halo surface brightness. These differences, however, are generally
small (typically $\la$0.2~dex for the foreground surface brightness) and not statistically
significant considering the sizes of the error bars. In general, the LB foreground models are
similar in overall spectral shape to the C14-SWCX models. As a result, the LB and C14-SWCX models
generally yield similar halo results. However, there are exceptions to this general rule, which
occur when the foreground surface brightness is relatively bright.  These exceptions include G236+38
and the \xmm\ observation of the filament. The LB model attributes more \OVII\ emission to the
foreground than the C14-SWCX model for these observations (compare the relevant panels in
Figures~\ref{fig:SpectraSWCX} and \ref{fig:SpectraLB}).  As a result, for these sight lines the LB
model yields a higher foreground surface brightness (by a factor of $\sim$2), a higher halo
temperature (because less \OVII\ emission is attributed to the halo), a lower halo emission measure,
and a lower halo surface brightness (although for G236+38 the differences are not statistically
significant). Another exception is MBM~16---with the LB model we detect halo emission, which we do
not with the C14-SWCX model. This is likely because, in order to fit the \OVII\ emission and the
softer emission, the LB component's temperature is such that there is little emission attributed to
the foreground above $\sim$0.6~\kev. Therefore, non-zero halo emission is needed to adequately fit
the spectra in this energy range. In contrast, the C14-SWCX model can fit the emission above
$\sim$0.6~\kev\ with the foreground component, leading to a better fit (note that MBM~16 is the only
observation for which we detect foreground \OVIII\ emission with the C14-SWCX model).

In summary, for most sight lines the different foreground models yield consistent results, within
the error bars. However, the ACX-SWCX model has a tendency to attribute less \OVII\ emission to the
foreground than the other foreground models. We will discuss why this is the case in
Section~\ref{subsec:ForegroundChoice}, but we will note here that it is likely an artifact of the
assumption that the solar wind ion distribution is described by a single temperature. Since this is
not a good description of the true solar wind ion distribution \citep{vonsteiger00}, the ACX-SWCX
results may be unreliable. In addition, for a few sight lines on which the foreground surface
brightness is relatively bright the LB and C14-SWCX foreground models yield discrepant results.  We
will discuss the implications of this for the halo results in
Section~\ref{subsubsec:HaloUncertaintyBias}.

\subsection{Results Obtained with Different Abundance Tables}
\label{subsec:ResultsDifferentAbundances}

\begin{figure*}
\centering
\plottwo{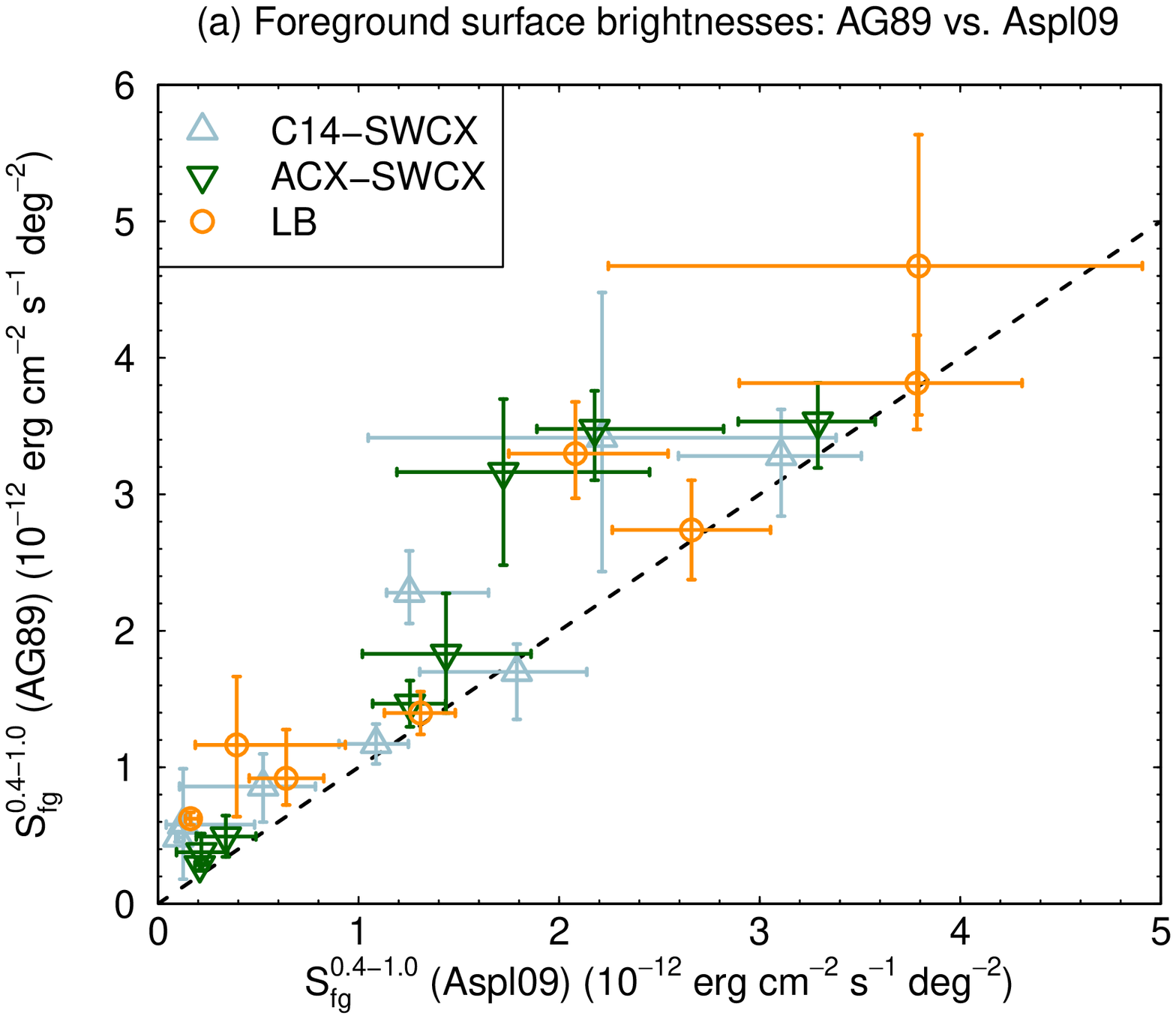}{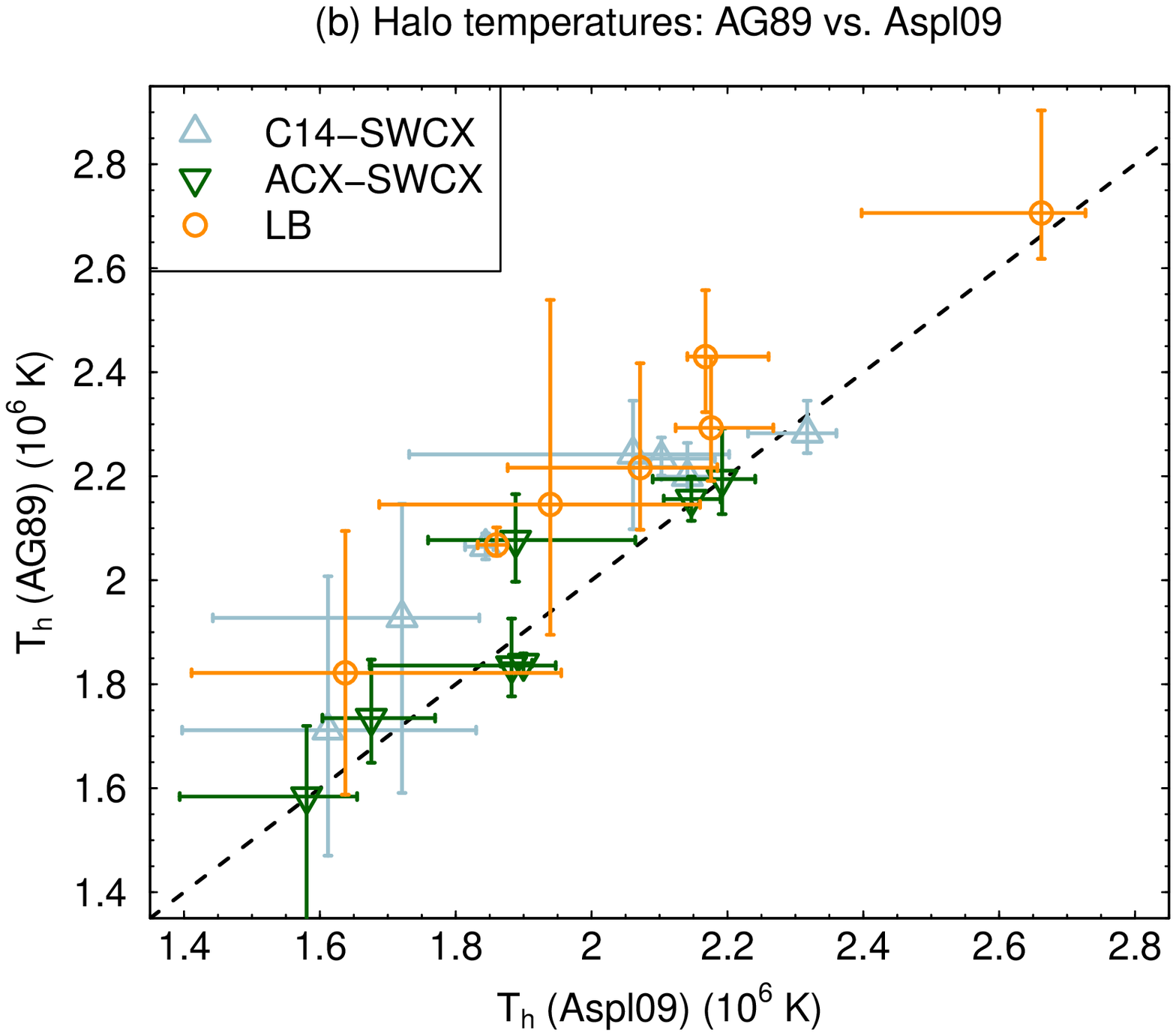} \\
\plottwo{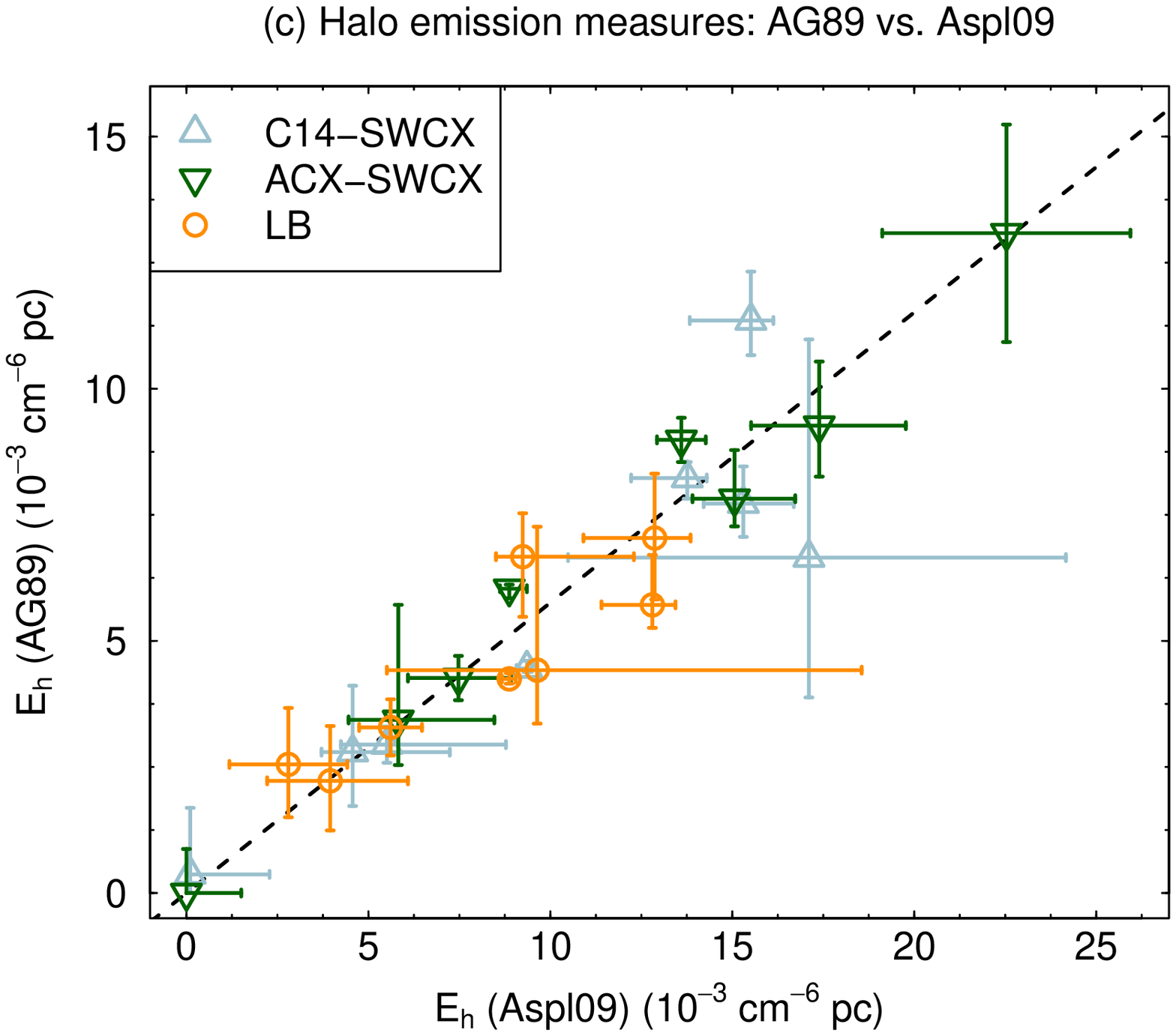}{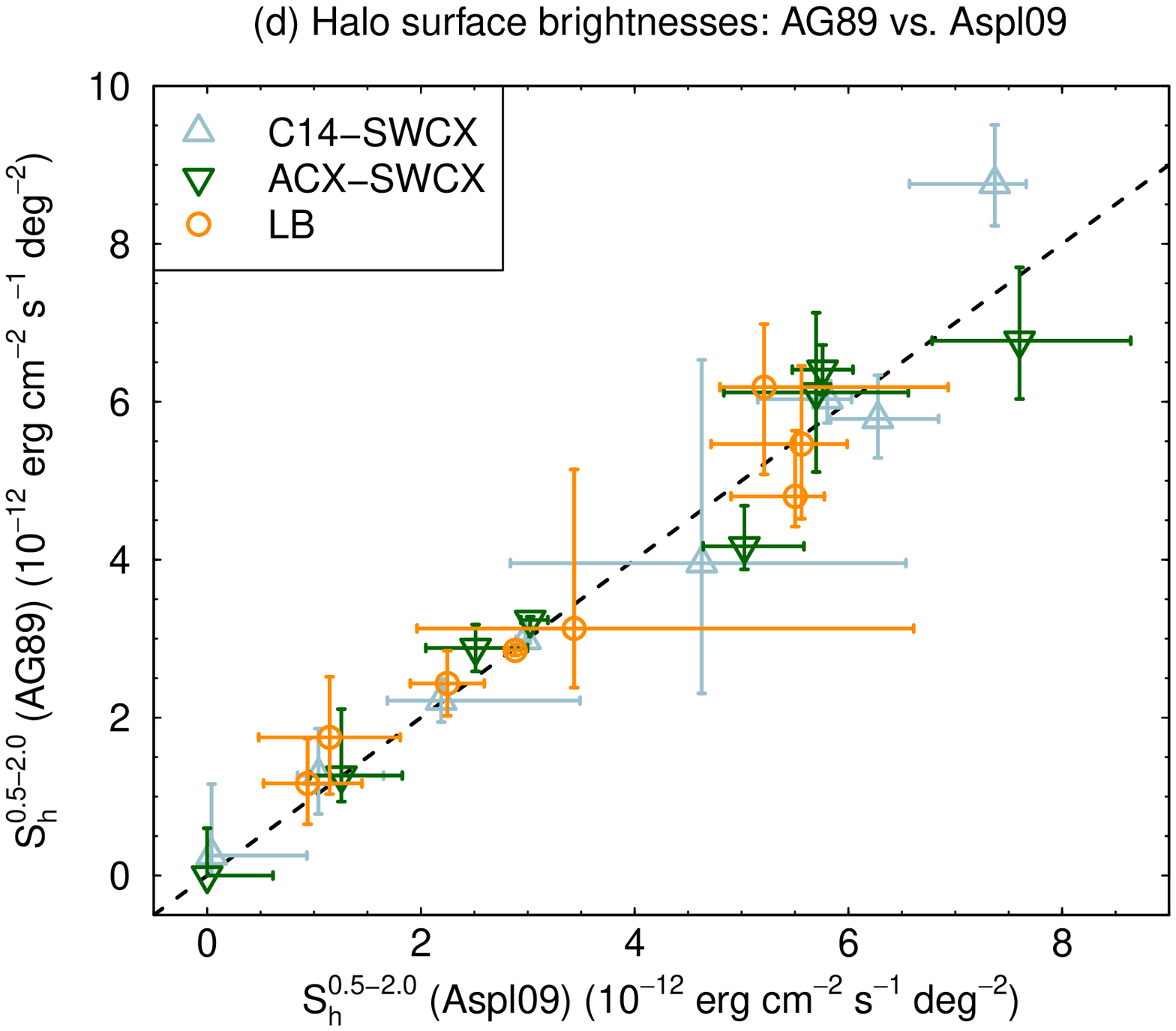}
\caption{Comparison of (a) the foreground 0.4--1.0~\kev\ surface brightnesses, (b) the halo
  temperatures, (c) the halo emission measures, and (d) the halo intrinsic 0.5--2.0~\kev\ surface
  brightnesses obtained using AG89 (ordinates) and Aspl09 (abscissae) abundances, for each of
  our three foreground models (see key).
  Note that the halo temperature plot (panel~(b)) does not include MBM~16, as the halo temperature
  was not a free parameter for this shadow (see Section~\ref{sec:SpectralModel}).
  In general, the dashed lines indicate equality. However, in the halo emission measure plot
  (panel~(c)), the gradient of the dashed line is equal to the ratio of the Aspl09 and AG89 oxygen
  abundances (0.576), rather than 1 (see footnote~\ref{fn:OxygenAbundanceGradient}).
  \label{fig:AsplvAG}}
\end{figure*}

\begin{figure*}
\centering
\plottwo{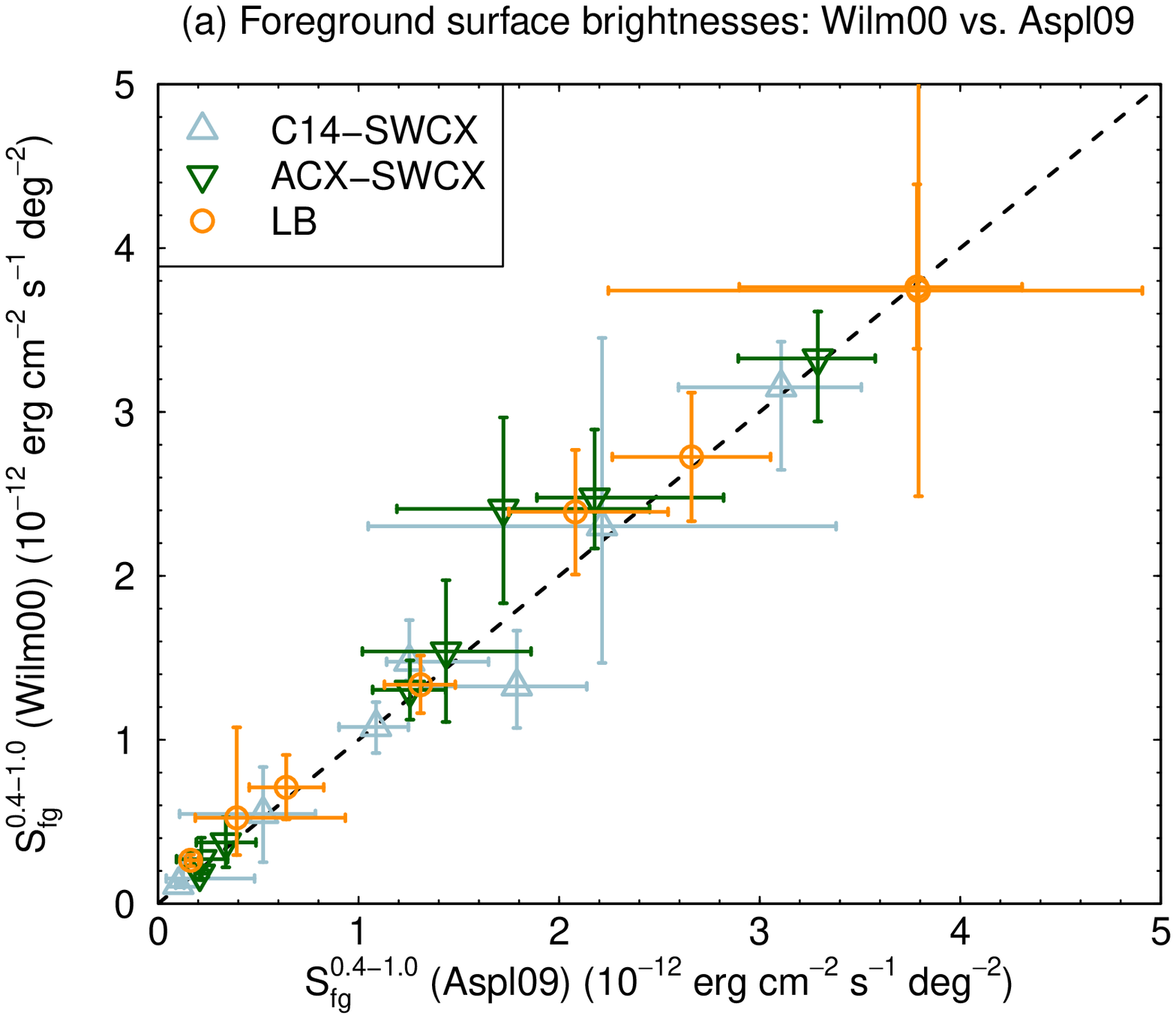}{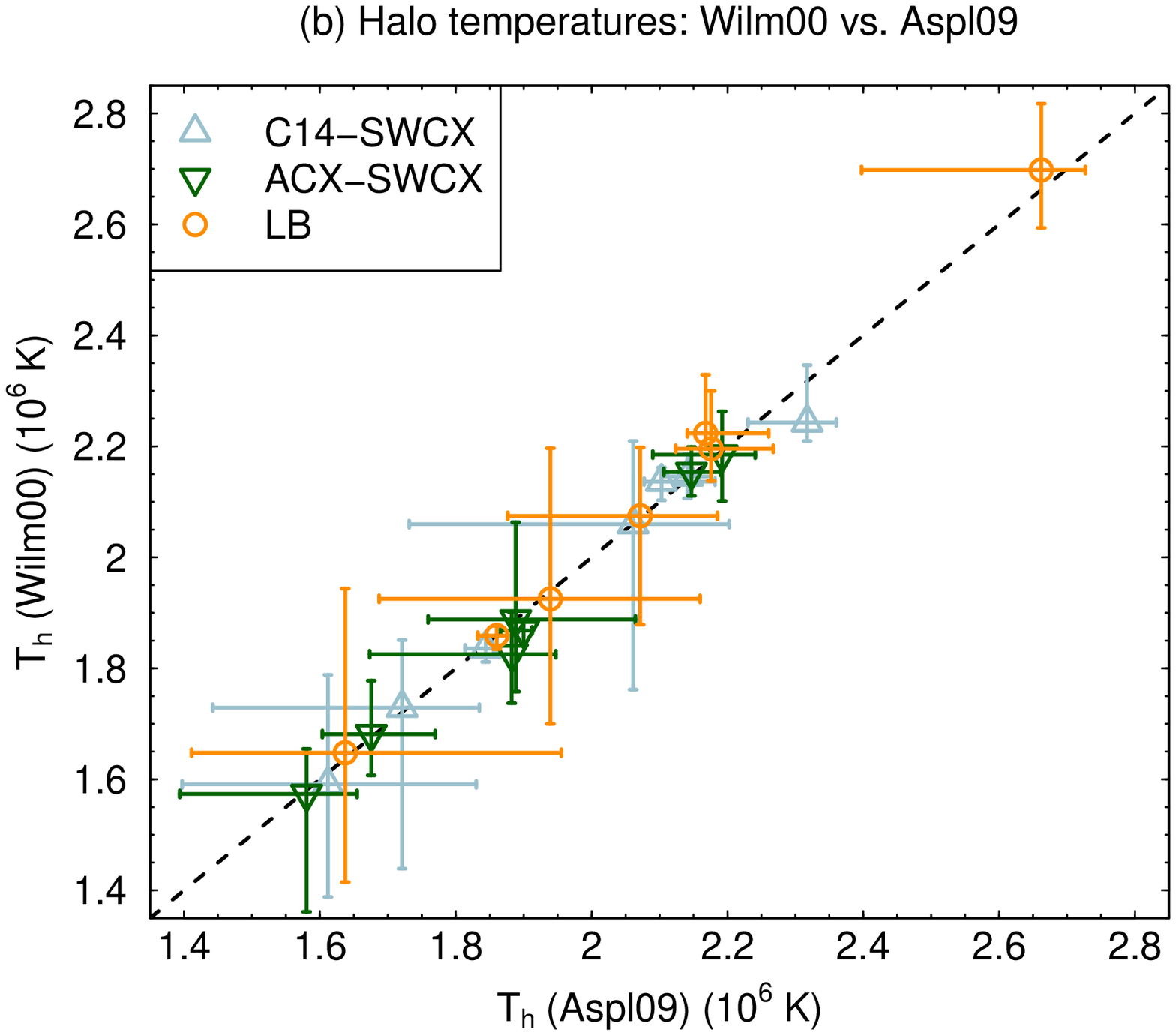} \\
\plottwo{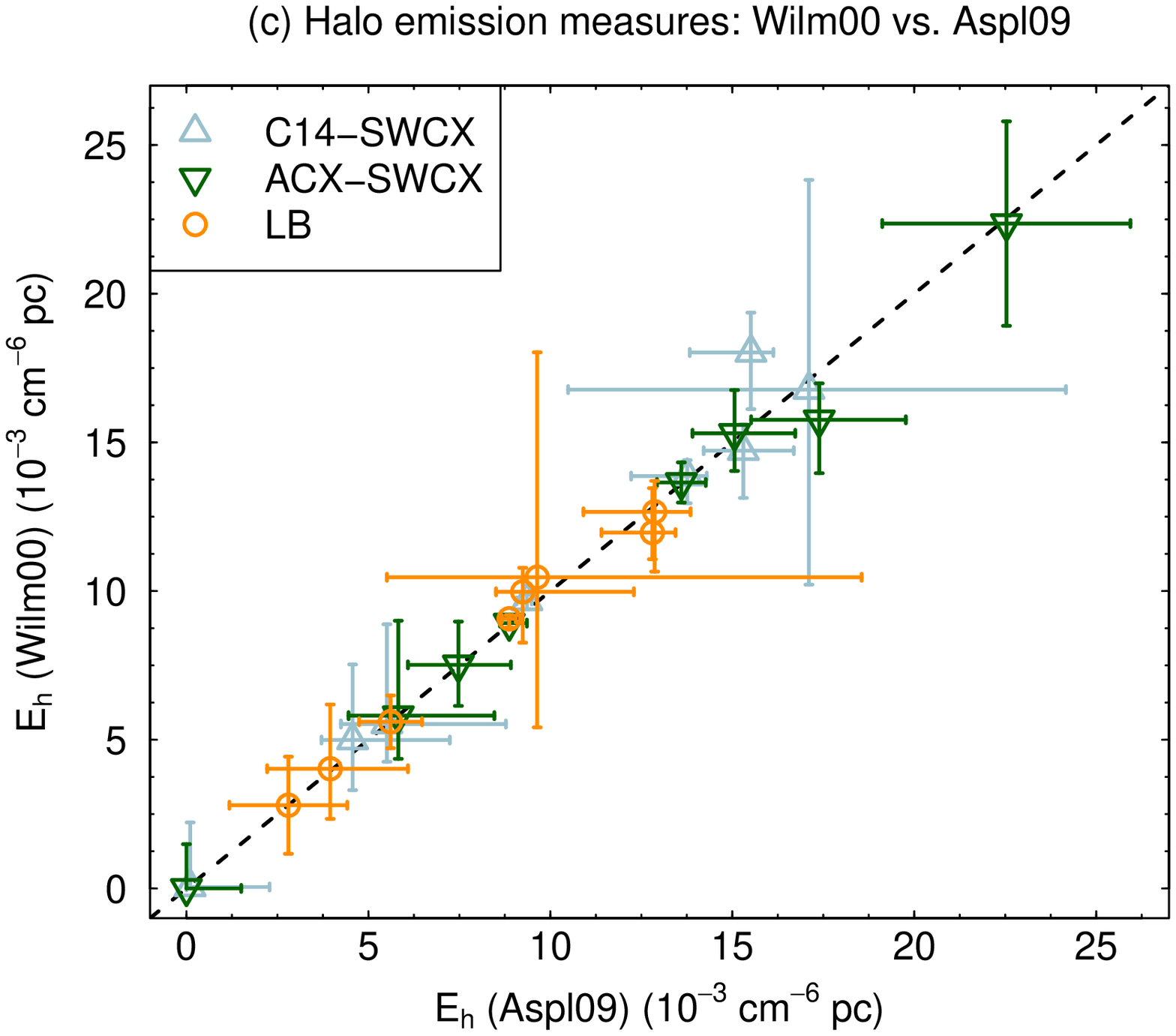}{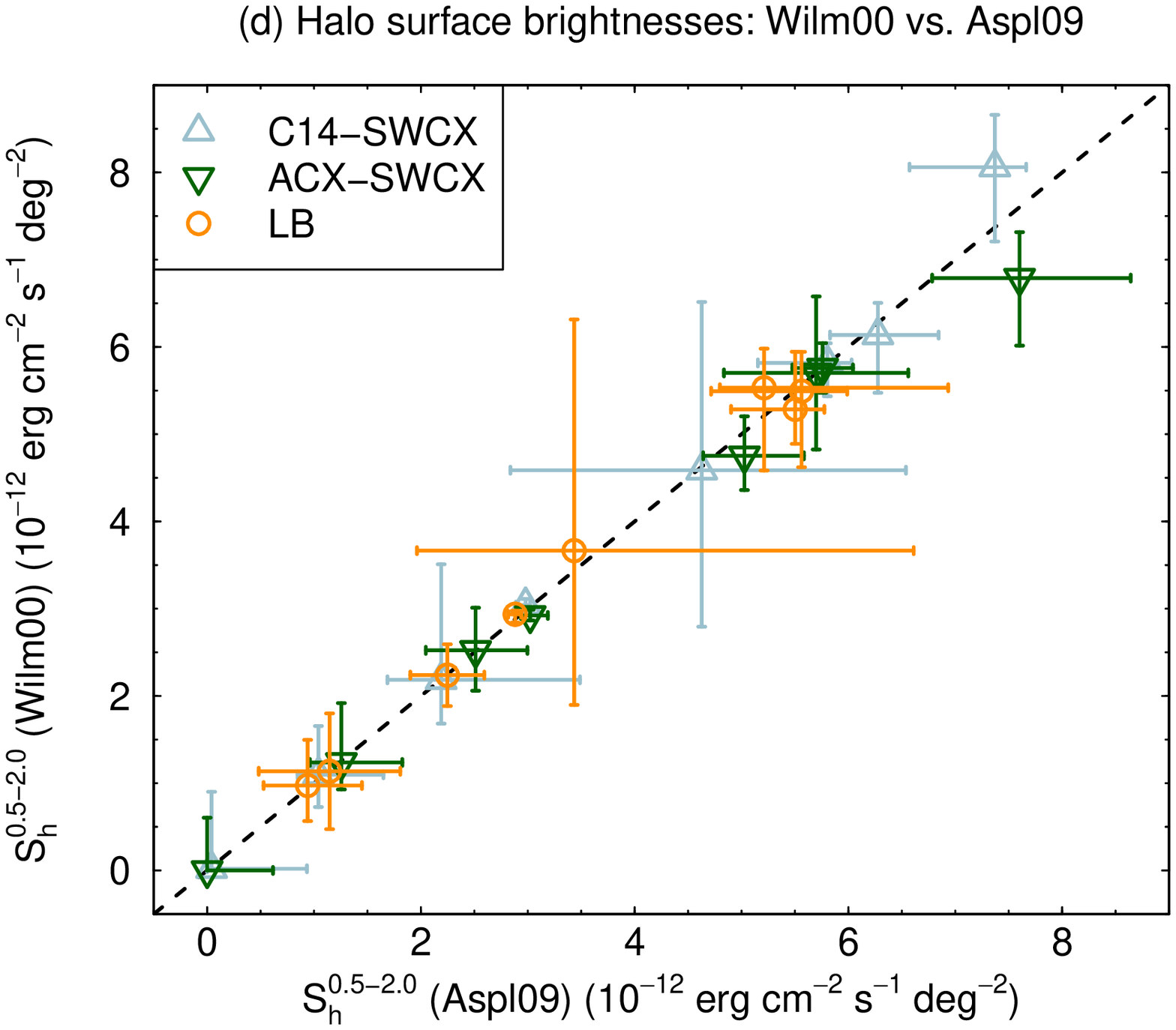}
\caption{Same as Figure~\ref{fig:AsplvAG}, but comparing results obtained with Wilm00 (ordinates)
  and Aspl09 (abscissae) abundances. However, unlike Figure~\ref{fig:AsplvAG}, the gradient of the
  dashed line in panel~(c) is 1, as the oxygen abundance is the same in the Wilm00 and Aspl09
  abundance tables.
  \label{fig:AsplvWilm}}
\end{figure*}

Figure~\ref{fig:AsplvAG} compares the foreground and halo results obtained using Aspl09 and AG89
abundances. The AG89 abundances result in systematically larger foreground surface brightnesses than
the Aspl09 abundances (Figure~\ref{fig:AsplvAG}(a)). This difference can be understood, at least
qualitatively, from the fact that the AG89 carbon, nitrogen, and oxygen abundances are higher than
the Aspl09 abundances (by 0.13, 0.22, and 0.24~dex, respectively), resulting in larger photoelectric
absorption cross-sections at the emission energies being considered. Increasing the absorption
cross-section increases the foreground surface brightness inferred from the observations, as we
shall now demonstrate.

If, for a given energy band, the foreground, halo, and extragalactic surface brightnesses are \SFG,
\SH, and \SEG, respectively, then the observed on- and off-shadow surface brightnesses are
\begin{equation}
  \Son = \SFG + \SH e^{-\sigma \NHon} + \SEG e^{-\sigma \NHon}
\end{equation}
and
\begin{equation}
  \Soff = \SFG + \SH e^{-\sigma \NHoff} + \SEG e^{-\sigma \NHoff},
\end{equation}
respectively, where $\sigma$ is the band-averaged absorption cross-section, and \NHon\ and
\NHoff\ are the on- and off-shadow hydrogen column densities, respectively.\footnote{Note that,
  because we are not considering absorption at a single energy, $\sigma$ will be a function of
  \NH\ and of the assumed background spectrum \citep{snowden94a}. However, Figure 4(b) in
  \citet{snowden94a} shows that, for a halo temperature of $\ga$$2 \times 10^6~\K$, the
  band-averaged absorption cross-section for the \rosat\ 3/4~\kev\ band does not vary strongly with
  \NH\ for the range of values considered here (i.e., $\mathrm{few} \times 10^{20}$ to $\mathrm{few}
  \times 10^{21}~\pcmsq$). Therefore, we do not expect $\sigma$ for the 0.4--1.0~\kev\ band to vary
  strongly over this \NH\ range.}  Solving the above equations for \SFG, we obtain
\begin{equation}
  \SFG = \frac{\Son e^{\sigma \NHon} - \Soff e^{\sigma \NHoff}}{e^{\sigma \NHon} - e^{\sigma \NHoff}},
  \label{eq:SFG}
\end{equation}
the derivative of which with respect to $\sigma$ is
\begin{equation}
  \frac{d \SFG}{d \sigma} = \frac{(\Soff - \Son) (\NHon - \NHoff) e^{\sigma(\NHon+\NHoff)}}{\left( e^{\sigma\NHon} - e^{\sigma\NHoff} \right)^2}.
  \label{eq:Deriv}
\end{equation}
Because $\Soff > \Son$ and $\NHon > \NHoff$, $d\SFG/d\sigma > 0$. Hence, we expect the larger
photoelectric absorption cross-sections resulting from the AG89 abundances to lead to larger
foreground surface brightnesses being inferred from the observed surface brightnesses \Son\ and
\Soff. Note, however, that the above analysis doesn't take into account changes in the shapes of the
spectral components that will also occur when the abundances are changed. In practice, the
differences between the foreground 0.4--1.0~\kev\ surface brightnesses measured using AG89 and
Aspl09 abundances (Table~\ref{tab:SB}) imply values of $d\SFG/d\sigma$ that are typically $\sim$2--3
times smaller than the values predicted by Equation~(\ref{eq:Deriv}) (the predicted values are
typically $\sim$$10^{-12}~\flux\ \pdegsq\ (10^{-22}~\cm^2)^{-1}$).\footnote{For these calculations
  we used cross sections of $5.15\times10^{-22}$ and $6.75\times10^{-22}~\cm^2$ for Aspl09 and AG89
  abundances, respectively, calculated for a $2.1\times10^6~\K$ plasma observed through an absorbing
  column density of $10^{21}~\pcmsq$.}

As the foreground emission is generally softer than the halo emission, increasing the foreground
surface brightness decreases the softer halo emission relative to the harder halo emission.  As a
result, the AG89 abundances generally yield higher halo temperatures than the Aspl09 abundances
(Figure~\ref{fig:AsplvAG}(b)), particularly for the C14-SWCX and LB foreground models.  Associated
with these higher halo temperatures are lower halo emission measures (lower than can be accounted
for by the difference in oxygen abundance;\footnote{As noted in
  footnote~\ref{fn:OxygenAbundanceG225}, we must account for differences in the oxygen abundance
  when comparing emission measures obtained using different abundance tables. Therefore, in the halo
  emission measure plot (Figure~\ref{fig:AsplvAG}(c)), the gradient of the dashed line is equal to
  the ratio of the Aspl09 and AG89 oxygen abundances (0.576). Setting the line gradient to this
  value helps us identify differences in the halo emission measures that are not simply due to the
  difference in the assumed oxygen abundance.\label{fn:OxygenAbundanceGradient}}
Figure~\ref{fig:AsplvAG}(c)).  Despite these differences, AG89 and Aspl09 abundances generally yield
similar halo surface brightnesses (Figure~\ref{fig:AsplvAG}(d)). The reason is likely the same as
that mentioned in Section~\ref{subsec:ResultsDifferentForegrounds}---the changes in the foreground
surface brightness resulting from changing the assumed abundances are generally small relative to
the total halo surface brightness.

Figure~\ref{fig:AsplvWilm} is similar to Figure~\ref{fig:AsplvAG}, but compares the Wilm00 results
with the Aspl09 results. In general, the two sets of results are in good agreement, which is likely
mainly due to the similarity of the carbon, nitrogen, and oxygen abundances, and hence of the
photoelectric absorption cross-sections.

In summary, while the Wilm00 and Aspl09 abundance tables generally yield very similar results, there
are systematic differences between these results and those obtained using AG89 abundances. These
differences result, at least in part, from differences in the photoelectric absorption
cross-sections. Since the AG89 carbon, nitrogen, and oxygen abundances are likely too high for the
sun and for the interstellar medium, the results obtained with this abundance table are likely
unreliable.

\subsection{Comparison with Previous Shadowing Studies}
\label{subsec:CompareWithPrevious}

\begin{figure}
\centering
\plotone{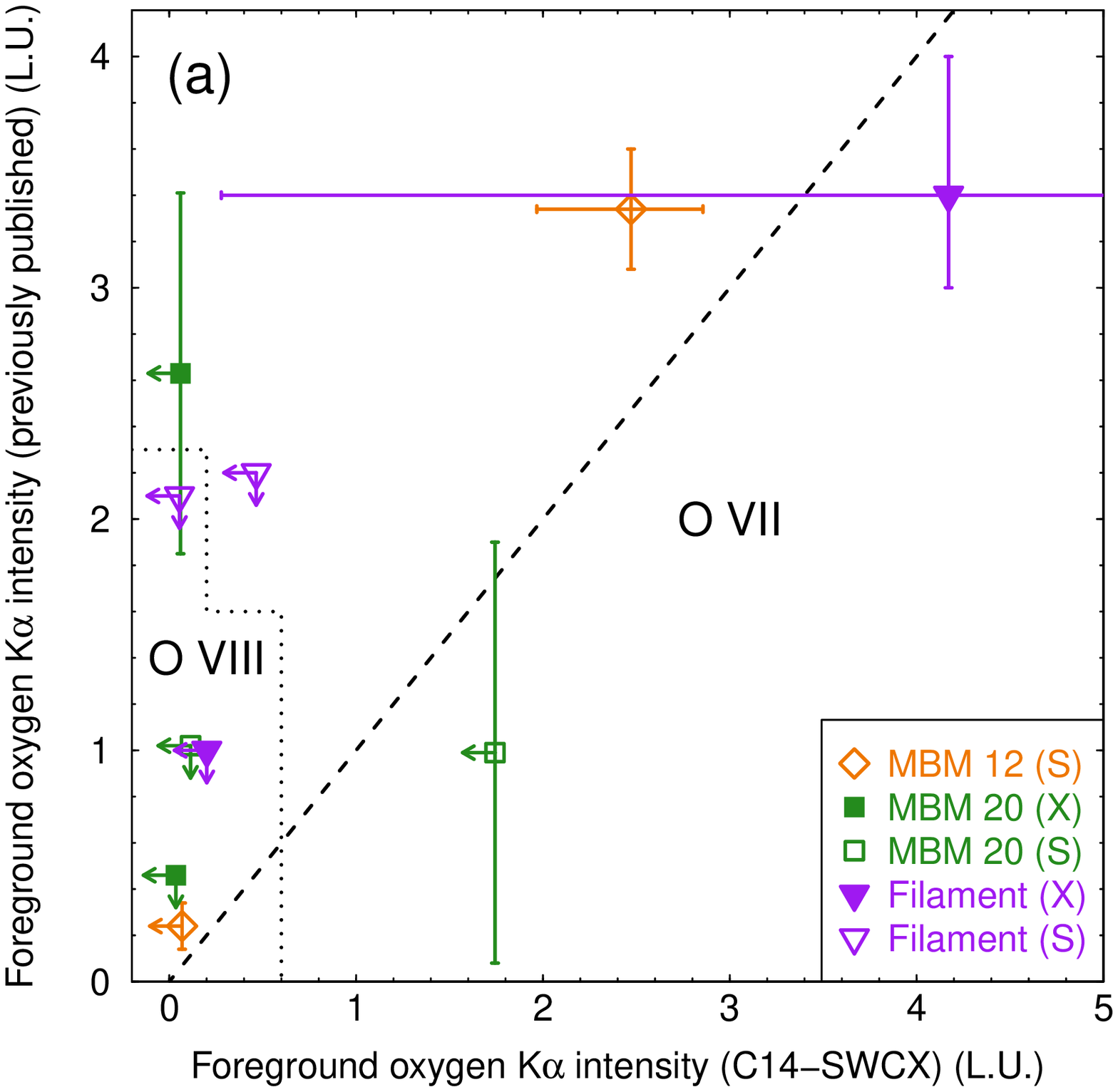} \\
\plotone{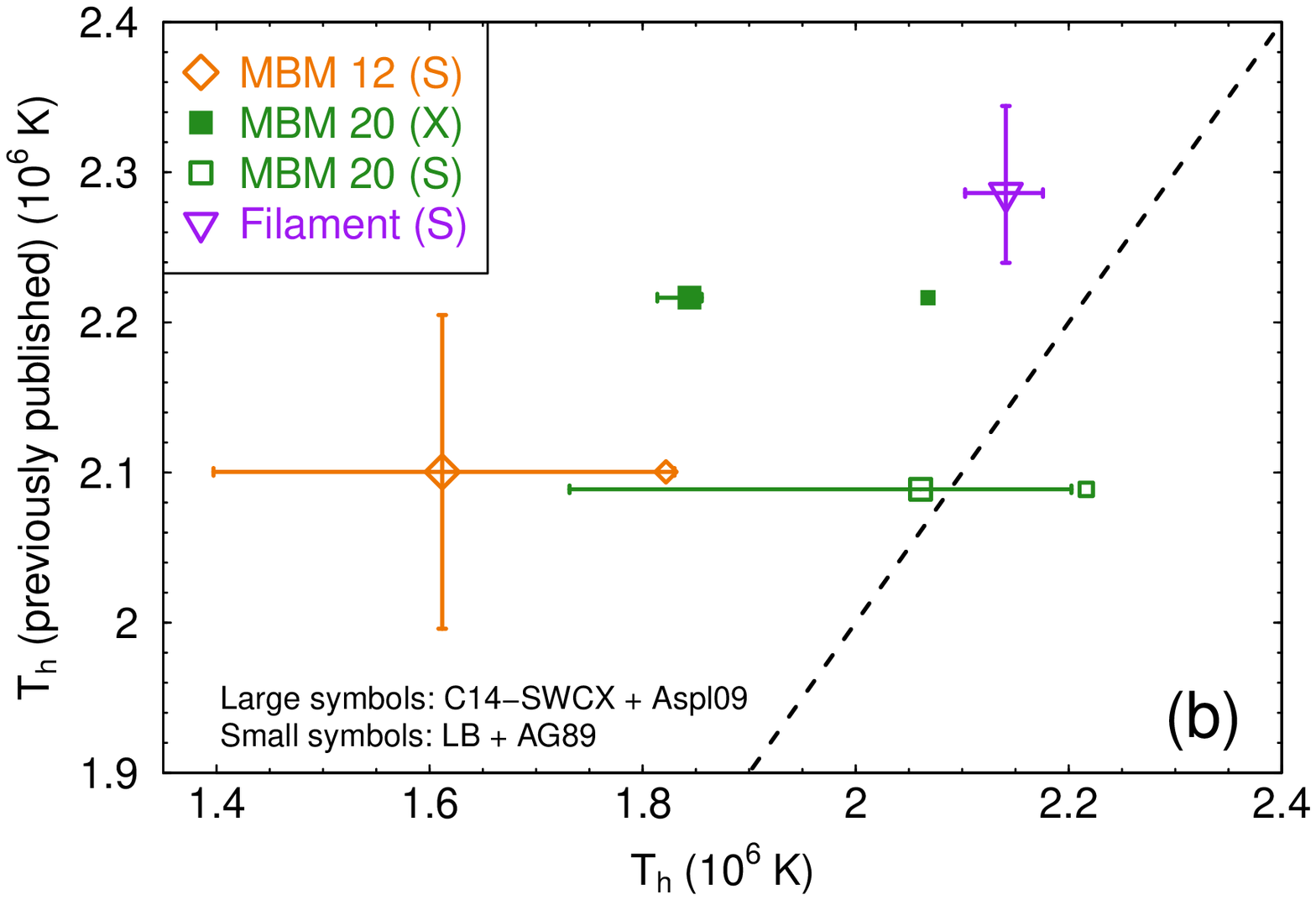}
\caption{Comparison of our measurements of (a) the foreground \OVII\ and
  \OVIII\ \Kalpha\ intensities and (b) the halo temperature with measurements from previous
  shadowing studies. In panel~(a), the dotted line separates the \OVIII\ results (lower left region)
  from the \OVII\ results.
  The symbols and colors for the various shadows match those used in Figure~\ref{fig:HaloResults}.
  The published results for MBM~12 and the \xmm\ and \suzaku\ observations of MBM~20 are from
  \citet{smith07a}, \citet{galeazzi07}, and \citet{gupta09b}, respectively. The published oxygen
  results for the \xmm\ observation of the filament are from \citet[note that they did not examine a
    $1T$ halo model]{henley07}, and the published results for the \suzaku\ observation of the
  filament are from \citet[oxygen intensities]{henley08a} and \citet[halo temperature]{lei09}.
  Note that the published oxygen measurements are summarized in Table~4 of \citet{gupta09b}.
  Our results are primarily from our reference model (C14-SWCX foreground and Aspl09 abundances).
  However, for shadows previously analyzed with AG89 abundances (MBM~12 and MBM~20), we also show in
  panel~(b) the temperatures we obtained using the LB foreground model with AG89 abundances, using
  smaller symbols without error bars.
  The dashed lines indicate equality.
  \label{fig:OtherShadows}}
\end{figure}

A number of the shadows in our sample have previously been analyzed. Here, we compare the results
from these earlier studies with our results.

Figure~\ref{fig:OtherShadows}(a) compares the foreground \OVII\ and \OVIII\ \Kalpha\ intensities
from our reference model with the previously published values. In general the results are
consistent, considering the large scatter and error bars.  However, we typically obtain tighter
upper limits on the foreground \OVIII\ \Lyalpha\ intensity. This is likely because our results were
obtained directly from the spectral fitting, whereas most of the previously published results were
obtained by measuring the total \OVIII\ intensities in the on- and off-shadow spectra independently,
and then using an equation similar to Equation~(\ref{eq:SFG}) to calculate the foreground intensity.
(This also accounts for our tighter upper limit on the foreground \OVII\ intensity from the
\suzaku\ observation of the filament.) In addition, our C14-SWCX foreground model includes emission
from \OVII\ \Kbeta\ (at 0.666~\kev, with an intensity $\sim$7\% of that from \Kalpha), which would
tend to reduce the intensity attributed to the foreground \OVIII\ (at 0.653~\kev).

For MBM~12 we obtain a lower foreground \OVII\ \Kalpha\ than \citet{smith07a}. This is due, at least
in part, to our using a lower on-cloud column density ($3.22\times10^{21}$ versus $4 \times
10^{21}~\pcmsq$) and to our assuming Aspl09 instead of AG89 abundances, lowering the absorption
cross-section at 0.57~\kev\ ($6.8\times10^{-22}$ versus $9.4\times10^{-22}~\cmsq$). Both of these
changes increase the transmissivity of the cloud to background \OVII\ emission, from 2\% to
11\%. \citet{smith07a} assumed that the contribution of the halo emission to the on-cloud
\OVII\ emission was negligible, whereas in our model $\sim$10\% of the on-cloud \OVII\ emission is
due to halo emission leaking through the cloud (see the MBM~12 plot in
Figure~\ref{fig:SpectraSWCX}).

For the MBM~20 observations, we obtain upper limits on the foreground \OVII\ \Kalpha\ intensity,
whereas \citet{galeazzi07} and \citet{gupta09b} quote detections. However, the reported detection
from the \suzaku\ observation \citep{gupta09b} is borderline---if the quoted error is $1\sigma$,
then at the 90\% level which we use for our errors, the line was not detected. Note also that
\citet{gupta09b} and we obtained almost identical best-fit foreground \OVII\ \Kalpha\ intensities
from this observation (0.99 and 0.97~\LU, respectively).

For the \xmm\ observation of MBM~20, we obtain a very tight upper limit on the foreground
\OVII\ intensity (0.06~\LU\ at the 90\%\ level). This is likely because the halo model is very
tightly constrained (note from the solid green squares in Figure~\ref{fig:HaloResults} that this
observation has the smallest uncertainties on the halo measurements) and can adequately explain the
observed \OVII\ emission alone, leaving little room for any foreground \OVII\ emission. We checked
the possibility that this tight upper limit on the foreground \OVII\ intensity is an artifact of our
assuming a $1T$ halo model, meaning that the halo spectrum may be artificially tightly constrained.
However, when we added a second thermal plasma component to our halo model, we obtained an upper
limit on the foreground \OVII\ intensity of 0.13~\LU. While this is not as tight a constraint as
with our original model, the foreground \OVII\ intensity is still more tightly constrained for this
observation than for any other. Therefore, the tight upper limit on the foreground \OVII\ intensity
for the \xmm\ observation of MBM~20 is not an artifact of our assuming a $1T$ halo model.

It is unclear exactly why we and \citet{galeazzi07} obtained different results from the
\xmm\ observation of MBM~20. However, the difference appears not to be due to the fact that
\citet{galeazzi07} analyzed only the pn data, assumed AG89 abundances, and used lower column
densities than us ($15.9\times10^{20}$ and $0.86\times10^{20}$ versus $17.6\times10^{20}$ and
$2.3\times10^{20}~\pcmsq$ (Table~\ref{tab:Observations})). If we re-fit the C14-SWCX model to the pn
data only, assuming AG89 abundances and \citeauthor{galeazzi07}'s column densities, we obtain a
foreground \OVII\ intensity of $0.85^{+0.26}_{-0.19}~\LU$ (this is similar to the value in
Section~(d) of Table~\ref{tab:Results}, obtained using the pn and MOS data and the column densities
from Table~\ref{tab:Observations}). While this is larger than the value obtained with our reference
model (shown in Figure~\ref{fig:OtherShadows}(a)), it is still smaller than the foreground
\OVII\ intensity reported by \citet{galeazzi07}, $2.63\pm0.78~\LU$.  It is possible that the
difference arises from the fact that \citet{galeazzi07} used a different method from us to calculate
the particle background, which may have affected the background-subtracted SXRB spectra, and thus
the fit results.

In general, our halo temperatures are in reasonable agreement with (although systematically lower
than) the previously published values.  The agreement is better if we use a model more similar to
those used in most previous studies (an LB foreground model with AG89 abundances, shown by the
smaller symbols in Figure~\ref{fig:OtherShadows}(b)). The largest temperature discrepancy is for
MBM~12. This is likely due to the higher transmissivity of the cloud to background \OVII\ in our
model (see above), resulting in more \OVII\ emission being attributed to the halo, and hence a lower
halo \OVIII/\OVII\ ratio and a lower halo temperature. For the \suzaku\ observation of the
filament, the 7\%\ difference between our and \citepossessive{lei09} halo temperatures is not due
to our assuming different column densities (see Section~\ref{sec:SpectralModel})---using
\citeauthor{lei09}'s column densities with our reference model shifts the best-fit temperature by
only 1\%. Instead, the discrepancy may be due to \citeauthor{lei09} fixing the foreground model
when they fitted a $1T$ halo model to the \suzaku\ spectra (see their Table~1).  Our halo
emission measures (not plotted) are also in reasonable agreement with the previously published
values, after accounting for differences in the assumed oxygen abundance (although for three of the
four shadowing observations considered here, our best-fit emission measures are larger than the
previously published values).

\section{DISCUSSION}
\label{sec:Discussion}

In our spectral analysis, we examined three different foreground models. For some sight lines, these
models yielded different results (Section~\ref{subsec:ResultsDifferentForegrounds}). Therefore, we
begin our discussion by examining the differences among the foreground models, how they are related
to the physics of the models, and how we come to favor one foreground model (C14-SWCX) over the
others (Section~\ref{subsec:ForegroundChoice}).

We then discuss our foreground measurements from the C14-SWCX model. Such measurements may be used
to test theoretical models of SWCX emission. We therefore compare our measurements with general
predictions from heliospheric SWCX emission models (Section~\ref{subsec:SWCX}). We also use our
measurements to place limits on the emission from the LB (Section~\ref{subsec:LB}).

Finally, we discuss our halo emission results.  In Section~\ref{subsubsec:HaloOtherStudies}, we
compare our halo measurements with those from non-shadowing studies of the halo, using \xmm\ (HS13)
and \suzaku\ (Yosh09).  In Section~\ref{subsubsec:HaloUncertaintyBias}, we discuss possible
uncertainties or biases in our halo measurements in addition to the statistical
uncertainties. Finally, we discuss the use of these (and similar) halo measurements in testing
models of the Galactic halo emission (Section~\ref{subsubsec:TestingHaloModels}).

\subsection{Our Preferred Foreground Model}
\label{subsec:ForegroundChoice}

\begin{figure}
\centering
\plotone{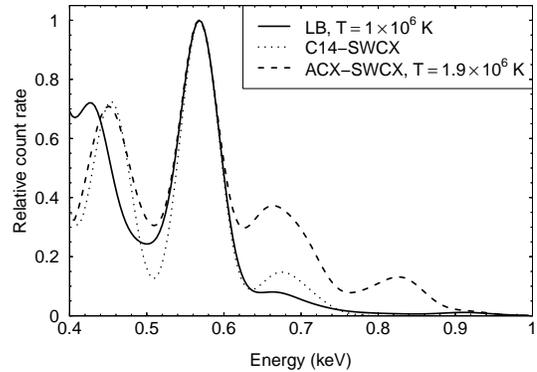}
\caption{Example model foreground spectra from the LB (solid), C14-SWCX (dotted), and ACX-SWCX
  (dashed) models, folded through the \suzaku\ response and normalized to the peak of the
  \OVII\ emission. The LB model's temperature was fixed at $T = 1 \times 10^6~\K$, and the other
  models' parameters were adjusted so that the lower-energy emission ($E \la 0.5~\kev$) was of
  similar brightness in all three models (see text for details).
  \label{fig:ModelFGSpectra}}
\end{figure}

Here we discuss the differences among the foreground models and how they are related to the physics
of the models. This discussion will lead us to pick one foreground model, C14-SWCX, as our preferred
model.

We noted in Section~\ref{subsec:ResultsDifferentForegrounds} that, for some sight lines, the
ACX-SWCX model attributes less \OVII\ emission to the foreground than the C14-SWCX and LB models,
resulting in a softer and fainter foreground. \citet{henley15a} also found this be true for
G225$-$66 (see their Figure~3). Attributing less of this emission to the foreground can affect the
measured halo properties.

The above difference between the models arises because the ACX-SWCX foreground model produces more
emission above $\sim$0.6~\kev\ relative to the \OVII\ emission than the other foreground models, as
shown in Figure~\ref{fig:ModelFGSpectra}. This plot shows model spectra from our three foreground
models. We used an LB model with a temperature of $1 \times 10^6~\K$ (a typical temperature for such
a model; solid line) as a reference. For the C14-SWCX model (dotted line), we disabled the
\CV\ lines (which lie below the energy range plotted) and the \OVIII\ lines (because the reference
LB model in Figure~\ref{fig:ModelFGSpectra} produces negligible \OVIII\ emission). We then adjusted
the \CVI\ intensity so that the emission below $\sim$0.5~\kev\ was similar in brightness (relative
to \OVII) to that from the LB model.\footnote{Note that the emission below $\sim$0.45~\kev\ is at
  different energies in the two models. The emission at $\sim$0.43~\kev\ in the LB model is from
  \NVI\ \Kalpha, which is not included in the C14-SWCX model. The emission at $\sim$0.45~\kev\ in
  the C14-SWCX model is from \CVI\ \Lybeta\ and \Lygamma.} We adjusted the temperature of the
ACX-SWCX model (dashed line) so that its emission below $\sim$0.5~\kev\ was also of similar
brightness relative to the \OVII\ emission---this was achieved at $T = 1.9 \times 10^6~\K$.  Having
thus adjusted the model spectra, we can immediately see that the emission above $\sim$0.6~\kev\ is
much brighter from the ACX-SWCX model than from the other two models. The broad feature at
$\sim$0.66~\kev\ is mainly due to \OVIII\ \Lyalpha, with contributions from \OVII\ \Kbeta--\Kdelta,
while the feature at $\sim$0.83~\kev\ is due to \OVIII\ \Lybeta--\Lydelta. This difference explains
why the ACX-SWCX foreground models that best fit the observations are sometimes softer than the
other foreground models: the fitting will not allow the ACX-SWCX models to produce as much
foreground \OVII\ as the other models, as doing so would result in too much higher-energy foreground
emission, worsening the fit.

Physically, this difference arises because different ions, with different temperature dependencies,
are responsible for the \OVII\ and \OVIII\ emission in the LB and ACX-SWCX models (note that in the
C14-SWCX model, the normalizations of the various ions' emission are completely independent, rather
than being governed by a single temperature parameter).  In the LB model, this emission is due to
collisional excitation of \Oplus{6} and \Oplus{7} ions, respectively, whereas in the ACX-SWCX model,
it arises from \Oplus{7} and \Oplus{8} ions, respectively, undergoing CX reactions with hydrogen and
helium. In a CIE plasma at $T = 1\times10^6~\K$ (the temperature of the LB model plotted in
Figure~\ref{fig:ModelFGSpectra}), \Oplus{7} is two orders of magnitude less abundant than \Oplus{6}
(ATOMDB v2.0.2; \citealt{foster12}), and so there is virtually no \OVIII\ emission from the LB
model. In contrast, at $T=1.9\times10^6~\K$ (the temperature of the ACX-SWCX model plotted in
Figure~\ref{fig:ModelFGSpectra}), \Oplus{8} is one-ninth as abundant as \Oplus{7} (ATOMDB), and so
the \OVIII\ CX emission is not negligible.

In general, when the foreground \OVII\ emission is independent of the foreground emission from other
ions (as with the C14-SWCX model), more \OVII\ emission is attributed to the foreground than with
the ACX-SWCX model. This suggests that the ACX-SWCX model may not accurately model the foreground
emission---the ACX-SWCX models may, in some cases, be artificially soft, which in turn results in
underestimated halo temperatures. The inaccuracy of the ACX-SWCX model in this context is likely due
to the fact that the ion populations in this model are governed by a single temperature, which is
not a good description of the true solar wind ion populations \citep{vonsteiger00}. This
shortcoming could in principle be overcome by using multiple independent ACX components to model
the foreground.  We briefly experimented with a two-component ACX foreground model---one component
representing the foreground carbon and nitrogen emission, and one the foreground emission from
oxygen and higher-$Z$ elements. We found that the results varied from shadow to shadow. In some
cases the ACX-SWCX foreground model was brought more into line with the C14-SWCX model, while in
other cases the discrepancies between the ACX-SWCX and C14-SWCX models were even
larger. Furthermore, for some observations one of the two ACX components was poorly
constrained. Fully exploring multitemperature ACX-SWCX foreground models is beyond the scope
of this paper.

In addition, we noted in Section~\ref{sec:SpectralModel} that the ACX model's simple analytical
expressions cannot be expected to reproduce the true CX $n$ and $l$ distributions for all relevant
reactions. This may result in inaccurate modeling of the higher-order lines relative to the
\Kalpha\ lines. While this may not be important in our spectra, it could be important when
modeling higher-quality spectra in which the higher-order lines are clearly seen. Therefore, even
in situations where a single ionization temperature is a good description for the ion populations,
the ACX model may have to be used with caution.

In contrast to the ACX-SWCX models, the LB foreground models are generally similar in spectral shape
and brightness to the C14-SWCX models, and thus generally yield similar halo results. However, a
pure-LB foreground model is likely unphysical.\footnote{Our main reason for including such a model
  is that it was widely used in earlier shadowing studies.} For example, such a model typically
implies an LB pressure several times larger than those of the cold interstellar clouds in the
vicinity of the sun \citep[e.g.,][]{jenkins09}. However, the similarity between the C14-SWCX and LB
foreground spectra implies that, although unphysical, a $\sim$$1\times10^6~\K$ thermal plasma model
may adequately model the foreground emission in a CCD-resolution X-ray spectrum, unless the
foreground is particularly bright (see Section~\ref{subsubsec:HaloUncertaintyBias}). (Unfortunately,
this similarity also means that one cannot separate spectroscopically the contributions of LB and
SWCX emission to the foreground.)

In summary, because the ACX-SWCX model may be unreliable (due to our assumption of a single
ionization temperature) and the LB model is likely unphysical, our preferred foreground model is
therefore the C14-SWCX model. It should be noted that this model does not include emission from
\NVI\ and \NVII---because of the relatively low resolution of the CCD spectrometers at low energies,
it is difficult to separate the emission from carbon, nitrogen, and oxygen. However, the relevant
nitrogen ions are less abundant in the solar wind than the corresponding carbon and oxygen ions
\citep{schwadron00}, and so omitting nitrogen should not seriously adversely affect our results.  We
will use the results from the C14-SWCX model in the following section, when we compare our
foreground measurements with some SWCX model predictions.

\subsection{Comparing the Foreground Emission with Heliospheric SWCX Model Predictions}
\label{subsec:SWCX}

As noted earlier, our measurements of the foreground emission may be used to test theoretical models
of SWCX emission. Generating detailed SWCX predictions for each observation, taking into account
contemporaneous solar wind conditions \citep{koutroumpa07,koutroumpa11,gupta09b}, is beyond the
scope of this paper. Instead, we will restrict our discussion to the general SWCX predictions of
\citet{koutroumpa06}, who describe how the heliospheric SWCX emission is expected to vary with solar
cycle and with ecliptic latitude, $\beta$. In particular, the SWCX emission depends on the path
lengths along a given sight line through the slow and fast solar winds---the slow solar wind is more
highly ionized than the fast solar wind, and contains more of the ions responsible for SWCX emission
in the \xmm/\suzaku\ band \citep{schwadron00}. At solar minimum, the slow solar wind is restricted
to within $\pm$20\degr\ of the solar equatorial plane, with the fast solar wind present at higher
latitudes, whereas at solar maximum, the solar wind at all latitudes as approximated as being in a
slow solar wind state \citep{koutroumpa06}. Therefore, broadly speaking, we expect the SWCX emission
to be fainter and softer on high-$\beta$ sight lines at solar minimum, compared with low-$\beta$
sight lines at solar minimum or with all sight lines at solar maximum, because high-$\beta$ sight
lines at solar minimum mainly sample the fast solar wind.

To test these expectations, we use the results from the C14-SWCX model (with Aspl09 abundances). We
also include results for G225$-$66 (\citealt{henley15a}; note that these results were obtained using
AG89 abundances). Since we are examining in particular how the SWCX emission depends on solar
activity and on ecliptic latitude, we tabulate the contemporaneous sunspot number (a measure a solar
activity) and $\beta$ for each of our observations in Table~\ref{tab:SWCXDetails}.

\begin{deluxetable}{llcc}
\tablecaption{Sunspot Numbers and Ecliptic Latitudes\label{tab:SWCXDetails}}
\tablehead{
                 &                       & \colhead{Sunspot}                 & \colhead{$\beta$\tablenotemark{b}} \\
\colhead{Shadow} & \colhead{Observatory} & \colhead{number\tablenotemark{a}} & \colhead{(deg)}
}
\startdata
Filament         & \xmm                  & 125.0                             & $-74.39$ \\
MBM 20           & \xmm                  &  37.8                             & $-36.20$ \\
MBM 12           & \suzaku               &   6.5                             &   +2.60  \\
Filament         & \suzaku               &   3.6                             & $-74.41$ \\
MBM 20           & \suzaku               &   1.8                             & $-36.23$ \\
G236+38          & \suzaku               &  47.9                             & $-12.21$ \\
G048+37          & \xmm                  &  36.7                             &  +49.23  \\
G225$-$66        & \xmm                  &  35.2                             & $-42.43$ \\
MBM 16           & \suzaku               &  65.5                             &  $-6.49$
\enddata
\tablecomments{In contrast to Table~\ref{tab:Observations}, the observations are tabulated in chronological order.}
\tablenotetext{a}{Sunspot data obtained from the National Geophysical Data Center
  (http://www.ngdc.noaa.gov/stp/\\ space-weather/solar-data/solar-indices/\\ sunspot-numbers/international/listings/).
  The tabulated values are averaged over 10~d either side of the on-shadow pointing.}
\tablenotetext{b}{Ecliptic latitude of the on-shadow pointing.}
\end{deluxetable}
\begin{figure}
\centering
\plotone{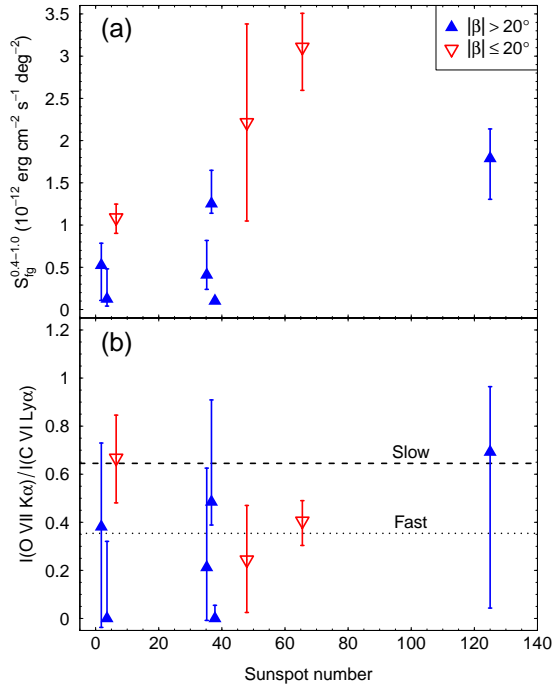}
\caption{Foreground (a) 0.4--1.0~\kev\ surface brightness and (b)
  \OVII\ \Kalpha/\CVI\ \Lyalpha\ intensity ratio plotted against sunspot number. Foreground results
  are for the C14-SWCX model with Aspl09 abundances.  Results for high and low ecliptic latitudes
  are shown with solid blue and open red triangles, respectively. The dotted and dashed lines in
  panel~(b) show the expected intensity ratios for the fast and slow solar wind, respectively (see
  text for details).
  \label{fig:ForegroundSunspot}}
\end{figure}

Figure~\ref{fig:ForegroundSunspot}(a) shows the 0.4--1.0~\kev\ surface brightness of the C14-SWCX
foreground model plotted against sunspot number. Note that we use different symbols to denote
results from high and low ecliptic latitudes. Among the observations toward higher latitudes, the
foreground surface brightness generally increases as the sunspot number increases. This is
qualitatively as expected, as the high latitude solar wind changes from a fast state at solar
minimum to a slow state at solar maximum.

Of the three observations taken near solar minimum (with sunspot number $<10$), the one toward low
ecliptic latitude (MBM~12) has the highest foreground surface brightness. Again, this is
(qualitatively) as expected---this observation samples only the slow solar wind, whereas the
observations toward higher ecliptic latitudes mainly sample the fast solar wind, which produces less
CX emission in the \xmm/\suzaku\ band \citep{koutroumpa06}.  Similarly, of the five observations
with intermediate sunspot numbers ($\sim$35--65), those toward low ecliptic latitudes tend to have
brighter foregrounds than those toward higher latitudes. Whether or not this is as expected depends
in part on the large-scale configuration of the solar wind at these intermediate sunspot
numbers. Surprisingly, one of these observations (G225$-$66) yielded a foreground surface brightness
of only $(0.41^{+0.41}_{-0.17}) \times 10^{-12}~\flux\ \pdegsq$ \citep{henley15a}, similar to the
values seen on high-ecliptic-latitude sight lines near solar minimum, despite being taken in 2013
February, only $\sim$9~months before solar maximum. However, as noted by \citet{henley15a}, the sun
was less active during this maximum than during preceding maxima (based on the sunspot number and
the solar 1--8~\angstrom\ flux; \citealt{winter14}), which may have affected the heliospheric SWCX
emission.

Figure~\ref{fig:ForegroundSunspot}(b) shows the ratio of the foreground \OVII\ \Kalpha\ and
\CVI\ \Lyalpha\ intensities, $I(\mbox{\OVII\ \Kalpha})/I(\mbox{\CVI\ \Lyalpha})$, against sunspot
number. We concentrate on this ratio because, in general, we do not detect foreground
\OVIII\ emission with this model, meaning that we can only place upper limits on the
\OVIII/\OVII\ ratio. However, it should be noted that \CVI\ \Lyalpha\ lies near the lower
limit of the \xmm\ and \suzaku\ bandpasses (in fact, for some fits, \CVI\ \Lyalpha\ lies below the
energy band used, and is constrained indirectly via the C14-SWCX line ratios). As the calibration
is uncertain at these low energies, it is possible that the uncertainties in
Figure~\ref{fig:ForegroundSunspot}(b) are underestimated.

Figure~\ref{fig:ForegroundSunspot}(b) also shows the intensity ratios expected
for the fast and slow solar winds, given by
\begin{equation}
  \frac{I(\mbox{\OVII\ \Kalpha})}{I(\mbox{\CVI\ \Lyalpha})} =
  \frac{n(\Oplus{7}) \sigma(\Oplus{7} + \mathrm{H}) y(\mbox{\OVII\ \Kalpha})}
       {n(\Cplus{6}) \sigma(\Cplus{6} + \mathrm{H}) y(\mbox{\CVI\ \Lyalpha})},
\end{equation}
where $n(X)$ is the density of ion $X$ in the solar wind, $\sigma(X + \mathrm{H})$ is the CX
cross-section for $X$ reacting with hydrogen,\footnote{We are ignoring CX interactions with
  helium. However, as helium is an order of magnitude less abundant than hydrogen (except
  within one or two AU of the Sun) and as CX cross-sections involving helium are typically
  smaller than those involving hydrogen \citep[e.g.,][Table~1]{koutroumpa06}, ignoring helium should
  not greatly affect our results. Note that the C14-SWCX model is based on line ratios for ions
  interacting with hydrogen only.} and
$y(L)$ is the yield for line $L$. To calculate the expected intensity ratios, we use ion abundances
from \citet[Table~1; $n(\Oplus{7})/n(\Cplus{6})=0.353$ and 0.629 for the fast and slow solar winds,
  respectively]{schwadron00} and cross-sections from \citet[Table~1;
  $\sigma(\Oplus{7}+\mathrm{H})/\sigma(\Cplus{6}+\mathrm{H})=0.817$ and 0.799 for the fast and slow
  solar winds, respectively]{koutroumpa06}. For the line yields, we use the CX line ratio data from
\citet{cumbee14}, assuming that the emission from higher-order lines not included in the model is
negligible: $y(\mbox{\OVII\ \Kalpha})=0.902$ and $y(\mbox{\CVI\ \Lyalpha})=0.719$.\footnote{These
  line ratios were calculated for a collision speed of $\sim$400~\kmps, appropriate for the slow
  solar wind \citep{smith03}. Relevant data for a collision speed appropriate for the fast solar
  wind ($\sim$700-800~\kmps) are not available.} Thus, we obtain
$I(\mbox{\OVII\ \Kalpha})/I(\mbox{\CVI\ \Lyalpha}) = 0.354$ and 0.645 for the fast and slow solar
winds, respectively.

Some of the observed ratios are as expected. For example, the observation with the highest sunspot
number, which was taken near solar maximum, and the low-ecliptic-latitude
observation that was taken near solar minimum are both expected to probe the slow solar
wind---both yield \OVII/\CVI\ emission ratios consistent with the slow solar wind value.
On the other hand, some observed emission ratios are not as expected. For example, at
intermediate sunspot numbers, two observations are toward low ecliptic latitudes, and are thus
expected to probe the slow solar wind. However, the \OVII/\CVI\ emission ratios for these two
observations are less than that expected for the slow solar wind. While these discrepancies
may simply be due to the calibration uncertainty near the \CVI\ lines, we note the possibility that
these discrepancies are due to localized variations in the solar wind (such as coronal mass
ejections; CMEs), crossing the line of sight, or to variations in the geocoronal SWCX emission.

In summary, only some aspects of our set of foreground emission measurements are as expected from a
model of heliospheric SWCX emission from a smooth solar wind \citep{koutroumpa06}. Deviations from
these expectations may be due to localized solar wind variations, such as CMEs, crossing
the line of sight \citep{smith05,koutroumpa07,henley08a,carter10}, or due to enhancements in the
geocoronal emission \citep{wargelin04,snowden04,fujimoto07,snowden09,ezoe10,wargelin14}. The details
of the latitudinal variation of the solar wind may also be important in explaining the discrepancies
between the \citet{koutroumpa06} model and the observations.  Our measurements of the foreground
SWCX emission will provide useful constraints on future SWCX models.

\subsection{Limits on the Local Bubble Emission}
\label{subsec:LB}

\begin{figure}
\centering
\plotone{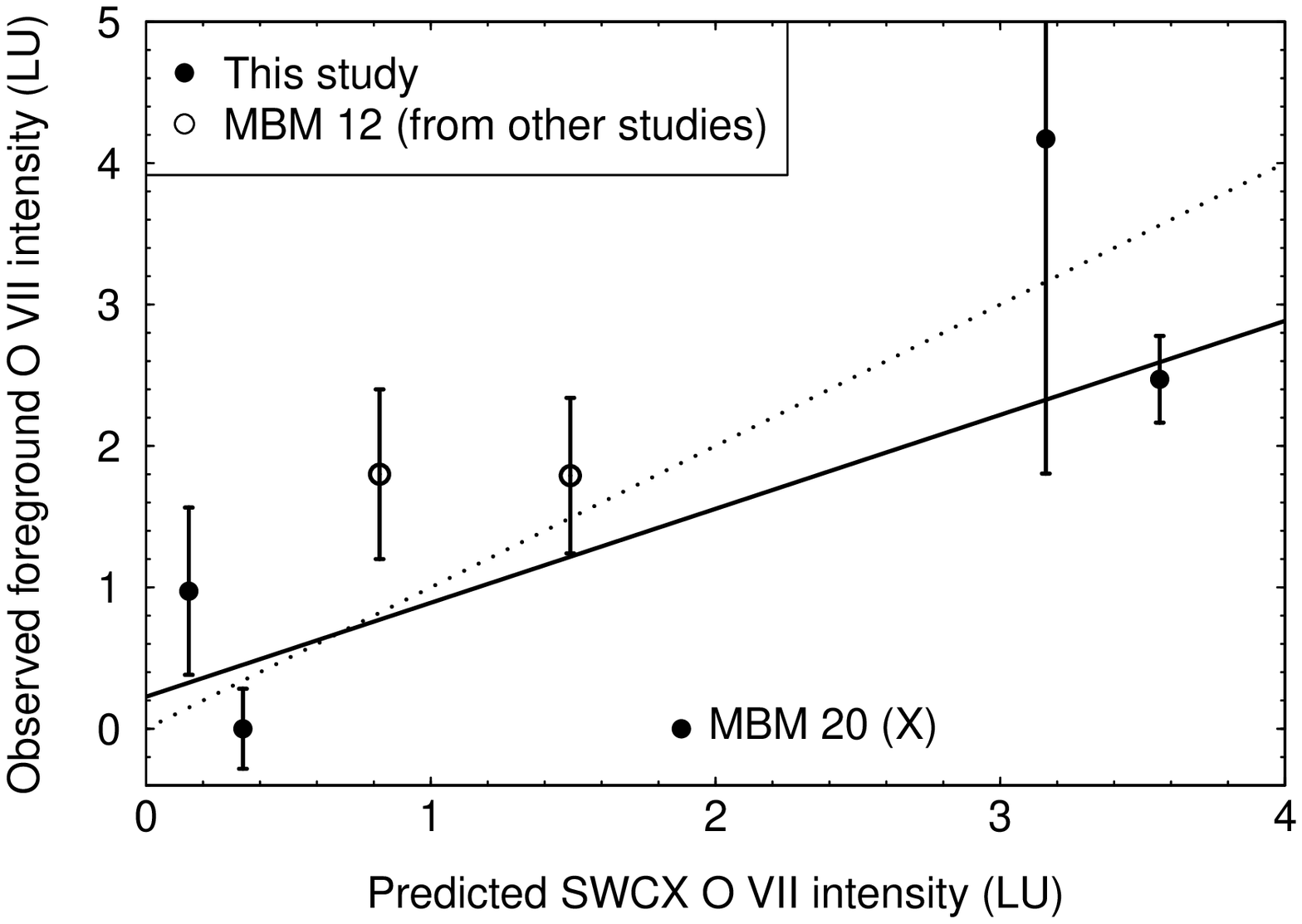}
\caption{Comparison of observed foreground \OVII\ intensity with heliospheric SWCX model predictions
  \citep[Table~5]{koutroumpa11}. Solid symbols indicate our measurements. For the purposes of this
  plot we show $1\sigma$ errors (see text for details). Open symbols represent measurements from
  \chandra\ and \xmm\ observations of MBM~12, not included in our sample
  \citep{smith05,koutroumpa11}. The dotted line indicates equality. The solid line is the best-fit
  linear regression line (excluding the \xmm\ observation of MBM~20, which is labeled; see text for
  details).
  \label{fig:SWCXLinReg}}
\end{figure}

\citet{koutroumpa07} developed the general \citet{koutroumpa06} heliospheric SWCX model (discussed
above) to take into account contemporaneous solar wind conditions during a given observation, and to
include (if necessary) the emission from localized solar wind enhancements crossing the line of
sight. Here, we follow \citet{koutroumpa07,koutroumpa11} and use such predictions to place limits on
the emission attributable to the LB.

Heliospheric SWCX predictions for a subset of our shadowing observations (specifically, the
observations of MBM~12, MBM~20, and the filament) are summarized in
\citet[Table~5]{koutroumpa11}. Figure~\ref{fig:SWCXLinReg} compares the predicted \OVII\ intensities
with our measurements. The figure also includes two additional measurements from \chandra\ and
\xmm\ observations of MBM~12, not included in our sample \citep{smith05,koutroumpa11}.  In general,
the predictions are in good agreement with the observations. The one exception is the
\xmm\ observation of MBM~20 (labeled in the plot), for which the model significantly overpredicts
the foreground \OVII\ intensity.

Following \citet{koutroumpa07,koutroumpa11}, we will use the data in
Figure~\ref{fig:SWCXLinReg} to place limits on the LB \OVII\ intensity. In this analysis, we are
assuming that the plasma filling the LB is uniform. Furthermore, we are assuming that the extent
of the LB toward the three clouds whose results are included in Figure~\ref{fig:SWCXLinReg} is
similar. From the maps of the Local Cavity in \citet{lallement03}, we estimate that the extent of
the Local Cavity (which we assume to be the same as that of the LB) is $\sim$100--150~\pc\ toward
these three clouds. This analysis involves fitting a straight line to the data in
Figure~\ref{fig:SWCXLinReg} \citep{koutroumpa07,koutroumpa11}. Since weighted-least-squares
linear regression requires symmetrical errors, for each data point we used the larger of the upper
and lower errors, rescaled from a 90\%\ confidence interval to a $1\sigma$ confidence interval
(these are the errors shown in Figure~\ref{fig:SWCXLinReg}). We excluded the \xmm\ observation of
MBM~20, as it is such an extreme outlier.

The best fit straight line (solid line in Figure~\ref{fig:SWCXLinReg}) has a gradient of
$0.66\pm0.18$ and an intercept of $0.23\pm0.38~\LU$ ($1\sigma$ errors). The gradient is consistent
with that obtained by \citet{koutroumpa11}, using the then-available measurements of the foreground
\OVII\ intensity ($0.65\pm0.16$). However, our intercept is significantly less than theirs
($1.07\pm0.47~\LU$). This is because some of the foreground \OVII\ intensities used in our fit
(obtained using the C14-SWCX model with Aspl09 abundances) are lower than those used by
\citet{koutroumpa11}. If we repeat our analysis using the foreground \OVII\ intensities obtained
using AG89 abundances, we obtain an intercept of $1.30\pm0.31~\LU$, consistent with
\citeauthor{koutroumpa11}'s value. However, as noted in
Section~\ref{subsec:ResultsDifferentAbundances}, results obtained using AG89 abundances may be
unreliable.

\citet{koutroumpa11} suggested that their non-zero intercept was evidence of \OVII\ emission from
the hot LB. However, our intercept is consistent with zero. Rescaling the $1\sigma$ confidence
interval on the intercept to a 90\%\ confidence interval, we find that the LB contributes
$\la$0.8~\LU\ to the foreground \OVII\ emission (for an LB path length of $\sim$100~\pc).

\subsection{Halo Emission}
\label{subsec:HaloDiscussion}

\subsubsection{Comparison with Previous Halo Studies}
\label{subsubsec:HaloOtherStudies}

We begin our discussion of the halo X-ray emission by comparing our measurements with those from
other recent studies of the halo with CCD-resolution spectra---see Figure~\ref{fig:HaloResults}.
The black crossed diamonds indicate the quartiles from an \xmm\ survey of the halo emission
(encompassing 110 high-Galactic-latitude ($|b|>30\degr$) sight lines; HS13). HS13 assumed an LB-like
foreground with $\Tfg = 1.2\times10^6~\K$. For each sight line the foreground emission measure
(typically $\sim$$(\mbox{3--6})\times10^{-3}~\emismeas$) was fixed using nearby shadowing
measurements of the foreground 1/4~\kev\ count rate \citep{snowden00}. Thus, HS13's foreground
modeling method involved extrapolating results from the 1/4~\kev\ band to the higher-energy
\xmm\ band, which may have introduced a bias to their halo measurements. The black stars show
results from a study of the halo emission using 12 high-latitude ($|b|>20\degr$)
\suzaku\ observations (Yosh09). Yosh09 used the same LB-like foreground model for all their sight
lines, with $\Tfg \sim 1.3\times10^6~\K$ and $\EMfg \sim 7\times10^{-3}~\emismeas$.\footnote{Note
  that HS13's and Yosh09's foreground emission measures are somewhat smaller than those we measured
  using the LB model with Aspl09 abundances (Table~\ref{tab:Results}, Section~(c)). This is partly
  because they assumed AG89 abundances, and partly because their assumed foreground temperatures are
  higher than some of our best-fit LB temperatures (both of which result in more emission in the
  \xmm/\suzaku\ band per unit emission measure). HS13 and Yosh09's foreground models have
  0.4--1.0~\kev\ surface brightnesses of $\sim$$(\mbox{0.6--1.1})\times10^{-12}$ (typically) and
  $\sim$$1.8\times10^{-12}~\flux\ \pdegsq$, respectively, placing them well within the range of
  foreground surface brightnesses that we measured (see, e.g., Figure~\ref{fig:ForegroundResults}).}
Note that, because Yosh09 and HS13 assumed AG89 abundances, we have rescaled their halo emission
measures in Figure~\ref{fig:HaloResults} to account for the different oxygen abundances in AG89 and
Aspl09 (see footnote~\ref{fn:OxygenAbundanceG225}).

The halo parameters that we obtain using our reference model are generally within the ranges of
values reported by HS13. When we compare our values with their lower and upper quartiles, we find
that some of our temperatures are on the low side compared to their values. This difference is not
due to HS13 using the \raymondsmith\ code, whereas we used APEC---for temperatures around
$2\times10^6~\K$, a given observed halo \OVIII/\OVII\ ratio implies a temperature $\sim$$10^5~\K$
lower with the \citeauthor{raymond77} code than with APEC. Nor is this difference due to HS13 using
AG89 abundances, which, in our analysis, resulted in higher halo temperatures than the Aspl09
abundances (Figure~\ref{fig:AsplvAG}(b))---the foreground model was fixed during HS13's fitting, and
so the discussion in Section~\ref{subsec:ResultsDifferentAbundances} is not applicable to their
results. However, it should be noted that the median halo temperature from the shadowing
measurements, $2.06 \times 10^6~\K$, lies between HS13's lower quartile and median, suggesting that
any difference between the two sets of results is not significant.

\begin{figure}
\centering
\plotone{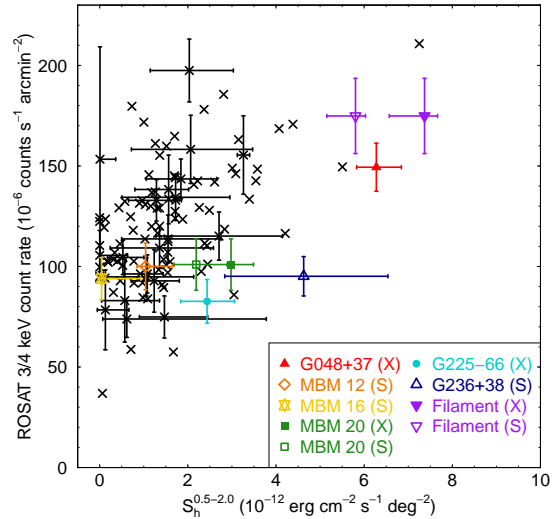}
\caption{Comparison of halo surface brightnesses (abscissae) from our shadowing measurements
  (colored symbols matching those used in Figure~\ref{fig:HaloResults}; see key) and from HS13
  (black diagonal crosses) with 3/4~\kev\ (R45) count rates for the same directions from the \rosat
  All-Sky Survey (ordinates; \citealt{snowden97}).  Note that, to avoid clutter, we only show error
  bars for a sample of the data points from HS13.
  \label{fig:R45vSB}}
\end{figure}

In contrast, the median halo surface brightness from the shadowing measurements, $2.98 \times
10^{-12}~\flux\ \pdegsq$, lies above the upper quartile from HS13, suggesting that the shadowing
measurements tend to yield systematically brighter halo emission than HS13's measurements. However,
the halo surface brightness is known to vary across the sky (Yosh09; HS13), and so it is possible
that the shadowing measurements happen to sample brighter-than-typical regions of the halo. To check
this, we compared the halo surface brightnesses from our measurements and from HS13 with the
3/4~\kev\ count rates for the same directions from the \rosat All-Sky Survey \citep{snowden97}. The
\rosat\ rates were extracted from circular regions of radius 0\fdg5, centered on each observation's
pointing direction (for the shadowing observations, we used the off-shadow pointing). The comparison
is shown in Figure~\ref{fig:R45vSB}. Note that, for most sight lines shown in the figure, the column
density is in the range $\sim$$\mbox{(1--3)} \times 10^{20}~\pcmsq$, over which range the observed
3/4~\kev\ count rate of a $\sim$$2 \times 10^6~\K$ halo plasma will decrease by $\la$20\%. Hence,
the total 3/4~\kev\ count rate (plotted in Figure~\ref{fig:R45vSB}) should be strongly correlated
with the halo's intrinsic 3/4~\kev\ count rate.

For lower 3/4~\kev\ count rates ($\la$$110\times10^{-6}~\rassrate$), the surface brightness
measurements from our shadowing measurements are similar to the values obtained by HS13. (The
best-fit surface brightness for G236+38 (blue triangle in Figure~\ref{fig:R45vSB}) is somewhat
outlying, but the uncertainty on this value is relatively large.)  Specifically, for sight lines
with 3/4~\kev\ count rates of $(\mbox{80--105})\times10^{-6}~\rassrate$, the median $\pm$
MADN\footnote{Normalized median absolute deviation---for a set of numbers
  $\boldsymbol{x}\equiv\{x_i\}$,
  $\mathrm{MADN}(\boldsymbol{x})=1.4826\times\mathrm{median}(|x_i-\mathrm{median}(\boldsymbol{x})|)$. The
  numerical factor ensures that the MADN of a normal distribution is equal to its standard deviation
  \citep[e.g.][]{feigelson12}.} surface brightnesses are $(2.31\pm1.44) \times 10^{12}$ and
$(0.95\pm0.75) \times 10^{-12}~\flux\ \pdegsq$ from our shadowing measurements and from HS13,
respectively. A Mann-Whitney $U$ test \citep[e.g.,][]{barlow89,wall03} indicates that the difference
in the median surface brightnesses is not statistically significant at the 1\%\ level ($U=49$,
two-sided $\pvalue=0.063$).

At higher 3/4~\kev\ count rates ($\ge$$145\times10^{-6}~\rassrate$), the surface brightnesses from
our shadowing measurements are larger than nearly all of the corresponding values from HS13. The
median $\pm$ MADN surface brightnesses are $(6.28\pm0.70) \times 10^{12}$ and $(2.22\pm1.40) \times
10^{-12}~\flux\ \pdegsq$, respectively. However, a Mann-Whitney $U$ test indicates that this
difference is also not statistically significant at the 1\%\ level ($U=2$, two-sided
$\pvalue=0.011$). Furthermore, two of the three shadowing measurements in this regime are of the
filament (purple triangles in Figure~\ref{fig:R45vSB})---this direction may not be representative of
the halo, as an extraplanar supernova remnant may lie in this direction \citep{lei09}.
Therefore, with the possible exception of the very brightest directions, there appears to be no
significant systematic difference between the halo surface brightnesses obtained from our shadowing
observations and those obtained by HS13.

When we compare our results with those from Yosh09, we find that the emission measures (after
accounting for the different assumed oxygen abundances) and surface brightnesses are in good
agreement, whereas our halo temperatures tend to be lower than Yosh09's (see
Figure~\ref{fig:HaloResults}). The median $\pm$ MADN temperatures from our and Yosh09's measurements
are $(2.06 \pm 0.22) \times 10^6$ and $(2.47 \pm 0.41) \times 10^6~\K$,
respectively.\footnote{Each of our data sets includes a sight line on which the halo temperature was
  fixed (MBM~16 in our data set, LH-2 in Yosh09's).  These sight lines were excluded from the
  calculation of the median temperatures.} A Mann-Whitney $U$ test indicates that this difference is
statistically significant at the 1\%\ level ($U=7$, two-sided $\pvalue=0.0026$).

The discrepancy between the two set of results may be due to Yosh09 assuming the same foreground
model for all their sight lines, whereas our results indicate that the foreground surface brightness
can vary significantly from observation to observation (see Figure~\ref{fig:ForegroundResults}). If
Yosh09's foreground model is too bright for a given observation, too much \OVII\ will be attributed
to the foreground; as a result, the halo \OVII/\OVIII\ ratio will be underestimated, and the halo
temperature overestimated. Note also that, if the foreground model is too bright, the halo surface
brightness will correspondingly be underestimated.  Thus, if Yosh09's foreground model is too bright
for some of their observations, we would expect to see an anticorrelation between their halo
temperatures and surface brightnesses.  In the lower panel of Figure~\ref{fig:HaloResults}, we can
see that the Yosh09 sight lines with higher halo temperatures tend to have lower surface
brightnesses. Although this anticorrelation is not statistically significant at the 1\%\ level
(Kendall's $\tau$ \citep[e.g.,][]{press92} is $-0.56$, with a \pvalue\ of 0.017), it does suggest
that Yosh09 may have overestimated the halo temperature on a few of their sight lines.

\subsubsection{Uncertainties or Biases in the Halo Measurements}
\label{subsubsec:HaloUncertaintyBias}

\begin{figure}
\centering
\plotone{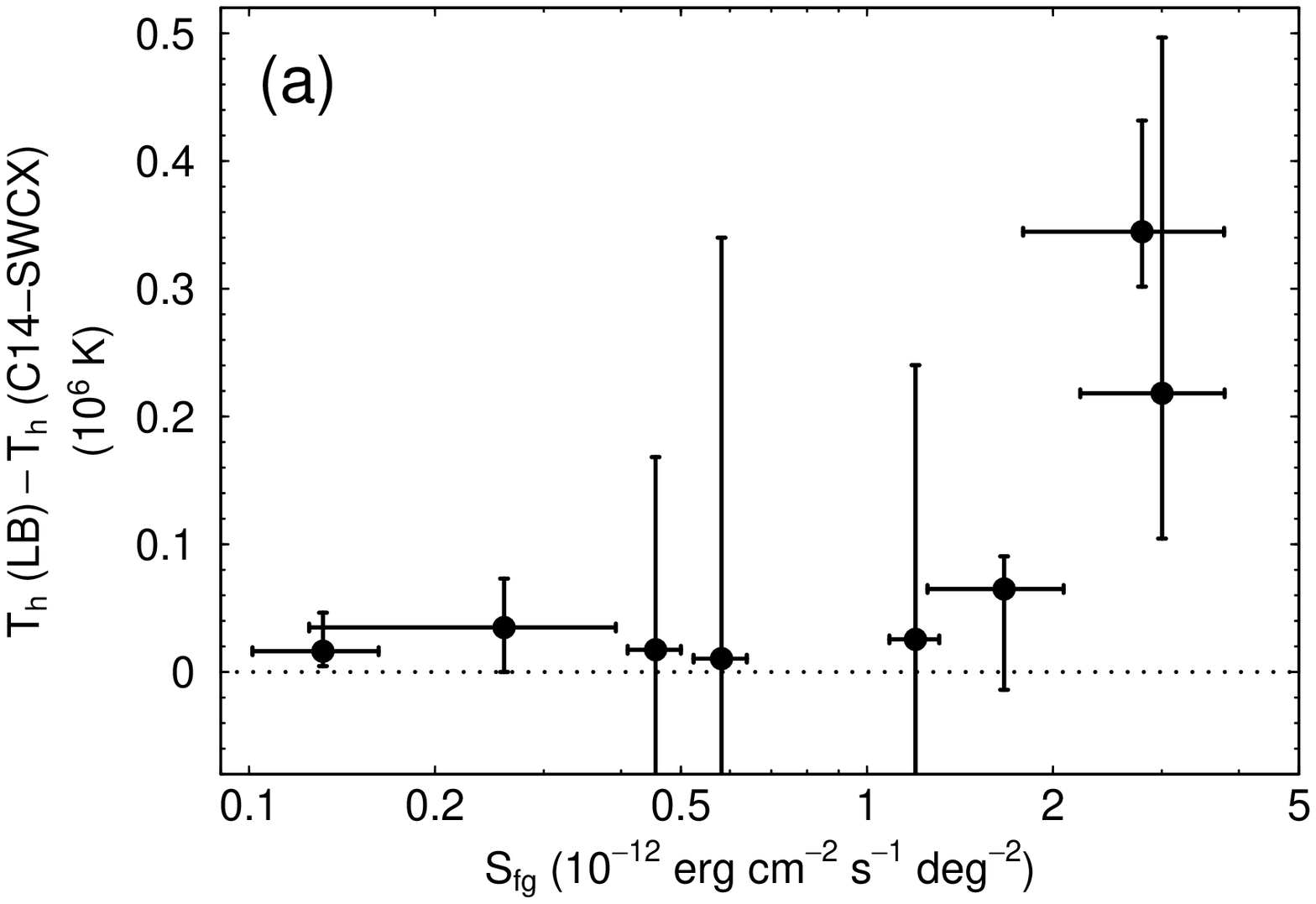} \\
\plotone{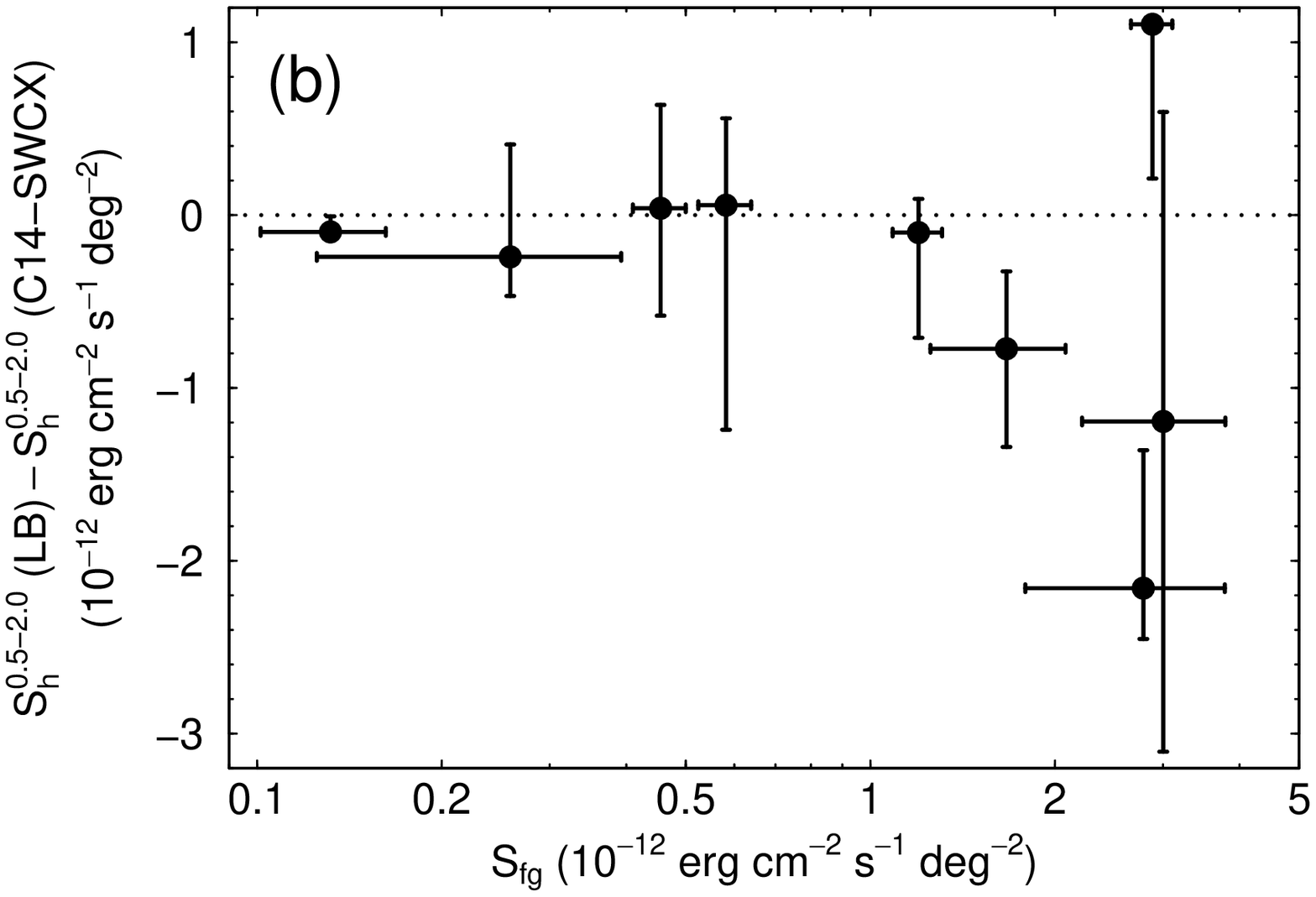}
\caption{Differences between in the halo (a) temperatures and (b) 0.5--2.0~\kev\ surface
  brightnesses obtained with the LB and C14-SWCX foreground models, as functions of the
  0.4--1.0~\kev\ foreground surface brightness.
  For each shadowing observation, the abscissa indicates the mean foreground surface
  brightness from the two foreground models, while the error bar indicates the
  lower and higher values.
  The ordinate shows the difference between the halo measurements obtained with the
  two models, while the error bar indicates the uncertainty on this difference due to
  the uncertainty on the measurement using the C14-SWCX model.
  Note that the plots include the results for G225$-$66 from \citet{henley15a}. The temperature plot
  does not include MBM~16, as the temperature was held fixed for this observation (see
  Section~\ref{sec:SpectralModel}).
  \label{fig:DiffvFG}}
\end{figure}

The uncertainties quoted in our tables of results (Tables~\ref{tab:Results} and \ref{tab:SB}) are
purely statistical uncertainties.  In this section, we discuss possible uncertainties or
biases in our halo measurements that may arise from the spectral modeling in addition
to the statistical uncertainties.

We described the results obtained with the different foreground models in
Section~\ref{subsec:ResultsDifferentForegrounds} and found that, in general, the C14-SWCX and LB
foreground models yielded consistent halo results. However, we noted some exceptions to this general
result. These occur when the foreground is relatively bright, as shown in Figure~\ref{fig:DiffvFG},
which shows the difference between the halo (a) temperatures and (b) 0.5--2.0~\kev\ surface
brightnesses obtained using the LB and C14-SWCX foreground models, as functions of the foreground
surface brightness. These differences may be interpreted in two ways. They may be interpreted as the
uncertainty in the halo parameters due to the fact that we do not know the relative contributions of
SWCX and LB emission to the \xmm/\suzaku\ foreground. Alternatively, if the LB emission is
negligible in the \xmm/\suzaku\ band, then these differences represent the bias in the halo
parameters due to assuming an LB-like foreground model.

When the foreground 0.4--1.0~\kev\ surface brightness is $\la$$1.5 \times 10^{-12}~\flux\ \pdegsq$,
the differences in the halo results obtained with the two foreground models are negligible:
$\la$$0.05\times10^6~\K$ and $\la$$0.2 \times 10^{-12}~\flux\ \pdegsq$ on the halo temperature and
surface brightness, respectively. For foreground surface brightnesses of
$\sim$$3\times10^{-12}~\flux\ \pdegsq$, however, these differences are
$\sim$$(\mbox{0.2--0.3})\times10^6~\K$ and $\sim$$(\mbox{1--2})\times10^{-12}~\flux\ \pdegsq$,
respectively (the observations in question are G236+38, MBM~16, and the \xmm\ observation of the
filament).  For G236+38, these differences are similar in size to the statistical uncertainties. For
MBM~16, the difference in the halo surface brightnesses from the two foreground models is of the
opposite sign to the other observations' differences (note, however, that halo emission was not
detected using the C14-SWCX for this observation, so a negative surface brightness difference is not
possible).

These results mean that the uncertainty in the true nature of the foreground X-ray emission has a
negligible effect on the halo measurements if the 0.4--1.0~\kev\ foreground surface brightness is
$\la 1.5 \times 10^{-12}~\flux\ \pdegsq$. If the foreground is brighter than this, then assuming an
LB-like foreground model when the true \xmm/\suzaku\ foreground is dominated by SWCX may bias the
halo temperature upward by $\sim$$(\mbox{0.2--0.3})\times10^6~\K$ and the halo surface brightness
downward by $\sim$$(\mbox{1--2})\times10^{-12}~\flux\ \pdegsq$, if the halo temperature is a free
parameter. Alternatively, if the true foreground in the \xmm/\suzaku\ band is a mixture of LB and
SWCX emission, then the uncertainty in this mixture results in an uncertainty of up to
$\sim$$(\mbox{0.2--0.3})\times10^6~\K$ and $\sim$$(\mbox{1--2})\times10^{-12}~\flux\ \pdegsq$ in the
temperature and surface brightness of the halo, respectively. Uncertainties in the CX line
ratios, which we are unable to quantify, could also contribute to the uncertainty in the halo
emission, if the foreground is relatively bright.

Our sample includes two shadows, MBM~20 and the filament, that have been observed twice, once each
with \xmm\ and \suzaku. For each shadow, the two observatories yield halo temperatures that differ
by $\sim$$0.2\times10^6~\K$ (using our reference model). However, there is not a systematic
difference between the two observatories---\xmm\ yields a lower temperature than \suzaku\ for
MBM~20, but a higher temperature for the filament. \xmm\ yields higher halo surface brightnesses
than \suzaku\ for both shadows, by $0.8\times10^{-12}$ and $1.6\times10^{-12}~\flux\ \pdegsq$ for
MBM~20 and the filament, respectively. Note that these differences are not statistically significant
for MBM~20, given the relatively large uncertainties on the \suzaku\ measurements, but are
statistically significant for the filament. The differences between the results from the two
satellites are unlikely to be due to our assumed column densities, since, for each cloud, we used
similar on- and off-shadow column densities for the \xmm\ and \suzaku\ observations.  Instead, the
differences may be due to our assuming that, for a given shadowing observation, the foreground is
identical in the on- and off-shadow spectra. If the SWCX emission changes between the two
pointings, the resulting halo measurement would be inaccurate. However, if this is the sole
explanation for the differences between the \xmm\ and \suzaku\ measurements, we might expect the
MBM~20 measurements to exhibit greater differences than the filament measurements, as the on- and
off-shadow MBM~20 pointings were further apart in time (for both the \xmm\ and the
\suzaku\ observations). Other possible explanations include halo structure on scales smaller than
the angular separation between the on- and off-shadow pointings (our spectral analysis also
assumes that the halo emission is identical in the two spectra), or inaccurate modeling of the
soft proton contamination in the \xmm\ spectra.  Regardless of the source of the differences
between the \xmm\ and \suzaku\ measurements of the halo emission in the directions of MBM~20 and
the filament, the existence of these differences implies that there may be uncertainties of up to
$\sim$$0.2\times10^6~\K$ and $\sim$25\%\ in the temperature and surface brightness of the halo,
respectively, in addition to the statistical uncertainties.

\subsubsection{Testing Models of the Galactic Halo Emission}
\label{subsubsec:TestingHaloModels}

We modeled the Galactic halo emission in the \xmm/\suzaku\ band using a $1T$ thermal plasma model,
assuming CIE. It should be noted that such a model is a characterization of the emission, and does
not necessarily reflect the true thermodynamic state of the halo gas. Our median halo temperature
and emission measure (obtained using our reference model) predict a halo 1/4~\kev\ \rosat\ count
rate of $\sim$$400\times10^{-6}~\rassrate$, $\sim$1/3 of the median halo count rate detected by
\citet{snowden00} in their study of X-ray shadows in the \rosat\ All-Sky Survey. This discrepancy
indicates the need for additional halo emission components. \citet{kuntz00} also showed that a $1T$
model cannot adequately model the halo emission down to $\sim$0.1~\kev, and used a two-temperature
halo model in their analysis of the \rosat\ All-Sky Survey. Other studies have used X-ray and
far-ultraviolet emission data to consider the distribution of hot halo gas down to
$\sim$$1\times10^5~\K$, using differential emission measure models
\citep{shelton07,lei09}. Recombination emission from overionized halo plasma may be important (e.g.,
\citealt{breitschwerdt94,avillez12a}; although see discussion in \citealt{henley15b}).  In addition,
\citet{henley15b} recently discussed the possibility that CX emission, rather than thermal emission
from hot gas, may contribute significantly to the observed halo emission.

Despite the uncertainties in the true nature of the Galactic halo emission, and of the hot halo gas,
measurements of the Galactic halo emission that use a $1T$ halo model can still be used to test
physical models of the halo emission, even if the emission is from multitemperature gas, or is from
gas out of ionization equilibrium, or is due to CX. As we have noted previously, the key is to
ensure that the predicted spectra are characterized in the same way as the observed spectra---the
X-ray temperatures, emission measures, and surface brightnesses that result from this
characterization can be compared with the observed values \citep{henley10b,henley15b}.

In Section~\ref{subsubsec:HaloOtherStudies}, we showed that, with the possible exception of the very
brightest directions, there appears to be no significant systematic difference between the halo
temperatures and surface brightnesses obtained from our shadowing observations and those obtained by
HS13. This further strengthens the conclusion of \citet{henley15a} that the latter results can be
used to test models of the halo X-ray emission \citep{henley15b}.  However, in
Section~\ref{subsubsec:HaloUncertaintyBias}, we discussed possible uncertainties or biases in the
halo emission in addition to the statistical uncertainties. When testing halo models, one
should take these potential uncertainties in the halo emission into account, which may limit the
ability of measurements of the halo X-ray emission to discriminate between models.

\section{SUMMARY AND CONCLUSIONS}
\label{sec:Summary}

We have analyzed eight \xmm\ and \suzaku\ observations of six shadowing clouds, including the first
such observations from the northern Galactic hemisphere (Section~\ref{sec:Reduction}). To our
results we added recently published results from a seventh cloud \citep{henley15a}. Ours is the
first uniform analysis of a sample of X-ray shadowing clouds with CCD-resolution spectra.

We tested a variety of foreground models---two that assume that the foreground emission is dominated
by SWCX (C14-SWCX and ACX-SWCX) and one that assumes that it is from the hot LB
(Section~\ref{sec:SpectralModel}). We also tested three abundance tables---AG89 (commonly used in
previous SXRB studies), Wilm00, and Aspl09. Our reference model was the C14-SWCX foreground model
with Aspl09 abundances. We modeled the Galactic halo emission using a $1T$ model---while this is a
simplification of the true halo emission, such a model can still usefully characterize the halo
emission (Section~\ref{subsubsec:TestingHaloModels}).

Our main results, and resulting conclusions, are as follows:
\begin{enumerate}
\setlength{\itemsep}{0pt}
\item The measured foreground surface brightnesses span more than an order of magnitude, from
  $\sim$$10^{-13}$ to $\sim 10^{-12}~\flux\ \pdegsq$ (0.4--1.0~\kev;
  Figure~\ref{fig:ForegroundResults}). The brighter foregrounds tend also to be harder.  The halo
  temperatures are all around $2\times10^6~\K$ ($(\mbox{1.6--2.3})\times10^6~\K$ for our reference
  model), with detected emission measures and 0.5--2.0~\kev\ surface brightnesses of
  $(\mbox{5--17})\times10^{-3}~\emismeas$ and $(\mbox{1--7})\times10^{-12}~\flux\ \pdegsq$,
  respectively (Figure~\ref{fig:HaloResults}). On one sight line (toward MBM~16) we do not detect
  halo emission with our reference model (a result previously reported by \citealt{ursino14}).
\item In general, the Wilm00 and Aspl09 abundances yield similar results, due to the similarity of
  the carbon, nitrogen, and oxygen abundances. However, the AG89 abundances tend to yield higher
  foreground surface brightnesses and higher halo temperatures. These differences are due to AG89's
  higher carbon, nitrogen, and oxygen abundances (Section~\ref{subsec:ResultsDifferentAbundances}).
  Since these abundances are almost certainly too high for the interstellar medium, using AG89
  abundances in shadowing analyses may yield inaccurate results.
\item The ACX-SWCX model tends to attribute less \OVII\ emission to the foreground than the other
  models, which can affect the halo measurements
  (Section~\ref{subsec:ResultsDifferentForegrounds}). This is due to the fact that this model tends
  to produce relatively more foreground emission above $\sim$0.6~\kev\ than the other foreground
  models; this in turn is likely an artifact of the assumption that the solar wind ion distribution
  can be described by a single temperature. Since this is not a good description of the true solar
  wind ion distribution, the results obtained with a single-temperature ACX-SWCX model may
  be unreliable (Section~\ref{subsec:ForegroundChoice}).
\item We compared our measurements of the foreground surface brightness and the foreground
  \OVII/\CVI\ intensity ratio with general predictions of heliospheric SWCX emission from a smooth
  solar wind \citep{koutroumpa06}. Only some aspects of our observations were as expected,
  indicating the importance of other processes (such as emission from localized solar wind
  enhancements, or enhancements in the geocoronal SWCX emission; Section~\ref{subsec:SWCX}).
\item By comparing our measurements of the foreground \OVII\ intensity with observation-specific
  SWCX predictions (and excluding the \xmm\ MBM~20 measurement), we placed an upper limit
  of $\sim$0.8~\LU\ (90\%\ confidence) on the LB \OVII\ emission (for an LB path length of
  $\sim$100~\pc; Section~\ref{subsec:LB}).
\item Our halo measurements are generally in good agreement with those from recent non-shadowing
  studies of the halo (HS13; Yosh09). However, by assuming a uniform foreground model, Yosh09 may
  have overestimated the halo temperature on some of their sight lines
  (Section~\ref{subsubsec:HaloOtherStudies}).
\item When the foreground is relatively faint ($\la$$1.5\times10^{-12}~\flux\ \pdegsq$ in the
  0.4--1.0~\kev\ band), the C14-SWCX and LB foreground models yield similar results for the halo. If
  the foreground is brighter than this, and is truly dominated by SWCX emission, then assuming an
  LB-like foreground model may bias the halo temperature upward and the halo surface brightness
  downward. Alternatively, if the foreground is in fact a mixture of SWCX and LB emission, the
  uncertainty in this mixture results in an uncertainty in the halo temperature and surface
  brightness (of up to $\sim$$(\mbox{0.2--0.3})\times10^6~\K$ and
  $\sim$$(\mbox{1--2})\times10^{-12}~\flux\ \pdegsq$, respectively;
  Section~\ref{subsubsec:HaloUncertaintyBias}).  Differences in the halo results obtained for the
  same shadow with \xmm\ and \suzaku\ suggest that there may be uncertainties in the halo
  temperature and surface brightness (of up to $\sim$$0.2\times10^6~\K$ and $\sim$25\%,
  respectively), in addition to the statistical uncertainties
  (Section~\ref{subsubsec:HaloUncertaintyBias}). Both types of uncertainty may limit the ability of
  X-ray measurements to discriminate between models of the Galactic halo emission.
\end{enumerate}

\acknowledgements
We thank Phillip Stancil for detailed comments on the $l$ distributions resulting from CX.
We also thank the anonymous referee, whose comments have helped improve this paper.
This research is based on observations obtained with \xmm, an ESA science mission with instruments
and contributions directly funded by ESA Member States and NASA, and with \suzaku, a collaborative
mission between JAXA and NASA.
We acknowledge use of the R software package \citep{R}.
This research was funded by NASA grant NNX13AF69G, awarded through the Astrophysics Data Analysis
Program.

\bibliography{references}

\clearpage
\setcounter{table}{\value{ResultsTable}}
{
  \LongTables
  \tabletypesize{\scriptsize}
  \setlength{\tabcolsep}{2.5pt}
  \begin{landscape}
    \begin{deluxetable}{lrr@{,}lrr@{,}lrr@{,}lrr@{,}lrr@{,}lrr@{,}lcrr@{,}lrr@{,}lr@{/}l}
\tablewidth{0pt}
\tabletypesize{\scriptsize}
\tablecaption{Spectral Fit Results\label{tab:Results}}
\tablehead{
                                  & \multicolumn{18}{c}{Foreground}                                                                                                                                                                                                                                                                                                         && \multicolumn{6}{c}{Halo} \\
\cline{2-19} \cline{21-26}
\colhead{Shadow\tablenotemark{a}} & \multicolumn{3}{c}{\Tfg}        & \multicolumn{3}{c}{Normalization\tablenotemark{b}} & \multicolumn{3}{c}{$I(\mbox{\CV})$\tablenotemark{c,e}} & \multicolumn{3}{c}{$I(\mbox{\CVI})$\tablenotemark{d,e}} & \multicolumn{3}{c}{$I(\mbox{\OVII})$\tablenotemark{c}} & \multicolumn{3}{c}{$I(\mbox{\OVIII})$\tablenotemark{d}} && \multicolumn{3}{c}{\Th}         & \multicolumn{3}{c}{\EMh}                  & \chisq & dof \\
                                  & \multicolumn{3}{c}{($10^6~\K$)} & \multicolumn{3}{c}{}                               & \multicolumn{3}{c}{(\LU)}                              & \multicolumn{3}{c}{(\LU)}                               & \multicolumn{3}{c}{(\LU)}                              & \multicolumn{3}{c}{(\LU)}                               && \multicolumn{3}{c}{($10^6~\K$)} & \multicolumn{3}{c}{($10^{-3}~\emismeas$)}                \\
\colhead{(1)}                     & \multicolumn{3}{c}{(2)}         & \multicolumn{3}{c}{(3)}                            & \multicolumn{3}{c}{(4)}                                & \multicolumn{3}{c}{(5)}                                 & \multicolumn{3}{c}{(6)}                                & \multicolumn{3}{c}{(7)}                                 && \multicolumn{3}{c}{(8)}         & \multicolumn{3}{c}{(9)}                   & \multicolumn{2}{c}{(10)}
}
\startdata
\cutinhead{(a) C14-SWCX foreground model, \citet{asplund09} abundances}
G048+37 (X)  &    \multicolumn{3}{c}{\nodata} &    \multicolumn{3}{c}{\nodata} &     0.00 &    (0.00 &    4.97) &     5.19 &    (4.45 &    6.01) &     2.52 &    (2.21 &    4.69) &     0.00 &    (0.00 &    0.25) & &     2.10 &    (2.08 &    2.18) &    15.30 &   (14.21 &   16.69) & 2010.05 & 1822 \\
MBM 12 (S)   &    \multicolumn{3}{c}{\nodata} &    \multicolumn{3}{c}{\nodata} &    12.59 &    (3.97 &   21.43) &     3.70 &    (2.90 &    4.41) &     2.47 &    (1.97 &    2.86) &     0.00 &    (0.00 &    0.07) & &     1.61 &    (1.40 &    1.83) &     4.57 &    (3.71 &    7.24) &  410.71 &  358 \\
MBM 16 (S)   &    \multicolumn{3}{c}{\nodata} &    \multicolumn{3}{c}{\nodata} &    \multicolumn{3}{c}{\nodata} &    13.05 &   (10.99 &   15.05) &     5.29 &    (4.24 &    6.01) &     0.75 &    (0.31 &    1.06) & & \multicolumn{3}{c}{2.10\tablenotemark{f}} &     0.10 &    (0.00 &    2.28) &  263.54 &  267 \\
MBM 20 (X)   &    \multicolumn{3}{c}{\nodata} &    \multicolumn{3}{c}{\nodata} &     3.83 &    (1.31 &    5.37) &     1.10 &    (0.73 &    1.30) &     0.00 &    (0.00 &    0.06) &     0.00 &    (0.00 &    0.04) & &     1.84 &    (1.81 &    1.86) &     9.36 &    (9.07 &    9.49) & 3973.74 & 3601 \\
MBM 20 (S)   &    \multicolumn{3}{c}{\nodata} &    \multicolumn{3}{c}{\nodata} &    37.82 &   (21.56 &   53.94) &     2.55 &    (1.39 &    3.71) &     0.97 &    (0.00 &    1.74) &     0.00 &    (0.00 &    0.11) & &     2.06 &    (1.73 &    2.20) &     5.51 &    (4.24 &    8.78) &  402.75 &  361 \\
G236+38 (S)  &    \multicolumn{3}{c}{\nodata} &    \multicolumn{3}{c}{\nodata} &    78.65 &    (0.00 &  156.63) &    13.64 &    (9.46 &   17.88) &     3.33 &    (0.52 &    6.24) &     0.00 &    (0.00 &    0.85) & &     1.72 &    (1.44 &    1.83) &    17.11 &   (10.49 &   24.17) &  263.35 &  230 \\
Filament (X) &    \multicolumn{3}{c}{\nodata} &    \multicolumn{3}{c}{\nodata} &     0.00 &    (0.00 &   20.51) &     6.03 &    (4.43 &    6.58) &     4.17 &    (0.28 &    5.38) &     0.00 &    (0.00 &    0.20) & &     2.32 &    (2.23 &    2.36) &    15.51 &   (13.83 &   16.13) &  736.50 &  673 \\
Filament (S) &    \multicolumn{3}{c}{\nodata} &    \multicolumn{3}{c}{\nodata} &     9.21 &    (1.21 &   18.13) &     1.45 &    (0.59 &    2.18) &     0.00 &    (0.00 &    0.47) &     0.00 &    (0.00 &    0.05) & &     2.14 &    (2.10 &    2.18) &    13.76 &   (12.22 &   14.30) &  550.18 &  398 \\
\cutinhead{(b) ACX-SWCX foreground model, \citet{asplund09} abundances}
G048+37 (X)  &     2.22 &    (2.13 &    2.43) &     1.53 &    (1.28 &    2.03) &    \multicolumn{3}{c}{\nodata} &    \multicolumn{3}{c}{\nodata} &    \multicolumn{3}{c}{\nodata} &    \multicolumn{3}{c}{\nodata} & &     1.88 &    (1.67 &    1.95) &    15.06 &   (13.91 &   16.73) & 1997.19 & 1824 \\
MBM 12 (S)   &     1.71 &    (1.61 &    1.80) &     1.69 &    (1.44 &    1.97) &    \multicolumn{3}{c}{\nodata} &    \multicolumn{3}{c}{\nodata} &    \multicolumn{3}{c}{\nodata} &    \multicolumn{3}{c}{\nodata} & &     1.58 &    (1.39 &    1.65) &     5.82 &    (4.45 &    8.46) &  428.35 &  360 \\
MBM 16 (S)   &     1.89 &    (1.78 &    1.96) &     3.42 &    (2.92 &    3.76) &    \multicolumn{3}{c}{\nodata} &    \multicolumn{3}{c}{\nodata} &    \multicolumn{3}{c}{\nodata} &    \multicolumn{3}{c}{\nodata} & & \multicolumn{3}{c}{2.10\tablenotemark{f}} &     0.00 &    (0.00 &    1.51) &  265.04 &  268 \\
MBM 20 (X)   &     0.98 &    (0.94 &    0.99) &     3.86 &    (3.18 &    4.29) &    \multicolumn{3}{c}{\nodata} &    \multicolumn{3}{c}{\nodata} &    \multicolumn{3}{c}{\nodata} &    \multicolumn{3}{c}{\nodata} & &     1.90 &    (1.87 &    1.91) &     8.87 &    (8.62 &    9.36) & 3975.93 & 3603 \\
MBM 20 (S)   &     0.81 &    (0.72 &    0.89) &    30.88 &   (16.22 &   57.93) &    \multicolumn{3}{c}{\nodata} &    \multicolumn{3}{c}{\nodata} &    \multicolumn{3}{c}{\nodata} &    \multicolumn{3}{c}{\nodata} & &     1.89 &    (1.76 &    2.06) &     7.47 &    (6.09 &    8.92) &  400.11 &  363 \\
G236+38 (S)  &     0.83 &    (0.76 &    0.91) &   102.70 &   (53.62 &  176.50) &    \multicolumn{3}{c}{\nodata} &    \multicolumn{3}{c}{\nodata} &    \multicolumn{3}{c}{\nodata} &    \multicolumn{3}{c}{\nodata} & &     1.68 &    (1.60 &    1.77) &    22.53 &   (19.12 &   25.94) &  261.95 &  232 \\
Filament (X) &     1.93 &    (1.14 &    2.46) &     1.73 &    (1.21 &    2.00) &    \multicolumn{3}{c}{\nodata} &    \multicolumn{3}{c}{\nodata} &    \multicolumn{3}{c}{\nodata} &    \multicolumn{3}{c}{\nodata} & &     2.19 &    (2.09 &    2.24) &    17.39 &   (15.52 &   19.77) &  733.06 &  675 \\
Filament (S) &     0.88 &    (0.74 &    1.05) &     8.92 &    (2.83 &   21.42) &    \multicolumn{3}{c}{\nodata} &    \multicolumn{3}{c}{\nodata} &    \multicolumn{3}{c}{\nodata} &    \multicolumn{3}{c}{\nodata} & &     2.15 &    (2.11 &    2.19) &    13.60 &   (12.93 &   14.27) &  547.94 &  400 \\
\cutinhead{(c) LB foreground model, \citet{asplund09} abundances}
G048+37 (X)  &     1.34 &    (1.26 &    1.40) &    12.01 &   (10.39 &   14.18) &    \multicolumn{3}{c}{\nodata} &    \multicolumn{3}{c}{\nodata} &    \multicolumn{3}{c}{\nodata} &    \multicolumn{3}{c}{\nodata} & &     2.17 &    (2.14 &    2.26) &    12.80 &   (11.41 &   13.44) & 1998.69 & 1824 \\
MBM 12 (S)   &     1.23 &    (1.12 &    1.30) &    10.15 &    (8.23 &   13.42) &    \multicolumn{3}{c}{\nodata} &    \multicolumn{3}{c}{\nodata} &    \multicolumn{3}{c}{\nodata} &    \multicolumn{3}{c}{\nodata} & &     1.64 &    (1.41 &    1.96) &     3.95 &    (2.22 &    6.09) &  395.81 &  360 \\
MBM 16 (S)   &     1.03 &    (0.86 &    1.13) &    42.29 &   (28.07 &   96.21) &    \multicolumn{3}{c}{\nodata} &    \multicolumn{3}{c}{\nodata} &    \multicolumn{3}{c}{\nodata} &    \multicolumn{3}{c}{\nodata} & & \multicolumn{3}{c}{2.10\tablenotemark{f}} &     2.80 &    (1.18 &    4.42) &  275.42 &  268 \\
MBM 20 (X)   &     0.77 &    (0.73 &    0.80) &    13.08 &   (10.50 &   15.57) &    \multicolumn{3}{c}{\nodata} &    \multicolumn{3}{c}{\nodata} &    \multicolumn{3}{c}{\nodata} &    \multicolumn{3}{c}{\nodata} & &     1.86 &    (1.83 &    1.88) &     8.87 &    (8.67 &    9.05) & 4011.09 & 3603 \\
MBM 20 (S)   &     0.66 &    (0.56 &    0.83) &   133.17 &   (38.30 &  362.44) &    \multicolumn{3}{c}{\nodata} &    \multicolumn{3}{c}{\nodata} &    \multicolumn{3}{c}{\nodata} &    \multicolumn{3}{c}{\nodata} & &     2.07 &    (1.88 &    2.19) &     5.61 &    (4.75 &    6.47) &  388.11 &  363 \\
G236+38 (S)  &     1.01 &    (0.64 &    1.10) &    66.87 &   (51.92 &  758.26) &    \multicolumn{3}{c}{\nodata} &    \multicolumn{3}{c}{\nodata} &    \multicolumn{3}{c}{\nodata} &    \multicolumn{3}{c}{\nodata} & &     1.94 &    (1.69 &    2.16) &     9.64 &    (5.51 &   18.56) &  257.71 &  232 \\
Filament (X) &     1.59 &    (1.48 &    1.70) &    13.06 &   (11.60 &   14.50) &    \multicolumn{3}{c}{\nodata} &    \multicolumn{3}{c}{\nodata} &    \multicolumn{3}{c}{\nodata} &    \multicolumn{3}{c}{\nodata} & &     2.66 &    (2.40 &    2.73) &     9.24 &    (8.51 &   12.29) &  727.95 &  675 \\
Filament (S) &     0.88 &    (0.65 &    1.29) &    12.87 &    (4.30 &   61.33) &    \multicolumn{3}{c}{\nodata} &    \multicolumn{3}{c}{\nodata} &    \multicolumn{3}{c}{\nodata} &    \multicolumn{3}{c}{\nodata} & &     2.18 &    (2.12 &    2.27) &    12.87 &   (10.91 &   13.85) &  548.85 &  400 \\
\cutinhead{(d) C14-SWCX foreground model, \citet{anders89} abundances}
G048+37 (X)  &    \multicolumn{3}{c}{\nodata} &    \multicolumn{3}{c}{\nodata} &     0.00 &    (0.00 &    5.57) &     7.28 &    (6.64 &    8.22) &     5.08 &    (4.79 &    5.62) &     0.07 &    (0.00 &    0.34) & &     2.23 &    (2.20 &    2.27) &     7.72 &    (7.06 &    8.46) & 2058.12 & 1822 \\
MBM 12 (S)   &    \multicolumn{3}{c}{\nodata} &    \multicolumn{3}{c}{\nodata} &    12.73 &    (4.11 &   21.37) &     3.93 &    (3.19 &    4.31) &     2.66 &    (2.27 &    3.05) &     0.00 &    (0.00 &    0.10) & &     1.71 &    (1.47 &    2.01) &     2.79 &    (1.73 &    4.11) &  401.16 &  358 \\
MBM 16 (S)   &    \multicolumn{3}{c}{\nodata} &    \multicolumn{3}{c}{\nodata} &    \multicolumn{3}{c}{\nodata} &    13.46 &   (11.44 &   15.48) &     5.37 &    (4.43 &    6.16) &     1.01 &    (0.60 &    1.35) & & \multicolumn{3}{c}{2.10\tablenotemark{f}} &     0.37 &    (0.00 &    1.69) &  260.91 &  267 \\
MBM 20 (X)   &    \multicolumn{3}{c}{\nodata} &    \multicolumn{3}{c}{\nodata} &     2.01 &    (0.37 &    4.46) &     2.60 &    (2.25 &    2.81) &     0.85 &    (0.73 &    1.02) &     0.00 &    (0.00 &    0.02) & &     2.06 &    (2.04 &    2.09) &     4.45 &    (4.38 &    4.61) & 4056.98 & 3601 \\
MBM 20 (S)   &    \multicolumn{3}{c}{\nodata} &    \multicolumn{3}{c}{\nodata} &    38.36 &   (17.25 &   50.01) &     3.34 &    (2.25 &    4.47) &     1.83 &    (1.15 &    2.47) &     0.00 &    (0.00 &    0.16) & &     2.24 &    (2.10 &    2.35) &     2.94 &    (2.58 &    3.30) &  411.30 &  361 \\
G236+38 (S)  &    \multicolumn{3}{c}{\nodata} &    \multicolumn{3}{c}{\nodata} &    67.21 &    (0.00 &  145.31) &    17.28 &   (13.32 &   21.18) &     6.16 &    (3.59 &    8.92) &     0.00 &    (0.00 &    1.03) & &     1.93 &    (1.59 &    2.15) &     6.65 &    (3.88 &   10.98) &  273.72 &  230 \\
Filament (X) &    \multicolumn{3}{c}{\nodata} &    \multicolumn{3}{c}{\nodata} &     1.51 &    (0.00 &   18.12) &     5.93 &    (5.16 &    7.31) &     3.75 &    (2.66 &    4.62) &     0.00 &    (0.00 &    0.15) & &     2.28 &    (2.24 &    2.35) &    11.35 &   (10.67 &   12.32) &  753.62 &  673 \\
Filament (S) &    \multicolumn{3}{c}{\nodata} &    \multicolumn{3}{c}{\nodata} &    12.30 &    (1.29 &   20.80) &     2.85 &    (2.03 &    3.57) &     1.07 &    (0.69 &    1.62) &     0.00 &    (0.00 &    0.06) & &     2.20 &    (2.17 &    2.26) &     8.23 &    (7.82 &    8.55) &  584.47 &  398 \\
\cutinhead{(e) ACX-SWCX foreground model, \citet{anders89} abundances}
G048+37 (X)  &     2.17 &    (2.11 &    2.22) &     1.60 &    (1.47 &    1.74) &    \multicolumn{3}{c}{\nodata} &    \multicolumn{3}{c}{\nodata} &    \multicolumn{3}{c}{\nodata} &    \multicolumn{3}{c}{\nodata} & &     1.84 &    (1.78 &    1.93) &     7.82 &    (7.27 &    8.78) & 2031.89 & 1824 \\
MBM 12 (S)   &     1.69 &    (1.61 &    1.77) &     1.29 &    (1.10 &    1.48) &    \multicolumn{3}{c}{\nodata} &    \multicolumn{3}{c}{\nodata} &    \multicolumn{3}{c}{\nodata} &    \multicolumn{3}{c}{\nodata} & &     1.58 &    (1.34 &    1.72) &     3.43 &    (2.54 &    5.71) &  412.79 &  360 \\
MBM 16 (S)   &     1.91 &    (1.80 &    2.00) &     2.22 &    (1.89 &    2.64) &    \multicolumn{3}{c}{\nodata} &    \multicolumn{3}{c}{\nodata} &    \multicolumn{3}{c}{\nodata} &    \multicolumn{3}{c}{\nodata} & & \multicolumn{3}{c}{2.10\tablenotemark{f}} &     0.00 &    (0.00 &    0.87) &  257.27 &  268 \\
MBM 20 (X)   &     1.18 &    (1.14 &    1.21) &     0.88 &    (0.75 &    1.01) &    \multicolumn{3}{c}{\nodata} &    \multicolumn{3}{c}{\nodata} &    \multicolumn{3}{c}{\nodata} &    \multicolumn{3}{c}{\nodata} & &     1.84 &    (1.82 &    1.86) &     6.03 &    (5.84 &    6.12) & 4062.64 & 3603 \\
MBM 20 (S)   &     0.83 &    (0.76 &    0.91) &    22.83 &   (12.99 &   37.52) &    \multicolumn{3}{c}{\nodata} &    \multicolumn{3}{c}{\nodata} &    \multicolumn{3}{c}{\nodata} &    \multicolumn{3}{c}{\nodata} & &     2.08 &    (2.00 &    2.17) &     4.26 &    (3.82 &    4.70) &  419.86 &  363 \\
G236+38 (S)  &     0.84 &    (0.78 &    0.91) &    79.57 &   (43.73 &  130.00) &    \multicolumn{3}{c}{\nodata} &    \multicolumn{3}{c}{\nodata} &    \multicolumn{3}{c}{\nodata} &    \multicolumn{3}{c}{\nodata} & &     1.73 &    (1.65 &    1.85) &    13.09 &   (10.93 &   15.24) &  280.79 &  232 \\
Filament (X) &     2.10 &    (1.85 &    2.21) &     1.58 &    (1.28 &    1.85) &    \multicolumn{3}{c}{\nodata} &    \multicolumn{3}{c}{\nodata} &    \multicolumn{3}{c}{\nodata} &    \multicolumn{3}{c}{\nodata} & &     2.19 &    (2.13 &    2.29) &     9.27 &    (8.26 &   10.54) &  750.33 &  675 \\
Filament (S) &     0.89 &    (0.81 &    1.00) &     9.38 &    (4.66 &   16.29) &    \multicolumn{3}{c}{\nodata} &    \multicolumn{3}{c}{\nodata} &    \multicolumn{3}{c}{\nodata} &    \multicolumn{3}{c}{\nodata} & &     2.16 &    (2.11 &    2.20) &     8.99 &    (8.55 &    9.42) &  580.75 &  400 \\
\pagebreak
\cutinhead{(f) LB foreground model, \citet{anders89} abundances}
G048+37 (X)  &     1.32 &    (1.28 &    1.37) &    12.31 &   (10.98 &   13.72) &    \multicolumn{3}{c}{\nodata} &    \multicolumn{3}{c}{\nodata} &    \multicolumn{3}{c}{\nodata} &    \multicolumn{3}{c}{\nodata} & &     2.43 &    (2.32 &    2.56) &     5.71 &    (5.26 &    6.70) & 2029.94 & 1824 \\
MBM 12 (S)   &     1.12 &    (1.06 &    1.22) &     9.23 &    (6.98 &   11.50) &    \multicolumn{3}{c}{\nodata} &    \multicolumn{3}{c}{\nodata} &    \multicolumn{3}{c}{\nodata} &    \multicolumn{3}{c}{\nodata} & &     1.82 &    (1.59 &    2.09) &     2.23 &    (1.24 &    3.31) &  383.39 &  360 \\
MBM 16 (S)   &     1.02 &    (0.82 &    1.09) &    29.61 &   (20.28 &   85.70) &    \multicolumn{3}{c}{\nodata} &    \multicolumn{3}{c}{\nodata} &    \multicolumn{3}{c}{\nodata} &    \multicolumn{3}{c}{\nodata} & & \multicolumn{3}{c}{2.10\tablenotemark{f}} &     2.55 &    (1.50 &    3.67) &  285.31 &  268 \\
MBM 20 (X)   &     1.07 &    (1.05 &    1.09) &     5.03 &    (4.68 &    5.43) &    \multicolumn{3}{c}{\nodata} &    \multicolumn{3}{c}{\nodata} &    \multicolumn{3}{c}{\nodata} &    \multicolumn{3}{c}{\nodata} & &     2.07 &    (2.05 &    2.10) &     4.25 &    (4.15 &    4.36) & 4052.79 & 3603 \\
MBM 20 (S)   &     0.70 &    (0.64 &    0.95) &    79.18 &   (17.41 &  157.54) &    \multicolumn{3}{c}{\nodata} &    \multicolumn{3}{c}{\nodata} &    \multicolumn{3}{c}{\nodata} &    \multicolumn{3}{c}{\nodata} & &     2.22 &    (2.10 &    2.42) &     3.28 &    (2.73 &    3.84) &  395.17 &  363 \\
G236+38 (S)  &     1.01 &    (0.81 &    1.07) &    51.74 &   (41.60 &  147.78) &    \multicolumn{3}{c}{\nodata} &    \multicolumn{3}{c}{\nodata} &    \multicolumn{3}{c}{\nodata} &    \multicolumn{3}{c}{\nodata} & &     2.15 &    (1.90 &    2.54) &     4.42 &    (3.36 &    7.26) &  265.02 &  232 \\
Filament (X) &     1.40 &    (1.34 &    1.54) &    11.43 &    (9.74 &   13.24) &    \multicolumn{3}{c}{\nodata} &    \multicolumn{3}{c}{\nodata} &    \multicolumn{3}{c}{\nodata} &    \multicolumn{3}{c}{\nodata} & &     2.71 &    (2.62 &    2.90) &     6.67 &    (5.48 &    7.53) &  730.96 &  675 \\
Filament (S) &     1.11 &    (0.84 &    1.28) &     8.16 &    (5.40 &   19.69) &    \multicolumn{3}{c}{\nodata} &    \multicolumn{3}{c}{\nodata} &    \multicolumn{3}{c}{\nodata} &    \multicolumn{3}{c}{\nodata} & &     2.29 &    (2.19 &    2.43) &     7.04 &    (5.82 &    8.32) &  575.47 &  400 \\
\cutinhead{(g) C14-SWCX foreground model, \citet{wilms00} abundances}
G048+37 (X)  &    \multicolumn{3}{c}{\nodata} &    \multicolumn{3}{c}{\nodata} &     0.00 &    (5.02 &    5.98) &     5.92 &    (5.23 &    6.79) &     3.06 &    (2.74 &    3.95) &     0.00 &    (0.00 &    0.19) & &     2.14 &    (2.10 &    2.16) &    14.72 &   (13.13 &   15.60) & 2012.41 & 1822 \\
MBM 12 (S)   &    \multicolumn{3}{c}{\nodata} &    \multicolumn{3}{c}{\nodata} &    13.08 &    (4.44 &   21.72) &     3.70 &    (2.95 &    4.45) &     2.43 &    (2.00 &    2.85) &     0.00 &    (0.00 &    0.07) & &     1.59 &    (1.39 &    1.79) &     4.99 &    (3.31 &    7.53) &  410.32 &  358 \\
MBM 16 (S)   &    \multicolumn{3}{c}{\nodata} &    \multicolumn{3}{c}{\nodata} &    \multicolumn{3}{c}{\nodata} &    13.24 &   (11.18 &   15.22) &     5.35 &    (4.31 &    6.05) &     0.77 &    (0.33 &    1.08) & & \multicolumn{3}{c}{2.10\tablenotemark{f}} &     0.05 &    (0.00 &    2.22) &  264.22 &  267 \\
MBM 20 (X)   &    \multicolumn{3}{c}{\nodata} &    \multicolumn{3}{c}{\nodata} &     4.49 &    (1.93 &    7.04) &     1.29 &    (1.07 &    1.63) &     0.00 &    (0.00 &    0.05) &     0.00 &    (0.00 &    0.05) & &     1.84 &    (1.81 &    1.86) &     9.67 &    (9.39 &    9.89) & 3959.80 & 3601 \\
MBM 20 (S)   &    \multicolumn{3}{c}{\nodata} &    \multicolumn{3}{c}{\nodata} &    39.41 &   (23.15 &   55.67) &     2.68 &    (1.52 &    3.85) &     1.01 &    (0.00 &    1.78) &     0.00 &    (0.00 &    0.13) & &     2.06 &    (1.76 &    2.21) &     5.53 &    (4.26 &    8.88) &  403.47 &  361 \\
G236+38 (S)  &    \multicolumn{3}{c}{\nodata} &    \multicolumn{3}{c}{\nodata} &    84.28 &   (10.66 &  151.69) &    14.02 &    (9.80 &   18.14) &     3.50 &    (0.60 &    6.35) &     0.00 &    (0.00 &    0.86) & &     1.73 &    (1.44 &    1.85) &    16.78 &   (10.22 &   23.83) &  263.32 &  230 \\
Filament (X) &    \multicolumn{3}{c}{\nodata} &    \multicolumn{3}{c}{\nodata} &     1.90 &    (0.00 &   20.33) &     4.78 &    (4.07 &    6.24) &     2.87 &    (0.82 &    3.68) &     0.00 &    (0.00 &    0.17) & &     2.24 &    (2.21 &    2.35) &    18.02 &   (16.12 &   19.36) &  738.18 &  673 \\
Filament (S) &    \multicolumn{3}{c}{\nodata} &    \multicolumn{3}{c}{\nodata} &    12.13 &    (4.28 &   20.02) &     1.78 &    (1.05 &    2.52) &     0.00 &    (0.00 &    0.48) &     0.00 &    (0.00 &    0.06) & &     2.15 &    (2.11 &    2.19) &    13.87 &   (12.95 &   14.40) &  553.17 &  398 \\
\cutinhead{(h) ACX-SWCX foreground model, \citet{wilms00} abundances}
G048+37 (X)  &     2.24 &    (2.16 &    2.37) &     1.73 &    (1.59 &    2.14) &    \multicolumn{3}{c}{\nodata} &    \multicolumn{3}{c}{\nodata} &    \multicolumn{3}{c}{\nodata} &    \multicolumn{3}{c}{\nodata} & &     1.83 &    (1.74 &    1.90) &    15.30 &   (14.04 &   16.76) & 1997.18 & 1824 \\
MBM 12 (S)   &     1.68 &    (1.58 &    1.77) &     1.87 &    (1.60 &    2.17) &    \multicolumn{3}{c}{\nodata} &    \multicolumn{3}{c}{\nodata} &    \multicolumn{3}{c}{\nodata} &    \multicolumn{3}{c}{\nodata} & &     1.57 &    (1.36 &    1.65) &     5.81 &    (4.36 &    9.00) &  428.47 &  360 \\
MBM 16 (S)   &     1.85 &    (1.74 &    1.92) &     3.69 &    (3.13 &    4.04) &    \multicolumn{3}{c}{\nodata} &    \multicolumn{3}{c}{\nodata} &    \multicolumn{3}{c}{\nodata} &    \multicolumn{3}{c}{\nodata} & & \multicolumn{3}{c}{2.10\tablenotemark{f}} &     0.00 &    (0.00 &    1.49) &  265.73 &  268 \\
MBM 20 (X)   &     0.92 &    (0.90 &    0.95) &     5.10 &    (4.40 &    6.12) &    \multicolumn{3}{c}{\nodata} &    \multicolumn{3}{c}{\nodata} &    \multicolumn{3}{c}{\nodata} &    \multicolumn{3}{c}{\nodata} & &     1.87 &    (1.84 &    1.88) &     8.92 &    (8.75 &    9.12) & 4008.11 & 3603 \\
MBM 20 (S)   &     0.80 &    (0.71 &    0.87) &    38.71 &   (21.13 &   71.01) &    \multicolumn{3}{c}{\nodata} &    \multicolumn{3}{c}{\nodata} &    \multicolumn{3}{c}{\nodata} &    \multicolumn{3}{c}{\nodata} & &     1.89 &    (1.76 &    2.06) &     7.52 &    (6.14 &    8.97) &  399.95 &  363 \\
G236+38 (S)  &     0.82 &    (0.75 &    0.89) &   129.10 &   (69.67 &  219.80) &    \multicolumn{3}{c}{\nodata} &    \multicolumn{3}{c}{\nodata} &    \multicolumn{3}{c}{\nodata} &    \multicolumn{3}{c}{\nodata} & &     1.68 &    (1.61 &    1.78) &    22.36 &   (18.92 &   25.80) &  261.45 &  232 \\
Filament (X) &     2.08 &    (1.83 &    2.28) &     2.02 &    (1.50 &    2.43) &    \multicolumn{3}{c}{\nodata} &    \multicolumn{3}{c}{\nodata} &    \multicolumn{3}{c}{\nodata} &    \multicolumn{3}{c}{\nodata} & &     2.19 &    (2.10 &    2.26) &    15.76 &   (13.96 &   16.98) &  736.11 &  675 \\
Filament (S) &     0.86 &    (0.75 &    0.98) &    13.76 &    (6.00 &   27.73) &    \multicolumn{3}{c}{\nodata} &    \multicolumn{3}{c}{\nodata} &    \multicolumn{3}{c}{\nodata} &    \multicolumn{3}{c}{\nodata} & &     2.15 &    (2.11 &    2.20) &    13.65 &   (12.98 &   14.33) &  549.84 &  400 \\
\cutinhead{(i) LB foreground model, \citet{wilms00} abundances}
G048+37 (X)  &     1.30 &    (1.24 &    1.36) &    15.21 &   (13.21 &   17.49) &    \multicolumn{3}{c}{\nodata} &    \multicolumn{3}{c}{\nodata} &    \multicolumn{3}{c}{\nodata} &    \multicolumn{3}{c}{\nodata} & &     2.22 &    (2.20 &    2.33) &    11.97 &   (11.07 &   13.46) & 1996.52 & 1824 \\
MBM 12 (S)   &     1.17 &    (1.09 &    1.28) &    12.16 &    (9.11 &   15.42) &    \multicolumn{3}{c}{\nodata} &    \multicolumn{3}{c}{\nodata} &    \multicolumn{3}{c}{\nodata} &    \multicolumn{3}{c}{\nodata} & &     1.65 &    (1.41 &    1.94) &     4.02 &    (2.34 &    6.19) &  394.00 &  360 \\
MBM 16 (S)   &     1.03 &    (0.86 &    1.13) &    42.95 &   (28.82 &   95.98) &    \multicolumn{3}{c}{\nodata} &    \multicolumn{3}{c}{\nodata} &    \multicolumn{3}{c}{\nodata} &    \multicolumn{3}{c}{\nodata} & & \multicolumn{3}{c}{2.10\tablenotemark{f}} &     2.80 &    (1.16 &    4.43) &  277.00 &  268 \\
MBM 20 (X)   &     0.81 &    (0.77 &    0.82) &    16.46 &   (13.63 &   18.05) &    \multicolumn{3}{c}{\nodata} &    \multicolumn{3}{c}{\nodata} &    \multicolumn{3}{c}{\nodata} &    \multicolumn{3}{c}{\nodata} & &     1.86 &    (1.83 &    1.88) &     9.07 &    (8.77 &    9.19) & 3989.33 & 3603 \\
MBM 20 (S)   &     0.66 &    (0.55 &    0.81) &   157.08 &   (48.14 &  442.71) &    \multicolumn{3}{c}{\nodata} &    \multicolumn{3}{c}{\nodata} &    \multicolumn{3}{c}{\nodata} &    \multicolumn{3}{c}{\nodata} & &     2.07 &    (1.88 &    2.20) &     5.61 &    (4.72 &    6.49) &  387.02 &  363 \\
G236+38 (S)  &     0.91 &    (0.64 &    1.08) &   104.82 &   (58.21 &  691.27) &    \multicolumn{3}{c}{\nodata} &    \multicolumn{3}{c}{\nodata} &    \multicolumn{3}{c}{\nodata} &    \multicolumn{3}{c}{\nodata} & &     1.93 &    (1.70 &    2.20) &    10.46 &    (5.42 &   18.03) &  257.36 &  232 \\
Filament (X) &     1.57 &    (1.52 &    1.62) &    13.42 &   (11.12 &   16.23) &    \multicolumn{3}{c}{\nodata} &    \multicolumn{3}{c}{\nodata} &    \multicolumn{3}{c}{\nodata} &    \multicolumn{3}{c}{\nodata} & &     2.70 &    (2.59 &    2.82) &     9.97 &    (8.26 &   10.78) &  725.84 &  675 \\
Filament (S) &     0.87 &    (0.66 &    1.24) &    18.48 &    (7.11 &   70.11) &    \multicolumn{3}{c}{\nodata} &    \multicolumn{3}{c}{\nodata} &    \multicolumn{3}{c}{\nodata} &    \multicolumn{3}{c}{\nodata} & &     2.20 &    (2.14 &    2.30) &    12.66 &   (10.65 &   13.70) &  550.06 &  400
\enddata
\tablecomments{Values in parentheses are the 90\%\ confidence intervals.}
\tablenotetext{a}{The X or S in parentheses indicates whether the shadow was observed with \xmm\ or \suzaku, respectively.}
\tablenotetext{b}{For the LB foreground model, this is the foreground emission measure, \EMfg, in units of $10^{-3}~\emismeas$.
  For the ACX-SWCX foreground model, this is the normalization of the foreground component, in units of $10^{-8}~\parcminsq$.}
\tablenotetext{c}{Foreground \Kalpha\ intensity. We have summed the intensities of the resonance,
  intercombination, and forbidden lines.}
\tablenotetext{d}{Foreground \Lyalpha\ intensity.}
\tablenotetext{e}{If this line lies below the energy range used for a given fit, its intensity is
  constrained by emission from that line spreading into the included energy range due to the
  finite spectral resolution, and/or by the higher-energy K lines, via the \citet{cumbee14} CX
  line ratios.}
\tablenotetext{f}{Value held fixed during fitting.}
\end{deluxetable}

    \clearpage
  \end{landscape}
}

\end{document}